\documentclass[11pt]{article}
\usepackage{eurosym}
\usepackage{amsfonts}
\usepackage{amssymb}
\usepackage{graphicx}
\usepackage{amsmath}
\usepackage{makeidx}
\usepackage{indentfirst}
\usepackage[T1]{fontenc}
\usepackage[utf8]{inputenc}

\newcounter{resultnum}[section]
\newcounter{conclusionnum}[section]
\newcounter{conditionnum}[section]
\newcounter{conjecturenum}[section]
\newcounter{examplenum}[section]
\newcounter{exercisenum}[section]
\newcounter{lemmanum}[section]
\newcounter{notationnum}[section]
\newcounter{theoremnum}[section]
\newcounter{definitionnum}[section]
\newcounter{corollarynum}[section]
\newcounter{remarknum}[section]
\newcounter{propositionnum}[section]
\newcounter{acknowledgementnum}[section]
\newcounter{algorithmnum}[section]
\newcounter{axiomnum}[section]
\newcounter{casenum}[section]
\newcounter{claimnum}[section]
\newcounter{summarynum}[section]
\newcounter{problemnum}[section]
\begin{document}
\title{Inconsistencies of nonmetric Einstein--Dirac-Maxwell theories and a cure for geometric flows of f(Q) black ellipsoid, toroid and wormhole solutions}
\date{March 4, 2025}
\author{ \textbf{Sergiu I. Vacaru} \thanks{emails: sergiu.vacaru@fulbrightmail.org ; sergiu.vacaru@gmail.com } \\
{\small \textit{\ Kocaeli University, Department of Physics, 41380, Izmit,
Turkey; }}\\
{\small \textit{Taras Shevchenko National University of Kyiv, Astronomical
Observatory, Kyiv, Ukraine;  }}\\
{\small \textit{Department of Physics, California State University at
Fresno, Fresno, CA 93740, USA }}
 }
\maketitle

\begin{abstract}
Many papers on modified gravity theories (MGTs), and metric-affine geometry have been published. New classes of black hole (BH), wormhole (WH), and cosmological solutions involving nonmetricity and torsion fields were constructed. Nevertheless, the fundamental problems of formulating nonmetric Einstein-Dirac-Maxwell  (EDM), equations, and study of important nonmetric gravitational, electromagnetic and fermion effects, have not been solved in MGTs. The main goal of this work is to elaborate on a model of nonmetric EDM theory as a generalization of f(Q) gravity. We develop our anholonomic frame and connection deformation method which allows us to decouple in general form and integrate nonmetric gravitational and matter fields equations. New classes of generated quasi-stationary solutions are defined by effective sources with Dirac and Maxwell fields, nonmetricity and torsion fields, and generating functions depending, in general, on all space-time coordinates. For respective nonholonomic parameterizations, such solutions describe nonmetric EDM deformations of BH and cosmological metrics. Variants of nonmetric BH, WH and toroid  solutions with locally anisotropic polarizations of the gravitational vacuum and masses of fermions, and effective electromagnetic sources, are constructed and analyzed. Such nonmetric deformed physical objects can't be characterized in the framework of the Bekenstein-Hawking paradigm if certain
effective horizon/ holographic configurations are not involved. We show how to define and compute other types of nonmetric geometric thermodynamic variables using generalizations of the concept of G. Perelman W-entropy.

\end{abstract}

\tableofcontents

%%%%%

%%%

\section{Introduction: Ambiguities with nonmetric spinors and EDM theories}

\label{sec1} Nonmetric geometric and gravity theories have been  known since 1918 when H. Weyl considered nonmetric spaces with the aim to unify gravity and electromagnetism \cite{weyl1918}. In such theories, the covariant derivative of the metric field is not zero. The direction of nonmetric gravity was
almost ignored for decades because of critics by A. Einstein and W. Pauli.
We cite \cite{hehl95,vmon05}, as early comprehensive reviews on so-called
metric-affine geometry and gravity (MAG) theories and \cite{lheis23} as a
recent review on $f(Q)$-gravity. The mentioned works are devoted to
geometric and physical models with effective nontrivial torsion and
nonmetricity fields. For (non) holonomic Lorentz manifolds and (co) tangent
Lorentz bundles, relativistic generalizations of the Einstein gravity theory
on metric and nonmetric geometric flows Finsler-Lagrange-Hamilton spaces
were elaborated \cite{vmon05,vplb10,sv12}. Those geometry and physics works
studied nonmetric generalizations of the Hamilton-Perelman theory of Ricci
flows \cite{perelman1,hamilton82,svnonh08}. References \cite%
{harko21,iosifidis22,khyllep23,ghil23,koussour23,jhao22,de22,lheis23}
contain original results and recent reviews on nonmetric modified gravity theories (MGTs), in general,
with nonzero torsion.

\vskip4pt There is a gap in modern literature on nonmetric MGTs concerning the problem of generalizing for metric-affine spaces the concept of spinors and formulating analogues of nonmetric Dirac equations; or other matter field equations involving fermionic fields; and gravitational field equations encoding nonmetricity with spinor variables. Such works have not been published and solutions describing nonmetric Einstein-Dirac  (ED) systems were not found. For Finsler gravity theories with nonmetric linear connections, the problem of
definition of nonmetric spinors was analyzed in \cite{vplb10}.

\vskip4pt Certain preliminary studies of the motion of spinning particles and variants of the Dirac equations in background spacetimes with torsion and nonmetricity (i.e. in non-Riemannian geometries) were performed during the last 20 years \cite{adak02,formiga12,berredo18}. Here we also note that  the monograph  \cite{vmon05} summarizes a series of our works on nonholonomic spinors and Dirac operators and in various types of nonmetric Finsler-Lagrange-Hamilton spaces constructed as (co) tangent Lorentz bundle generalizations of the Einstein gravity and MGTs. Similar geometric and analytic constructions can be performed, for instance, for the $f(Q)$-gravity effectively modelled on nonholonomic Lorentz manifolds. The main issue of \cite{adak02,formiga12} is that those "early" works have not analyzed and solved the problem of incompatibility of tetradic and gamma-matrix decompositions of spacetime metrics with parallel transports with arbitrary affine linear connections and respective spin coefficients. In \cite{berredo18}, the ambiguity in the definition of Fock spin coefficients for arbitrary affine connections is stated but there is no proposal on how to cure it. None of the self-consistent variants of nonmetric generalized Einstein-Dirac equations and their physically important solutions have been considered in the mentioned papers. We discuss and solve such problems in sections 1.1, 2.2, 2.3, 3.2 and 4 of this work.

\vskip4pt The main goal of this work is to elaborate on nonmetric generalizations of the Einstein-Dirac-Maxwel (EDM) theory for a class of MAGs but not considering Finsler MGTs (which will be studied in our future partner works). We shall develop  new geometric methods for constructing exact and parametric solutions of nonmetric EDM equations and their geometric flows and provide explicit examples of quasi-stationary solutions describing black hole (BH), black ellipsoid (BE), black torus (BT), and wormhole (WH), configurations encoding nonmetricity and Dirac-Maxwell interactions.

\vskip4pt In this paper, we do not provide a historical (comprehensive) list of publications on nonmetricity MGTs and non-standard particle physics nor discuss recent attempts to consider applications, for instance, of $f(Q)$ gravity in modern accelerating cosmology and dark energy (DE), and dark matter (DM), physics. For details, motivations and nonholonomic geometric methods to be used in the next (sub) sections, we cite our recent partner works \cite{lb24epjc1,lb24epjc2,lb24grg} and references therein. %%%%%

\subsection{Problems with the definition of spinors and Diract operators on nonmetric spacetimes}

Let us state in explicit form the problem of definition of nonmetric Dirac spinor equations: For a general metric-affine structure $(g=\{g_{\beta \gamma }\},D=\{\Gamma _{\ \beta \gamma }^{\alpha }\})$ on a four-dimensional, 4-d, manifold $V$ of necessary smooth class, a nontrivial nonmetricity field $Q=\{Q_{\alpha \beta \gamma }:=D_{\lambda }g_{\beta\gamma}\}\neq 0$ is treated as an additional fundamental geometric object. It has to be considered together with the metric and linear connection structure and respective, in general, nontrivial torsion and curvature tensors, which will be defined in the next section. We use standard notations for the general relativity (GR) and MGTs but follow additionally the conventions on nonholonomic MAGs introduced in \cite{lb24epjc1,lb24epjc2} considering nonlinear connections. Corresponding nonholonomic decompositions and distortions of linear connections allow us to apply the anholonomic frame and connection deformation method, AFCDM. We elaborated such a geometric and analytic method for constructing exact and parametric off-diagonal solutions in  GR and MAGs with nontrivial torsion and nonmetricity structures. %%%

\vskip4pt For a MAG theory generalizing the Einstein gravity theory, we consider a symmetric pseudo-Riemannian metric $g=\{g_{\beta \gamma }\}$ of Lorentz signature $(+++-)$. A corresponding Levi-Civita (LC) connection $\nabla = \{\nabla _{\lambda }\}$ can be considered as an additional one
defined only by $g$ and which via a distortion tensor is related to a prescribed  affine/linear connection 
$D=\{ D_\beta \}$. Here we note that a $\nabla$ is uniquely defined from the conditions of zero torsion, 
$T_{\ \beta \lambda }^{\alpha }[\nabla ]=0,$ and metric compatibility, 
$Q[\nabla ]=\{\nabla _{\lambda }g_{\beta \gamma }\}=0$. We shall formulate our nonmetric theories as effective models on 4-d Lorentz manifolds supposing that for $Q=0$ and $T=0$  the Einstein gravity theory is obtained. We shall also omit abstract or coordinate indices for geometric objects if that does not result in ambiguities. The Einstein convention on summarizing up-low repeated indices will be used. We consider
that readers are familiar with the main concepts of mathematical relativity, Clifford and spinor structures, Dirac equations in GR, and exact solutions provided in monographs \cite{misner,hawking73,wald82,kramer03}.

\vskip4pt Using a nonsingular Lorentz metric structure and its inverse metric
tensor, $g^{\mu \nu }$, we can introduce tetradic fields $e=\{e_{\quad 
\underline{\mu }}^{\mu }\},$ when $e_{\quad \underline{\mu }}^{\mu }e_{\quad 
\underline{\nu }}^{\nu } \eta ^{\underline{\mu }\underline{\nu }}=g^{\mu
\nu} $ and define respective Dirac gamma matrices $\gamma ^{\mu }=\gamma ^{%
\underline{\mu }}e_{\quad \underline{\mu }}^{\mu }$. We have 
\begin{equation}
\gamma ^{\mu }\gamma ^{\nu }+\gamma ^{\nu }\gamma ^{\mu }=-2g^{\mu \nu }I,
\label{gammamatr}
\end{equation}%
where $I$ is the unity $4\times 4$ matrix and $\eta ^{\underline{\mu }%
\underline{\nu }}=diag(1,1,1,-1)$.\footnote{\label{fngamma}A choice for such
gamma matrices is $\gamma _{\underline{i}}=(\gamma _{i^{\prime }}=\left( 
\begin{array}{cc}
O & \sigma _{i^{\prime }} \\ 
-\sigma _{i^{\prime }} & O%
\end{array}%
\right) ,\gamma _{4}=\left( 
\begin{array}{cc}
O & I \\ 
I & O%
\end{array}%
\right) ),$ where $\sigma _{i^{\prime }}$ (for $i^{\prime }=1,2,3$) are the
Pauli matrices $\sigma _{1}=\left( 
\begin{array}{cc}
0 & 1 \\ 
1 & 0%
\end{array}%
\right) ,$ $\sigma _{2}=\left( 
\begin{array}{cc}
0 & -i \\ 
i & 0%
\end{array}%
\right) $ and $\sigma _{3}=\left( 
\begin{array}{cc}
1 & 0 \\ 
0 & -1%
\end{array}%
\right) ;$ and, respectively, $I$ and $O$ are the identity and zero $2\times
2$ matrices. We can consider $\gamma _{\underline{i}}\gamma _{\underline{j}%
}+\gamma _{\underline{j}}\gamma _{\underline{i}}=-2\eta _{\underline{i}%
\underline{j}}$ for the Minkowski spacetime metric $\eta _{\underline{i}%
\underline{j}}=diag(1,1,1,-1).$} The condition $Q[\nabla ]=0$ allows us to
define the gamma matrices and Dirac 4-spinors $\psi $ in a metric compatible
form when the transports are defined and computed by using the Dirac
operator $\mathcal{\breve{D}}_{\alpha }[\nabla ]$ determined by certain
geometric data $\left( e,\nabla \right) .$ This way, we can introduce as in
GR a well-defined Dirac equation 
\begin{equation}
\lbrack i\ \hbar \gamma ^{\alpha }\mathcal{\breve{D}}_{\alpha }-m_{0}]\ \psi
=0  \label{diracgr}
\end{equation}%
for a general relativistic fermion field $\Psi $ with mass $m_{0},$ where $%
\hbar $ is the Planck constant. Such a $\mathcal{\breve{D}}$ does not
involve a general affine connection $D$.

\vskip4pt For a general metric-affine structure $(g,D)$, with $Q\neq 0,$ the formula for a gamma-matrix decomposition of a metric tensor (\ref{gammamatr}) is not preserved under covariant transports by an arbitrary $D$. So, we have a conceptual problem of how to define in a unique way (for general nonmetric spaces) an analogue of the Dirac equation (\ref{diracgr}), see details in section \ref{sscanondir}. If $D$ is with a nontrivial torsion, but metric compatible, the problem can be solved for various types of nonassociative and noncommutative and generalized Finsler/ Einstein-Cartan/ string MGTs etc., as it was discussed in \cite{lb24grg} and references therein. Unfortunately, to formulate certain general geometric and physical
principles which would allow to define in a unique form the Dirac equations for general MAG theories  is not possible. We have to consider additional mathematical constructions and physically motivated assumptions for elaborating self-consistent and viable nonmetric generalizations of the Einstein-Dirac (ED), theory. During the last decade hundreds of papers on nonmetric MAG theories, including $f(Q)$-modifications, were published and devoted to the study of possible implications in modern cosmology and DE and DM physics (see above-cited papers and reviews). Nevertheless, up till the present, the issue of nonmetric generalizations of the model of standard particle physics, with fermions, has not been analyzed and solved by experts in the mentioned directions of mathematical and theoretical physics and geometry.

\subsection{The aims, main hypothesis, and structure of the paper}

Let us state the \textit{general goal of our research program} \cite%
{vmon05,vplb10,sv12,lb24epjc1,lb24epjc2} on constructing nonmetric geometric flow models and MGTs, and elaborating possible applications in modern cosmology and astrophysics: \textsf{To formulate certain general principles for constructing geometric and quantum information flow models and physical viable classical and quantum models of nonmetric Einstein-Yang-Mills-Dirac-Higgs (EYMDH), theories in a self-consistent mathematical form.} We shall also propose and analyze analytic and geometric computational tests and explanations of new physical effects which can be performed using exact and parametric off-diagonal solutions of respective systems of nonlinear partial differential equations, PDEs.

\vskip4pt In this work, the \textit{main purposes} are to perform a self-consistent generalization of the Dirac equation in GR (\ref{diracgr}) to certain nonmetric variants with Dirac operators on MAGs (for instance, in 
$f(Q)$ gravity ) and study related effective models of nonmetric Einstein-Dirac-Maxwell (EDM) theories. For such nonlinear systems of PDEs involving nonmetric and nonholonomic spinor and Abelian gauge field
structures, we shall construct and study the physical properties of new classes of exact off-diagonal solutions. Respective quasi-stationary solutions describe nonmetric  BHs (and ellipsoid/rotoid configurations; in brief, BEs); nonmetric BT configurations; and nonmetric WHs.

\vskip4pt Nonmetric modifications of gravity and matter field theories can be performed if distortions of affine connections considered in GR or in metric-compatible MGTs with nontrivial torsion fields are considered. For an LC-connection $\nabla [g]$, we can construct a metric-affine space endowed with a linear bi-connection structure $(\nabla, D)$. This involves also a distortion tensor $Z=\{Z_{\ \beta \gamma }^{\alpha }\}$, when 
\begin{equation}
\left( \nabla \lbrack g],D=\nabla +Z;\ Q=Dg\neq 0\right) .  \label{bicon}
\end{equation}
A $\nabla \lbrack g]$ can be always defined geometrically on a metric-affine manifold $V$ for a given metric field $g$. The main issue for constructing a physically important nonmetric MGT is to decide on what type of generalized affine connection $D$ (and corresponding nontrivial torsion, $T[D],$ and nonmetricity, $Q[D]$ fields) we use for elaborating our geometric and physical models. Such an MGT can be formulated following generalized abstract geometric and variational principles as in \cite{misner} but in a form explaining, for instance, new observational data in modern cosmology and DE and DM physics. Using a distortion tensor $Z,$ nonmetric MGTs can be modelled by effective geometric constructions on a Lorentz manifold. In such
an approach, some (effective) matter sources for modified Einstein equations contain additional terms determined by $Z$ and $Q$ fields and standard or distorted matter fields. For instance, we can consider nonmetric versions of scalar and spinor fields, various types of nonmetric gauge fields etc. and elaborate on nonmetric modifications both of the standard model of particle physics and GR. The geometric deformations of gravitational and matter field equations with $\nabla \rightarrow D=\nabla +Z $ are defined in some
straightforward forms if we do not consider spinors/ fermion fields.

\vskip4pt To introduce in a self-consistent form certain Clifford and spinor structures in MAGs we have to formulate additional geometric and physical principles which allow to define physical viable nonmetric distortions of the Einstein-Dirac equations. Such issues appear always if certain Lorentz manifolds or (co) tangent Lorentz bundles are enabled with additional nonmetric connection structures \cite{vplb10,vmon05}. To decide on how the distortions $D=\nabla +Z$ and $Q$-fields have to be included in a physically viable nonmetric EDM theory, we can search for certain well-defined solutions of respective physically important nonlinear systems of nonlinear PDEs. Analyzing corresponding physical effects, we can decide on viability
of certain type of geometric distortions and effective theories.

\vskip4pt We found new classes of off-diagonal solutions for describing BH, WH and locally anisotropic cosmological configurations modified by $Q$-fields by applying the anholonomic frame and connection deformation method, AFCDM \cite{lb24epjc1,lb24epjc2}. In this work, we generalize those geometric methods for constructing exact and parametric solutions of nonmetric EDM equations. We prove that nontrivial $Q$-fields result in locally anisotropic modifications of the mass of fermions and nonmetric
effective sources for the Maxwell equations. Considering corresponding nonlinear symmetries, we model nonmetric off-diagonal polarizations of the gravitational vacuum, which result also in certain effective modifications of some (effective) cosmological constants. Such physical effects are different, for instance, from those considered as a possible origin of mass in gauge theories of scale invariance due to non-metricity \cite{ghil23}. We cite \cite{bubuianu17,vacaru18,vacaru20} for recent reviews of the AFCDM and
developments for MGTs and GR. Those results and methods are for constructing metric-compatible off-diagonal solutions on Lorentz manifolds or (co) tangent Lorentz bundles and to study possible applications in modern cosmology, astrophysics and (non) standard particle physics.

\vskip4pt For MAGs, we can formulate an EDM theory using the so-called canonical d-connection which is metric-compatible. Such constructions are provided in \cite{lb24grg} in a more general form including nonassociative and noncommutative structures determined by R-flux deformations in string theory. In this work, we consider nonmetric deformations instead of nonassociative R-flux deformations. The article is also formulated as a status report on recent generalizations of the AFCDM and applications for constructing off-diagonal solutions defining nonmetric geometric flow models of EDM systems and MGTs. This requests an increase in the number of self-citations when references to the works by other authors will be provided only if certain their very important results are developed in this paper.

\vskip4pt Generic off-diagonal solutions in nonmetric geometric flow theory and nonmetric EDM do not involve, in general, any hypersurface configurations or certain duality and holographic properties. Such solutions do not have a thermodynamic interpretation in the framework of the Bekenstein-Hawking paradigm \cite{bek2,haw2}. Nevertheless, this is possible in the framework of nonmetric and nonholonomic geometric flow and EDM theories \cite{sv12,lb24epjc1,lb24epjc2,lb24grg,svnonh08,vacaru20} by generalizing the concept of G. Perelman W-entropy \cite{perelman1}.

\vskip4pt We generalize for nonmetric EDM systems the main \textbf{Hypothesis} formulated in certain different forms in our partner works \cite{vmon05,vplb10,sv12,lb24epjc1,lb24epjc2} and structured here in such a form:
\begin{enumerate}
\item \textsf{Mathematically self-consistent nonmetric geometric flow theories involving physically viable nonmetric EDM configurations (we can consider more general nonmetric EYMDH systems) can be constructed for $\mathit{Q}$--deformations of geometric and physical objects on nonholonomic
Lorentz manifolds and cotangent Lorentz bundles. Such nonmetric geometric generalizations are performed by using distortions of connections for classes of well-defined metric-compatible EYMDH theories, which may involve, or not, nontrivial torsion structures. }
\item \textsf{Using nonholonomic dyadic variables and canonically adapted (non) linear connection structures and applying the AFCDM, the corresponding nonmetric geometric evolution flow equations, dynamical nonmetric EYMHD equations, or metric and nonmetric EDM equations, can be integrated in
certain general off-diagonal forms. Respective classes of quasi-stationary or locally anisotropic cosmological solutions are determined by generating functions and generating sources depending on all spacetime coordinates. }
\item \textsf{New classes of off-diagonal solutions for nonmetric geometric flow, nonholonomic Ricci soliton configurations for EDM systems etc. are characterized by generalized G. Perelman thermodynamic variables. They describe new nonmetric physical effects and observational features of the DE and DM theories and non-standard models of particle physics with nonmetricity fields. This allows us to analyze and decide on the issue of what type of nonmetric Dirac equations and nonmetric Einstein-Dirac systems
are physically well-motivated. }
\end{enumerate}

\vskip5pt The goals and structure of the article are organized as follow:\ 

In section \ref{sec2}, we outline the nonholonomic geometry of nonmetric MGTs and $f(Q)$ gravity using geometric constructions with metrics and affine connections adapted to a nonlinear connection structure. Then we introduce the nonholonomic canonical Dirac operators and formulate a model of metric-compatible EDM theory defined in nonholonomic canonical variables which can be extended for $f(Q)$ gravity. This consists of the \textbf{first aim} of this paper. A model of nonmetric EDM theory is constructed in the third subsection (the \textbf{second aim}).

Two main purposes of this work are stated for section \ref{sec3}. We formulate the theory of nonmetric geometric flows of EDM systems in the first subsection (consisting of the \textbf{third aim}). We define there the $f(Q)$ deformed G. Perelman functionals and thermodynamic variables and derive the metric geometric flow equations for nonmetric EDM systems. The \textbf{fourth aim} of this work (in the second subsection) is to generalize the AFCDM to generate exact and parametric off-diagonal solutions for quasi-stationary nonmetric EDM configurations and their geometric flow evolution. We also show how to compute nonmetric-induced anisotropic polarizations of the masses of Dirac fermions and nonmetric-induced electromagnetic sources. Then certain general formulas for computing G. Perelman thermodynamic variables of nonmetric quasi-stationary EDM solutions with effective cosmological constants are formulated.

We construct explicit classes of exact and parametric quasi-stationary solutions (the \textbf{fifth aim}) in section \ref{sec4}. The first class of solutions describes off-diagonal nonmetric deformations of Kerr de Sitter
solution to ellipsoidal EDM configurations. The second class of solutions (with different variables and nonlinear symmetries)  is considered for nonmetric locally anisotropic WHs. Then, we study  toroid configurations and respective black torus (BT) solutions  encoding nonmetric EDM data. For the mentioned classes of quasi-stationary solutions, the nonmetric $f(Q)$ and Dirac-Maxwell contributions are encoded into respective off-diagonal terms, generating functions and effective sources which are related to corresponding classes of nonlinear symmetries.

Finally, we shall conclude the results and discuss further perspectives for nonmetric EDM theories with extensions to nonmetric EYMDH models and applications in modern cosmology and quantum information theories in section \ref{sec5}.

\section{Nonholonomic (non) metric Einstein and Dirac-Maxwell equations}

\label{sec2}

In this section, we elaborate on a model of nonmetric Einstein-Dirac-Maxwell, EDM, theory. The geometric constructions are performed in nonholonomic (2+2) variables and using distortions of affine
connections for the $f(Q)$ gravity as in \cite{lb24epjc2}. We cite \cite{lb24epjc1,lb24grg} for details on nonmetric geometric flows and nonholonomic MAGs and similar nonassociative generalizations of the EDM
theory. Such a formulation allows to application of the AFCDM and construction of quasi-stationary off-diagonal solutions for nonmetric EDM systems in section \ref{sec3}.

\subsection{Nonlinear connections and nonmetric deformations of the Einstein equations}

Let us consider a four dimensional, 4-d, and of necessary smooth class \textit{metric-affine } spacetime manifold $\ _{1}^{3}\mathcal{V}:=(g,D=\nabla +Z)$. In general, this defines a nonmetric extension of the
concept of curved spacetime in GR. In these formulas, we use an LC-connection structure 
$\nabla =\{\breve{\Gamma}_{\ \beta \gamma }^{\alpha }(u)\}$ and a distortion tensor 
$Z=\{Z_{\ \beta \gamma }^{\alpha }(u)\}$, for a respective bi-connection structure (\ref{bicon}). The left abstract labels for $\ _{1}^{3}\mathcal{V} $ emphasize that we obtain a  4-d Lorentz manifold 
$\ _{1}^{3}V$ defined by a symmetric metric tensor $g=\{g_{\alpha \beta}(u^{\gamma })\}$ of signature $(+,+,+,-)$ if the distortion tensor vanishes, $Z=0.$ The indices of geometric objects on 
$\ _{1}^{3}\mathcal{V}$ can be abstract or coordinate ones.\footnote{We label local coordinates as $u^{\alpha }=(u^{\acute{\imath}},u^{4}=t)=(x^{i},y^{a}),$ for $\acute{\imath}=1,2,3;$ and follow such
conventions: for local frames/ coordinates/ indices that typical 3-d space indices run values of type $\acute{\imath}=1,2,3;$ for $u^{4}=y^{4}=ct$; the light velocity constant $c$ can be always fixed as $c=1$ for corresponding systems of unities and coordinates. Indices of type $i=1,2$ and $a=3,4$ will
be used for a conventional 2+2 splitting when, in brief, we shall write correspondingly $u=(x,t)=(x,y).$%
\par
The local frames $e_{\alpha }=e_{\ \alpha }^{\alpha ^{\prime }}(u)\partial
_{\alpha ^{\prime }}$ and (dual) frames, or co-frames, $e^{\beta }=e_{\beta
^{\prime }}^{\ \beta }(u)d^{\beta ^{\prime }}$ can be related to respective
coordinate (co) frames $\partial _{\alpha ^{\prime }}=\partial /\partial
u^{\alpha ^{\prime }}$ and $d^{\beta ^{\prime }}=du^{\beta ^{\prime }}$
using certain matrices $e_{\ \alpha }^{\alpha ^{\prime }}(u)$ and $e_{\beta
^{\prime }}^{\ \beta }(u)$ defining some tetradic (equivalently, vierbein
coefficients). For various types of underlined/ primed indices, we may use
similar conventions and necessary 3+1 and/or 2+2 splitting as in \cite%
{misner,lb24epjc1,lb24epjc2,lb24grg}.}

In general, we can define $D=\{\Gamma _{\ \beta \gamma }^{\alpha }(u)\}$ as an independent linear (affine) connection structure on $\ _{1}^{3}\mathcal{V} $ and formulate various types of metric-affine geometric flow and gravity models, i.e. MAG theories, determined by geometric data $(g,D).$ Nevertheless, for elaborating on self-consistent and physical important geometric of physical models with deformations/ distortions of geometric objects in GR by nonzero torsion and/or nontrivial nonmetric fields, we have to consider theories with bi-connection structure $(\nabla \lbrack g],D).$ Using additional assumptions involving certain abstract geometric principles, variational calculi for respective couplings and effective
Lagrangians/ Hamiltonians, we can define various types of geometric objects (curvatures, torsions, nonmetricity fields, etc.). Then, we can postulate in abstract geometric forms, or derive variationally, some corresponding physical important systems of nonlinear PDEs for modified geometric flow and generalized Einstein and matter field equations.

Studying and comparing the physical implications of different  MGTs by constructing and analysing the physical properties of various classes of physical important solutions  is important. For instance, various generalized  BH,  WH, and locally anisotropic cosmological solutions \cite{harko21,iosifidis22,khyllep23,ghil23,koussour23,jhao22,de22,lheis23,lb24epjc1,lb24epjc2} were generated. In most general forms, such solutions are described by generic off-diagonal metrics\footnote{%
a metric $g$ is generic off-diagonal if it can't be diagonalized by some coordinate transform in a finite spacetime region} and generalized (non) linear connections, when the corresponding coefficients are found in exact or parametric forms by applying the AFCDM. In this work, we develop and apply such geometric methods for elaborating on nonmetric EDM theories.

\subsubsection{N-connections on metric-affine spaces}

We can always define on $\ _{1}^{3}\mathcal{V}$ a nonlinear connection, N-connection, structure, which can be defined in global form as a Whitney (direct) sum using the tangent bundle $T\ _{1}^{3}\mathcal{V}$: 
\begin{equation}
\mathbf{N}:\ T\ _{1}^{3}\mathcal{V}=h\ _{1}^{3}\mathcal{V}\oplus v\ _{1}^{3}%
\mathcal{V}.  \label{ncon}
\end{equation}%
In local form, an N-connection is given by its coefficients $N_{i}^{a}$ when 
$\mathbf{N}=N_{i}^{a}(x,y)dx^{i}\otimes \partial /\partial y^{a}$. A N-connection (\ref{ncon}) states a conventional horizontal, h, and vertical, v, splitting ( h- and v--decomposition) into respective 2-d subspaces, $hV$ and $vV.$ Such a nonholonomic structure with 2+2 splitting can be defined by local bases, $\mathbf{e}_{\nu },$ and (dual) co-bases, $\mathbf{e}^{\mu },$ 
\begin{align}
\mathbf{e}_{\nu }& =(\mathbf{e}_{i},e_{a})=(\mathbf{e}_{i}=\partial
/\partial x^{i}-\ N_{i}^{a}(u)\partial /\partial y^{a},\ e_{a}=\partial
_{a}=\partial /\partial y^{a}),\mbox{ and  }  \label{nader} \\
\mathbf{e}^{\mu }& =(e^{i},\mathbf{e}^{a})=(e^{i}=dx^{i},\ \mathbf{e}%
^{a}=dy^{a}+\ N_{i}^{a}(u)dx^{i}).  \label{nadif}
\end{align}%
The term nonholonomic (equivalently, anholonomic) is used because an N-elongated basis (\ref{nader}) satisfies certain anholonomy relations 
\begin{equation}
\lbrack \mathbf{e}_{\alpha },\mathbf{e}_{\beta }]=\mathbf{e}_{\alpha }%
\mathbf{e}_{\beta }-\mathbf{e}_{\beta }\mathbf{e}_{\alpha }=W_{\alpha \beta
}^{\gamma }\mathbf{e}_{\gamma }.  \label{anhrel}
\end{equation}%
In these formulas, the nontrivial anholonomy coefficients are computed 
$W_{ia}^{b}=\partial _{a}N_{i}^{b},W_{ji}^{a}=\Omega _{ij}^{a}=
\mathbf{e}_{j}\left( N_{i}^{a}\right) -\mathbf{e}_{i}(N_{j}^{a}),$ where $\Omega _{ij}^{a}$ defines the coefficients of the so-called N-connection curvature. If all $W_{ia}^{b}$ are zero for a 
$\mathbf{e}_{\alpha },$ we say that such a N-adapted base is holonomic; we can write such bases using partial derivatives $\partial _{\alpha }$ with $N_{i}^{a}=0.$ The coefficients $N_{j}^{a}$ may be nontrivial even all $W_{\alpha \beta }^{\gamma }=0.$

Decomposing geometric objects (tensors, connections etc.) on  $\ _{1}^{3}\mathcal{V}$ with respect to N-adapted bases (\ref{nader}) and (\ref{nadif}), we formulate a nonholonomic dyadic formalism for metric-affine geometry. In such cases, the terms d-objects, d-tensors, d-connections etc. can be used if we have to emphasize that the geometric constructions are performed in a distinguished form (by an N-connection structure), in brief, a d-form. For nonholonomic metric-affine manifolds enabled with N-connection structure
(\ref{nader}), we shall use boldface symbols and write, for instance, $\mathbf{V}$ instead of 
 $\ _{1}^{3}\mathcal{V}$ (in particular, instead of  $\ _{1}^{3}V$) considering that such generalized spacetimes can be enabled with double conventional $2+2$ and $3+1$ splitting.\footnote{The dyadic variables are important for finding solutions but the ADM type (3+1) decompositions \cite{misner} are useful for elaborating thermodynamic and statistical models.} The geometric constructions can be performed in
N-adapted forms for respective nonholonomic tangent, $T\mathbf{V,}$ and cotangent, $T^{\ast}\mathbf{V}$, bundles; their tensor products, $T\mathbf{V\otimes }T^{\ast }\mathbf{V,}$ etc. To emphasize that certain geometric objects are N-adapted we shall use boldface symbols. For instance, we can
write a d--vector as $\mathbf{X}=(hX,vX)$ and a second rank d-tensor as $\mathbf{F}=(hhF,hvF,vhF,vvF),$ see details in \cite{vmon05,lb24epjc1,lb24epjc2,lb24grg}. In this subsection, we summarize necessary definitions and formulas for nonholonomic dyadic decompositions of fundamental geometric d-objects on a nonholonomic \textit{metric-affine }manifold $\mathbf{V}=\{\ _{1}^{3}\mathcal{V\}}$ enabled with N-connection structure $\mathbf{N}$ (\ref{ncon}).

\subsubsection{Metrics and affine connections adapted to N-connection structures}

A metric structure defined by a symmetric tensor $g\in T^{\ast }(\ _{1}^{3}%
\mathcal{V)}\otimes T^{\ast }(\ _{1}^{3}\mathcal{V)}$ can be written as a
d-tensor $\mathbf{g}\in T^{\ast }\mathbf{V\otimes }T^{\ast }\mathbf{V}$ and
parameterized in three equivalent forms: 
\begin{align}
g& =g_{\alpha ^{\prime }\beta ^{\prime }}(u)e^{\alpha ^{\prime }}\otimes
e^{\beta ^{\prime }},\mbox{ with respect to a coframe }e^{\alpha ^{\prime
}}\in T^{\ast }(\ _{1}^{3}\mathcal{V)};  \notag \\
& =\mathbf{g}=(hg,vg)=\ g_{ij}(x,y)\ e^{i}\otimes e^{j}+\ g_{ab}(x,y)\ 
\mathbf{e}^{a}\otimes \mathbf{e}^{b},  \label{dm} \\
& \qquad \mbox{ in N-adapted form with }hg=\{\ g_{ij}\},vg=\{g_{ab}\}; 
\notag \\
& =\underline{g}_{\alpha \beta }(u)du^{\alpha }\otimes du^{\beta },%
\mbox{
for a coordinate coframe }du^{\beta }\in T^{\ast }(\ _{1}^{3}\mathcal{V)}%
\mbox{ and }\underline{g}_{\alpha \beta }=\left[ 
\begin{array}{cc}
g_{ij}+N_{i}^{a}N_{j}^{b}g_{ab} & N_{j}^{e}g_{ae} \\ 
N_{i}^{e}g_{be} & g_{ab}%
\end{array}%
\right] .  \label{offdiagm}
\end{align}%
A metric structure $g=\{\underline{g}_{\alpha \beta }\}$ is generic
off--diagonal if the anholonomy coefficients $W_{\alpha \beta }^{\gamma }$
are not trivial.

A general metric-affine manifold $(g,D)$ is defined by an independent linear
connection structure $D=\{\Gamma _{\ \beta \lambda }^{\alpha }\},$ which may
be, or not, adapted to an N-connection structure. For N-adapted geometric
constructions we shall use boldface metric-metric affine symbols $(\mathbf{%
g,D})$, when $\mathbf{D}=(hD,vD)$ is an arbitrary distinguished connection,
d-connection, stated as an affine connection preserving under parallelism
the N--connection splitting (\ref{ncon}). Such a d-connection is
characterized by respective h- and v-indices ($i,j,k, ..., = 1,2;$ and $%
a,b,c,...=3,4$), 
\begin{equation}
\mathbf{D}=\{\mathbf{\Gamma }_{\ \alpha \beta }^{\gamma }=(L_{jk}^{i},\acute{%
L}_{bk}^{a};\acute{C}_{jc}^{i},C_{bc}^{a})\},\mbox{ where }hD=(L_{jk}^{i},%
\acute{L}_{bk}^{a})\mbox{ and }vD=(\acute{C}_{jc}^{i},C_{bc}^{a}),
\label{hvdcon}
\end{equation}%
when the coefficients are computed with respect to N-adapted frames (\ref%
{nader}) and (\ref{nadif}).

For any d-connection $\mathbf{D}$ (\ref{hvdcon}) and d-vectors $\mathbf{X}$ and $\mathbf{Y}$, we can define in abstract geometric form the fundamental geometric d-objects (respectively, the torsion, curvature and nonmetricity d-tensors), 
\begin{equation*}
\mathcal{T}(\mathbf{X,Y}):=\mathbf{D}_{\mathbf{X}}\mathbf{Y}-\mathbf{D}_{%
\mathbf{Y}}\mathbf{X}-[\mathbf{X,Y}];\mathcal{R}(\mathbf{X,Y}):=\mathbf{D}_{%
\mathbf{X}}\mathbf{D}_{\mathbf{Y}}-\mathbf{D}_{\mathbf{Y}}\mathbf{D}_{%
\mathbf{X}}-\mathbf{D}_{\mathbf{[X,Y]}};\mbox{ and }\mathcal{Q}(\mathbf{X}):=%
\mathbf{D}_{\mathbf{X}}\mathbf{g},
\end{equation*}%
and compute the corresponding coefficients with respect to N-adapted frames (\ref{nader}) and (\ref{nadif}), 
\begin{equation}
\mathcal{T}=\{\mathbf{T}_{\ \alpha \beta }^{\gamma }\},\mathcal{R}=\mathbf{%
\{R}_{\ \beta \gamma \delta }^{\alpha }\},\mathcal{Q}=\mathbf{\{Q}_{\gamma
\alpha \beta }\}.  \label{fundgeomncoef}
\end{equation}%
Contracting respective indices of the curvature d-tensor and using the
inverse d-tensor $\mathbf{g}^{\alpha \beta }$, we obtain: 
\begin{equation}
\mathbf{R}ic=\{\mathbf{R}_{\ \beta \gamma }:=\mathbf{R}_{\ \beta \gamma
\alpha }^{\alpha }\},\mbox{ the Ricci d-tensor};\ \mathbf{R}sc=\mathbf{g}%
^{\alpha \beta }\mathbf{R}_{\ \alpha \beta },\mbox{
the scalar curvature}.  \label{dricci}
\end{equation}%
In similar forms, we can define and compute other types of geometric d-objects for nonholonomic metric-affine manifolds and their (co) tangent bundles.

\subsubsection{Canonical nonholonomic metric-affine structures}

Any geometric and gravity model studied in \cite
{hehl95,harko21,iosifidis22,khyllep23,ghil23,koussour23,jhao22,de22,lheis23}
(for metric-affine generalizations of Finsler-Largange-Hamilton theories,
see \cite{vmon05,vplb10,sv12,svnonh08}) can be formulated in N-adapted form
using the geometric objects (\ref{dm}) -(\ref{dricci}). Such formulations in
canonical nonholonomic variables are important for decoupling and
integrating corresponding modified gravitational and geometric flow
equations using the AFCDM \cite{lb24epjc1,lb24epjc2,lb24grg}.

For any metric or d-metric structure $\mathbf{g}$ (\ref{dm}), we can define
two important linear connection structures: 
\begin{equation}
(\mathbf{g,N})\rightarrow \left\{ 
\begin{array}{cc}
\mathbf{\nabla :} & \mathbf{\nabla g}=0;\ _{\nabla }\mathcal{T}=0,\ 
\mbox{\
LC--connection }; \\ 
\widehat{\mathbf{D}}: & \widehat{\mathbf{Q}}=0;\ h\widehat{\mathcal{T}}=0,v%
\widehat{\mathcal{T}}=0,\ hv\widehat{\mathcal{T}}\neq 0,%
\mbox{ the canonical
d-connection}.%
\end{array}%
\right.  \label{twocon}
\end{equation}%
Such (auxiliary) linear connections are related via a corresponding
canonical distortion relation, 
\begin{equation}
\widehat{\mathbf{D}}[\mathbf{g}]=\nabla \lbrack \mathbf{g}]+\widehat{\mathbf{%
Z}}[\mathbf{g}],  \label{cdist}
\end{equation}%
when the canonical distortion d-tensor, $\widehat{\mathbf{Z}}[\mathbf{g}]$,
and $\nabla \lbrack \mathbf{g}]$ are determined by the same metric structure 
$\mathbf{g}$. We use "hat" labels for geometric d-objects defined by $%
\widehat{\mathbf{D}}$ (\ref{twocon}) and $\widehat{\mathbf{Z}}$ (\ref{cdist}%
) and when the N-adapted coefficients of $\widehat{\mathbf{D}}=\{\widehat{%
\mathbf{\Gamma }}_{\ \alpha \beta }^{\gamma }=(\widehat{L}_{jk}^{i},\widehat{%
L}_{bk}^{a},\widehat{C}_{jc}^{i},\widehat{C}_{bc}^{a})\}$ are computed 
\begin{eqnarray}
\widehat{L}_{jk}^{i} &=&\frac{1}{2}g^{ir}(\mathbf{e}_{k}g_{jr}+\mathbf{e}%
_{j}g_{kr}-\mathbf{e}_{r}g_{jk}),\widehat{L}_{bk}^{a}=e_{b}(N_{k}^{a})+\frac{%
1}{2}g^{ac}(\mathbf{e}_{k}g_{bc}-g_{dc}\ e_{b}N_{k}^{d}-g_{db}\
e_{c}N_{k}^{d}),  \notag \\
\widehat{C}_{jc}^{i} &=&\frac{1}{2}g^{ik}e_{c}g_{jk},\ \widehat{C}_{bc}^{a}=%
\frac{1}{2}g^{ad}(e_{c}g_{bd}+e_{b}g_{cd}-e_{d}g_{bc}).  \label{coefcandcon}
\end{eqnarray}%
If a general affine d-connection is prescribed $\mathbf{D}$ (\ref{hvdcon}),
we can consider $\nabla $ and $\widehat{\mathbf{D}}$ as auxiliary linear
connections defined only by the metric and N-connection structures.

In our approach, we shall work with the canonical d-connection $\widehat{%
\mathbf{D}}$ (\ref{coefcandcon}) with the nonholonomically induced d-torsion
(it can be considered as an auxiliary one to other types of torsion and
nonmetricity structures on a metric-affine manifold)%
\begin{equation*}
\widehat{\mathcal{T}}=\{\widehat{\mathbf{T}}_{\beta \gamma }^{\alpha }=%
\widehat{\mathbf{\Gamma }}_{\beta \gamma }^{\sigma }-\widehat{\mathbf{\Gamma 
}}_{\gamma \beta }^{\sigma }+w_{\beta \gamma }^{\sigma }\},
\end{equation*}%
which is completely determined by the coefficients of $\ \mathbf{g}$ and $%
\mathbf{N}$ and anholonomy coefficients $w_{\beta \gamma }^{\sigma }$ from (%
\ref{anhrel}). Such a canonical d-torsion $\widehat{\mathcal{T}}$ can be
irreducibly decomposed into $h$- and $v$-parts as 
\begin{eqnarray}
\widehat{\mathbf{T}}_{\beta \gamma }^{\alpha } &=&\ ^{1}\widehat{\mathbf{T}}%
_{\beta \gamma }^{\alpha }-\frac{1}{3}(\delta _{\beta }^{\alpha }\ ^{2}%
\widehat{\mathbf{T}}_{\gamma }-\delta _{\gamma }^{\alpha }\ ^{2}\widehat{%
\mathbf{T}}_{\beta })+\mathbf{g}^{\alpha \sigma }\epsilon _{\beta \gamma
\sigma \rho }\ ^{3}\widehat{\mathbf{T}}^{\rho },\mbox{ where }
\label{candtors} \\
\ ^{1}\widehat{\mathbf{T}}_{\beta \alpha }^{\alpha } &=&0,\epsilon ^{\beta
\gamma \sigma \rho }\ ^{1}\widehat{\mathbf{T}}_{\gamma \sigma \rho }=0,\ ^{2}%
\widehat{\mathbf{T}}_{\beta }:=\widehat{\mathbf{T}}_{\beta \alpha }^{\alpha
},\ ^{3}\widehat{\mathbf{T}}^{\rho }:=\frac{1}{6}\epsilon ^{\rho \beta
\gamma \sigma }\ \widehat{\mathbf{T}}_{\beta \gamma \sigma },  \notag
\end{eqnarray}%
where $\epsilon ^{\beta \gamma \sigma \rho }$ is completely antisymmetric.
The contortion s-tensor is defined 
\begin{equation}
\ \widehat{\mathbf{K}}_{\beta \gamma \sigma }=\widehat{\mathbf{T}}_{\beta
\gamma \sigma }+\widehat{\mathbf{T}}_{\gamma \sigma \beta }+\widehat{\mathbf{%
T}}_{\sigma \gamma \beta }.  \label{cotors}
\end{equation}%
Such values with "hats" are induced by a N-connection structure and written
in N-adapted forms.\footnote{They include nonholonomic torsion components but in a from which is
different from, for instance, the Riemann-Cartan theory or string gravity with torsion. In those gravity models, there are considered algebraic equations for motivating torsion fields as generated by certain spin like
fluids with nontrivial sources or certain other completely anti-symmetric torsion fields.}

The canonical distortion relation (\ref{cdist}) can be generalized as 
\begin{equation}
\mathbf{D}=\nabla +\mathbf{L}=\widehat{\mathbf{D}}+\widehat{\mathbf{L}},%
\mbox{ where }\widehat{\mathbf{L}}=\mathbf{L}-\widehat{\mathbf{Z}},
\label{disf}
\end{equation}%
for an additional disformation d-tensor $\mathbf{L}=\{\mathbf{L}_{\ \beta
\lambda }^{\alpha }=\frac{1}{2}(\mathbf{Q}_{\ \beta \lambda }^{\alpha }-%
\mathbf{Q}_{\ \beta \ \lambda }^{\ \alpha }-\mathbf{Q}_{\ \lambda \ \beta
}^{\ \alpha })\},$ with $\mathbf{Q}_{\alpha \beta \lambda }:=\mathbf{D}%
_{\alpha }\mathbf{g}_{\beta \lambda },$ for $\widehat{\mathbf{Q}}_{\alpha
\beta \lambda }=\widehat{\mathbf{D}}_{\alpha }\mathbf{g}_{\beta \lambda }=0.$
For nonmetric geometric constructions with disformations (\ref{disf}) (in
N-adapted, or not, forms), we can use respective nonmetricity d-vectors and
vectors:%
\begin{align}
\mathbf{Q}_{\alpha }& =\mathbf{g}^{\beta \lambda }\mathbf{Q}_{\alpha \beta
\lambda }=\mathbf{Q}_{\alpha \ \lambda }^{\ \lambda },\ ^{\intercal }\mathbf{%
Q}_{\beta }=\mathbf{g}^{\alpha \lambda }\mathbf{Q}_{\alpha \beta \lambda }=%
\mathbf{Q}_{\alpha \beta }^{\quad \alpha };\mbox{ and, correspondingly, } 
\notag \\
Q_{\alpha }& =g^{\beta \lambda }Q_{\alpha \beta \lambda }=Q_{\alpha \
\lambda }^{\ \lambda },\ ^{\intercal }Q_{\beta }=g^{\beta \lambda }Q_{\alpha
\beta }^{\quad \alpha }=Q_{\alpha \beta }^{\quad \alpha },  \label{nmdv}
\end{align}%
where the coefficients are computed correspondingly for N-adapted frames (%
\ref{nader}), (\ref{nadif}) (and any $e_{\alpha }$ and $e^{\beta }$) see
details in \cite{lb24epjc2}. In our works, we shall use also such canonical
geometric d-objects: the nonmetricity conjugate d-tensor and tensor,%
\begin{equation}
\widehat{\mathbf{P}}_{\ \ \alpha \beta }^{\gamma }=\frac{1}{4}(-2\widehat{%
\mathbf{L}}_{\ \alpha \beta }^{\gamma }+\mathbf{Q}^{\gamma }\mathbf{g}%
_{\alpha \beta }-\ ^{\intercal }\mathbf{Q}^{\gamma }\mathbf{g}_{\alpha \beta
}-\frac{1}{2}\delta _{\alpha }^{\gamma }\mathbf{Q}_{\beta }-\frac{1}{2}%
\delta _{\beta }^{\gamma }\mathbf{Q}_{\alpha }),  \label{nmcjdt}
\end{equation}%
and the nonmetricity scalar for respective d-connections and LC-connection,%
\begin{equation}
\widehat{\mathbf{Q}}=-Q_{\alpha \beta \lambda }\widehat{\mathbf{P}}^{\alpha
\beta \lambda },Q=-Q_{\alpha \beta \lambda }\mathbf{P}^{\alpha \beta \lambda
}\mbox{ and }Q=-Q_{\alpha \beta \lambda }P^{\alpha \beta \lambda }.
\label{nmsc}
\end{equation}

Considering distortion relations (\ref{cdist}) and (\ref{disf}) relating
certain $\nabla =\{\breve{\Gamma}_{\ \beta \gamma }^{\alpha }(u)\},$ $%
\mathbf{D}=\{\Gamma _{\ \alpha \beta }^{\gamma }\},\widehat{\mathbf{D}}%
=\{\Gamma _{\ \alpha \beta }^{\gamma }\}$ and $D=\{\Gamma _{\ \alpha \beta
}^{\gamma }\},$ we can compute corresponding distortion relations for
fundamental d-tensors (\ref{fundgeomncoef}), which is necessary for
formulating in canonical nonholonomic variables of the $f(\widehat{\mathbf{Q}%
})$ gravity as in section 2 of \cite{lb24epjc2}. For convenience, we provide
here respective parameterizations for distortions of the Ricci d-tensor and
respective Ricci scalars from (\ref{dricci}): 
\begin{align}
\mathbf{R}ic& =\breve{R}ic+\breve{Z}ic=\widehat{\mathbf{R}}ic+\widehat{%
\mathbf{Z}}ic,\mbox{ for respective coefficients }  \label{driccidist} \\
& \mathbf{R}ic=\{\mathbf{R}_{\beta \gamma }=\mathbf{R}_{\ \beta \gamma
\alpha }^{\alpha }\};\breve{R}ic=\{\breve{R}_{\beta \gamma }=\breve{R}_{\
\beta \gamma \alpha }^{\alpha }\},\breve{Z}ic=\{\breve{Z}_{\beta \gamma }=%
\breve{Z}_{\ \beta \gamma \alpha }^{\alpha }\};  \notag \\
& \widehat{\mathbf{R}}ic=\{\widehat{\mathbf{R}}_{\beta \gamma }=\widehat{%
\mathbf{R}}_{\ \beta \gamma \alpha }^{\alpha }\},\widehat{\mathbf{Z}}ic=\{%
\widehat{\mathbf{Z}}_{\beta \gamma }:=\widehat{\mathbf{Z}}_{\ \beta \gamma
\alpha }^{\alpha }\};\mbox{ and }  \notag \\
\mathbf{R}sc& =\mathbf{g}^{\alpha \beta }\mathbf{R}_{\ \alpha \beta }=\breve{%
R}sc+\breve{Z}sc=\widehat{\mathbf{R}}sc+\widehat{\mathbf{Z}}sc,\mbox{where }
\notag \\
&\breve{R}sc =g^{\beta \gamma }\breve{R}_{\beta \gamma },\breve{Z}%
sc=g^{\beta \gamma }\breve{Z}_{\beta \gamma };\widehat{\mathbf{R}}sc=\mathbf{%
g}^{\alpha \beta }\widehat{\mathbf{R}}_{\alpha \beta },\widehat{\mathbf{Z}}%
sc=\mathbf{g}^{\alpha \beta }\widehat{\mathbf{Z}}_{\alpha \beta }.  \notag
\end{align}

Above geometric d-objects and objects can be used for elaborating and
analyzing physical properties of various models of nonmetric geometric flows
and MGTs. The motivation and priority of the canonical "hat" variables is
that they allow to decouple and integrate in certain general off-diagonal
forms corresponding physically important systems of nonlinear PDEs.

\subsubsection{Nonmetric $f(Q)$ gravity in canonical nonholonomic variables}

In canonical N-adapted form, we can elaborate a model of nonmetric gravity
using a gravitational Lagrange density $\ ^{g}\widehat{\mathcal{L}}=\frac{1}{%
2\kappa }\widehat{f}(\widehat{\mathbf{Q}})$ and for the matter fields $^{m}%
\widehat{\mathcal{L}}$ defined in "hat" variables. The action is postulated
in the form:%
\begin{equation}
\widehat{\mathcal{S}}=\int \sqrt{|\mathbf{g}_{\alpha \beta }|}\delta ^{4}u(\
^{g}\widehat{\mathcal{L}}+\ ^{m}\widehat{\mathcal{L}}).  \label{actnmc}
\end{equation}%
In this formula, hats state that all Lagrange densities and geometric
objects are written in nonholonomic dyadic form, with boldface indices;
using $\widehat{\mathbf{D}}$ and disformations (\ref{disf}) and the measure $%
\sqrt{|\mathbf{g}_{\alpha \beta }|}\delta ^{4}u,$ for $\delta
^{4}u=du^{1}du^{2}\delta u^{3}\delta u^{4}$ with $\delta u^{a}=\mathbf{e}%
^{a} $ (\ref{nadif}). Choosing $\widehat{\mathcal{S}}=\mathcal{S}$ with a $%
\mathcal{S}$ of type $f(Q)$, we obtain equivalent classes of nonmetric
gravity theories but formulated for different types of nonholonomic or
holonomic variables.

In N-adapted variational form \cite{vmon05,bubuianu17,vacaru18,lb24epjc2},
or applying "pure" nonholonomic geometric methods as in \cite{misner}, we
can derive such nonmetric gravitational field equations: 
\begin{align}
\frac{2}{\sqrt{|\mathbf{g}|}}\widehat{\mathbf{D}}_{\gamma }(\sqrt{|g|}%
\widehat{f_{Q}}\widehat{\mathbf{P}}_{\ \ \alpha \beta }^{\gamma })+\frac{1}{2%
}\widehat{f}\mathbf{g}_{\alpha \beta }+\widehat{f_{Q}}(\widehat{\mathbf{P}}%
_{\beta \mu \nu }\mathbf{Q}_{\alpha }^{\ \mu \nu }-2\widehat{\mathbf{P}}%
_{\alpha \mu \nu }\mathbf{Q}_{\quad \beta }^{\mu \nu })& =\kappa \widehat{%
\mathbf{T}}_{\alpha \beta }  \label{cfeq3a} \\
\mbox{ and }\widehat{\mathbf{D}}_{\alpha }\widehat{\mathbf{D}}_{\beta }(%
\sqrt{|\mathbf{g}|}\widehat{f_{Q}}\widehat{\mathbf{P}}_{\ \ \gamma }^{\alpha
\beta })& =0.  \label{cfeq3b}
\end{align}
 In these formulas, for $\widehat{f_{Q}}:=\partial \widehat{f}/\partial 
\widehat{Q};$ with $\widehat{\mathbf{P}}_{\ \ \alpha \beta }^{\gamma }$ and $%
\mathbf{Q}_{\ \ \alpha \beta }^{\gamma }$ defined as in formulas (\ref{nmdv}) and (\ref{nmsc}); the energy-momentum d-tensor $\widehat{\mathbf{T}}%
_{\alpha \beta }$ is defined by N-adapted variations for $\ ^{m}\widehat{%
\mathcal{L}}$ on the d-metric; and all equations being written with respect
to N-elongated frames (\ref{nader}) and (\ref{nadif}).

Using distortion relations (\ref{driccidist}), we can write the system (\ref%
{cfeq3a}) in a form provided in \cite{jhao22} for the Einstein tensor, $%
\breve{E}:=\breve{R}ic-\frac{1}{2}g\breve{R}sc$, computed for $\nabla .$
There were studied also possible physical implications of some equivalent
effective Einstein equations with effective energy-momentum tensor which may
describe dark energy \cite{jhao22,koussour23}.\footnote{%
In this work, we follow a different system of notations and chose an
opposite sign before $\ ^{m}T_{\alpha \beta }.$} Such effective
gravitational field equations can be written in the form%
\begin{align}
\breve{E}_{\alpha \beta }& =\frac{\kappa }{f_{Q}}\ ^{m}T_{\alpha \beta }+\
^{Q}T_{\alpha \beta }=\kappa \breve{T}_{\alpha \beta },\mbox{ or }
\label{gfeq2a} \\
\breve{R}_{\alpha \beta }& =\breve{\Upsilon}_{\alpha \beta },\mbox{ where }%
\breve{\Upsilon}_{\alpha \beta }=\kappa (\breve{T}_{\alpha \beta }-\frac{1}{2%
}g_{\alpha \beta }\breve{T}),\mbox{ for }\breve{T}=g^{\alpha \beta }\breve{T}%
_{\alpha \beta },  \label{gfeq2b} \\
\ ^{Q}T_{\alpha \beta }& =\frac{1}{2}g_{\alpha \beta }(\frac{f}{f_{Q}}-Q)+2%
\frac{f_{QQ}}{f_{Q}}\nabla _{\gamma }(QP_{\ \ \alpha \beta }^{\gamma }),
\label{deemt} \\
\ ^{m}T_{\alpha \beta }& :=-\frac{2}{\sqrt{|g|}}\frac{\delta (\sqrt{|g|}\
^{m}\mathcal{L})}{\delta g^{\alpha \beta }}=\ ^{m}\mathcal{L}g_{\alpha \beta
}+2\frac{\delta (\ ^{m}\mathcal{L})}{\delta g^{\alpha \beta }}.
\label{emtlc}
\end{align}%
The formula (\ref{emtlc}) \ holds if $\ ^{m}\mathcal{L}$ does not depend in
explicit form on $\Gamma _{\ \alpha \beta }^{\gamma }$ (it has to be
modified, for instance, for spinor fields, see next subsections). The
nonmetric modifications of GR are encoded into the effective energy-momentum
tensor $\ ^{Q}T_{\alpha \beta }$ (\ref{deemt}).

The compatibility with the Einstein equations in GR for zero distortions and
zero nonmetricity can be guaranteed if we compute the distortions of Ricci
d-tensors and tensors using formulas (\ref{driccidist}). The covariant
representation of (\ref{cfeq3a}) can be formulated by analogy to the
formulas with $\breve{E}$ (\ref{gfeq2a}) but working with the Einstein d-tensor $%
\widehat{\mathbf{E}}:=\widehat{\mathbf{R}}ic-\frac{1}{2}\mathbf{g}\widehat{%
\mathbf{R}}sc$ for $\widehat{\mathbf{D}}.$ In this paper, we consider a form
of nonmetric gravitational equations with $\widehat{\mathbf{R}}ic=\{\widehat{%
\mathbf{R}}_{\alpha \beta}\}$ in the left side and certain effective sources 
$\widehat{\yen }ic=\{\widehat{\yen }_{\alpha \beta }\}$ in the right side.
Such (effective) sources may encode $\widehat{f}(\widehat{\mathbf{Q}})$%
-deformations of the canonical and LC data. Such a representation is
convenient for applying the AFCDM for constructing off-diagonal solutions.

We can distort the equations (\ref{gfeq2a}), for $\nabla \rightarrow 
\widehat{\mathbf{D}}=\nabla +\widehat{\mathbf{Z}}$ (\ref{cdist}) and (\ref%
{disf}), or redefine in N-adapted form the variational calculus on
metric-affine spacetime. This way, we derive the $\widehat{f}(\widehat{%
\mathbf{Q}})$ gravitational equations in canonical dyadic variables with
effective sources (encoding both nonmetric geometric distortions and matter fields), 
\begin{align}
\widehat{\mathbf{R}}_{\alpha \beta }& =\widehat{\yen }_{\alpha \beta },%
\mbox{ for }  \label{cfeq4a} \\
\widehat{\yen }_{\alpha \beta }& =\ ^{e}\widehat{\mathbf{Y}}_{\alpha \beta
}+\ ^{m}\widehat{\mathbf{Y}}_{\alpha \beta }.  \label{ceemt}
\end{align}%
The source $\widehat{\yen }_{\alpha \beta }$ (\ref{ceemt}) for (\ref{cfeq4a}) is defined by two d-tensors: the first one, $\ ^{e}\widehat{\mathbf{Y}}_{\alpha \beta }=\breve{Z}ic_{\alpha \beta }-\widehat{\mathbf{Z}}ic_{\alpha
\beta }$, is of geometric distorting nature, which can be computed in
explicit form using (\ref{driccidist}). The second one is the
energy-momentum d-tensor $\ ^{m}\widehat{\mathbf{Y}}_{\alpha \beta }$ of the
matter fields encoding also contributions of the nonmetricity scalar $%
\widehat{\mathbf{Q}}$ and d-tensor $\widehat{\mathbf{P}}_{\ \ \alpha \beta
}^{\gamma }$ defined respectively by formulas (\ref{nmdv}), (\ref{nmcjdt})
and (\ref{nmsc}).

The nonlinear systems of PDEs (\ref{cfeq3a}) \ or (\ref{gfeq2a}) can not be
decoupled and integrated in certain general forms with off-diagonal metrics.
This is possible if we apply the AFCDM for the nonmetric and canonically
modified Einstein equations (\ref{cfeq4a}) considered as a nonholonomic
dyadic formulation of (\ref{gfeq2b}). Such nonmetric gravitational field
equations can be extended for EDM systems which can be also decoupled and
integrated in off-diagonal form (we prove in the next section).

We note that the solutions of (\ref{cfeq3a}) or (\ref{cfeq4a}) with $\widehat{\mathbf{D}}$ do not consist solutions of the system (\ref{gfeq2a}) with $\nabla .$ Imposing additional nonholonomic constraints of type 
$ \widehat{\mathbf{D}}_{|\mathcal{T}=0}=\nabla ,$ when the canonical nonholonomic torsion induced by N-coefficients become zero, $\widehat{\mathbf{T}}_{\ \alpha \beta }^{\gamma },$ we can extract exact/parametric solutions for (\ref{gfeq2a}). Such conditions can be satisfied by restricting respectively the class of generating functions and generating effective sources as we proved in 
\cite{bubuianu17,vacaru18,vacaru20,lb24epjc1,lb24epjc2}.

\subsection{Nonholonomic Dirac d-operators and metric compatible EDM systems}

The Dirac operator on space-time Lorentz manifolds was defined by considering tetradic (equivalently, vierbeind) decompositions of the metric structure. In such an approach, the relativistic gamma matrix formalism for 4-d Minkowski spacetime is extended to curved spacetimes using tetrads and the Dirac equations in GR are defined by certain spin covariant operators encoding the LC connection. Similar constructions can be performed for other type metric compatible linear connections defined in an effective form on any Lorenz manifold or (co) bundle (phase) spaces as we explained for formulas (\ref{gammamatr}) and (\ref{diracgr}), see details and discussions in \cite{vplb10,vacaru18,lb24grg}. In this subsection, we shall use the canonical d-operator $\widehat{\mathbf{D}}=\{\Gamma _{\ \alpha \beta}^{\gamma }\}$ (\ref{coefcandcon}) which allows us a metric compatible definition of the Dirac d-operator. Then a respective generalization of the modified Einstein equations (\ref{cfeq4a}) to nonmetric EDM systems, when nonmetricity contributions are encoded into $\ ^{e}\widehat{\mathbf{Y}}_{\alpha \beta }$ from $\widehat{\yen }_{\alpha \beta }$ (\ref{ceemt}). The nonmetricity field can be also encoded via anisotropic effective polarizations of masses of fermions or as an induced mass by additional nonlinear terms in modified
Dirac equations.

\subsubsection{The nonholonomic canonical Dirac d-operator}

\label{sscanondir}Any d-metric $\mathbf{g}=\mathbf{g}_{\alpha \beta }\ 
\mathbf{e}^{\alpha }\mathbf{e}^{\beta }$ (\ref{dm}) can be decomposed with
respect to N-adapted (2+2)-frames $\mathbf{e}_{\ \mu }^{\underline{\mu }%
}=(e_{\ i}^{\underline{i}},e_{\ a}^{\underline{a}}),$ when $\ \mathbf{e}_{\
}^{\underline{\mu }}=\mathbf{e}_{\ \alpha }^{\underline{\mu }}\ \mathbf{e}%
^{\alpha }$ for a $\mathbf{e}^{\alpha }$ of type (\ref{nadif}) and $\ 
\mathbf{e}_{\ \mu }^{\underline{\mu }}\ \mathbf{e}_{\underline{\nu }}^{\ 
\underline{\mu }}=\ \mathbf{\delta }_{\underline{\nu }}^{\underline{\mu }},$
where $\ \mathbf{\delta }_{\underline{\nu }}^{\underline{\mu }}$ is the
Kronecker symbol. For such frames, we can consider corresponding h- and
v-splitting of all geometric d-objects, for instance, $hg=\{g_{ij}=e_{\ i}^{%
\underline{i}}e_{\ j}^{\underline{j}}\eta _{\underline{i}\underline{j}}\}$
and $vg=\{g_{ab}=e_{\ a}^{\underline{a}}e_{\ b}^{\underline{b}}\eta _{%
\underline{a}\underline{b}}\},$ for $\eta _{\underline{i}\underline{j}%
}=diag(1,1)$ and $\eta _{\underline{a}\underline{b}}=diag(1,-1).$ So, with
respect to N-adapted frames, Clifford d-structures $\ _{s}^{\shortmid }%
\mathcal{C}l(\ _{1}^{3}\mathcal{V})$ on $\ _{1}^{3}\mathcal{V}$ are defined
by corresponding h- and v-gamma matrices $\ \gamma _{\mu }=(\gamma
_{i},\gamma _{a}),$ when $\ \gamma _{\mu }=\mathbf{e}_{\ \mu }^{\underline{%
\mu }}\gamma _{\underline{\mu }},$ when $\ ^{\shortmid }\gamma _{\underline{%
\mu }}=(\gamma _{\underline{i}},\ ^{\shortmid }\gamma ^{\underline{a}})$ are
subjected to the conditions $\gamma _{\underline{i}}\gamma _{\underline{j}%
}+\gamma _{\underline{j}}\gamma _{\underline{i}}=-2\eta _{\underline{i}%
\underline{j}}$ and $\gamma _{\underline{a}}\gamma _{\underline{b}}+\gamma _{%
\underline{b}}\gamma _{\underline{a}}=-2\eta _{\underline{a}\underline{b}}.$
A choice for such h-gamma matrices corresponds to that from the footnote \ref%
{fngamma}. In particle physics, it is considered also the $\gamma ^{5}$%
-matrix defined by the property that $\ \gamma _{i^{\prime }}\gamma
^{5}=\{\left( 
\begin{array}{cc}
O & I \\ 
-I & O%
\end{array}%
\right) ,\left( 
\begin{array}{cc}
\sigma _{i^{\prime }} & O \\ 
O & -\sigma _{i^{\prime }}%
\end{array}%
\right) \}$, for $i^{\prime },j^{\prime },...=1,2,3.$

A Dirac spinor field $\psi (x^{i^{\prime }},t)$ on $\ _{1}^{3}\mathcal{V}$
is defined as a complex 4-d vector field $\psi (u)=(%
\begin{array}{c}
\psi _{1} \\ 
\psi _{2}%
\end{array}%
),$ where $\psi _{1}(x^{i^{\prime }},t)$ and $\psi _{2}(x^{i^{\prime }},t)$
are 2-d complex fields. We shall consider also the Dirac conjugate d-spinor
field $\overline{\psi }:=-\psi ^{\dagger }\gamma ^{4},$ where $\dagger $
means the Hermitian conjugation. We note that all local constructions with
spinor fields on metric-affine spaces can be performed using the metric and
corresponding tetradic structure as in GR but the main difference is that
the nonmetricity field modified the rules of covariant derivation and
transport along curves of spinor fields.

The covariant on $\ _{1}^{3}\mathcal{V}$ spinor derivative $\ \mathcal{D}$
acting on $\psi $ and $\overline{\psi }$ can be defined in canonical
N-adapted form if we use the canonical d-connection: 
\begin{eqnarray}
\ \widehat{\mathcal{D}} &=&\{\ \ \widehat{\mathcal{D}}_{\alpha }=\mathbf{e}%
_{\alpha }-\widehat{\mathbf{\Gamma }}_{\alpha }\},\mbox{ where }\ \ 
\label{candirac} \\
&&\widehat{\mathbf{\Gamma }}_{\alpha }=-\frac{1}{4}\widehat{\varsigma }%
_{\alpha \underline{\beta }\underline{\gamma }}\gamma ^{\underline{\beta }%
}\gamma ^{\underline{\gamma }},\mbox{ with }\ \widehat{\varsigma }_{\alpha \ 
\underline{\nu }}^{\ \underline{\beta }}=\mathbf{e}_{\ \beta }^{\underline{%
\beta }}\ \mathbf{e}_{\underline{\nu }}^{\ \nu }\ \widehat{\mathbf{\Gamma }}%
_{\ \alpha \nu }^{\beta }-\mathbf{e}_{\underline{\nu }}^{\ \nu }\mathbf{e}%
_{\ \alpha }(\mathbf{e}_{\ \nu }^{\underline{\beta }})+\widehat{K}_{\alpha \ 
\underline{\nu }}^{\ \underline{\beta }},  \notag
\end{eqnarray}%
which includes the contorsion d-tensor $\ \widehat{K}_{\beta \alpha \nu }$ (%
\ref{cotors}) for $\ \widehat{\mathbf{D}}=\{\widehat{\mathbf{\Gamma }}_{\
\alpha \nu }^{\beta }\}.$ The formulas (\ref{candirac}) define the \textbf{%
canonical Dirac d-operator} $\widehat{\mathcal{D}}$ on $_{1}^{3}\mathcal{V}.$

Using the same metric structure (which can be considered in N-adapted form, $%
\mathbf{g}$ (\ref{dm}), or equivalent off-diagonal form, $g=\{\underline{g}%
_{\alpha \beta }\}$ (\ref{offdiagm})), we can define another Dirac operator $%
\ \mathcal{\breve{D}}=\{\mathcal{\breve{D}}_{\alpha }=\mathbf{e}_{\alpha }-%
\breve{\Gamma}_{\alpha }\}$ constructed as above for some arbitrary vierbein
fields $\ \mathbf{e}_{\ \beta }^{\underline{\beta }}$ and $\ \mathbf{e}_{%
\underline{\nu }}^{\ \nu }$ but working with the coefficients of the
LC-connection $\nabla =\{\breve{\Gamma}_{\ \beta \gamma }^{\alpha }\}$
instead of $\ \widehat{\mathbf{\Gamma }}_{\ \alpha \nu }^{\beta }.$ This
covariant Dirac operator $\mathcal{\breve{D}}$ is not a d-operator.
Nevertheless, we can always compute a canonical distortion d-adapted
relation $\ \widehat{\mathcal{D}}=\ \mathcal{\breve{D}}+\widehat{\mathcal{Z}}%
,$ with $\widehat{\mathcal{Z}}$ determined by the canonical d-connection
distortion $\widehat{\mathbf{D}}=\nabla +\widehat{\mathbf{Z}}$ (\ref{cdist}).

\subsubsection{Metric compatible EDM equations in canonical dyadic variables}

The action of the canonical Dirac d-operator $\ \widehat{\mathcal{D}}%
_{\alpha }$ (\ref{candirac}) on spinor s-fields is defined in the form 
\begin{equation*}
\ \widehat{\mathcal{D}}_{\alpha }\psi =\mathbf{e}_{\alpha }\psi -\widehat{%
\mathbf{\Gamma }}_{\alpha }\psi \mbox{ and }\ \widehat{\mathcal{D}}_{\alpha }%
\overline{\psi }=\mathbf{e}_{\alpha }\overline{\psi }-\overline{\psi }%
\widehat{\mathbf{\Gamma }}_{\alpha }.
\end{equation*}%
We consider that for zero nonmetricity we obtain a standard EDM theory of
coupled $U(1)$-gauge relativistic fermions with equal mass $m_{0} $ and
physical constants $G=c=1$ as in GR \cite{misner}. The main assumption is that for
distortions of linear connections resulting in a metric compatible
d-connection we can consider the same mass term $\ m_{0} $ even the Dirac
operator is modified by possible nontrivial torsion coefficients. The
interaction constant $q$ of the U(1) gauge potential $\mathbf{A}_{\alpha }$
is used for extending the canonical Dirac d-operator $\widehat{\mathcal{D}}%
_{\alpha }\rightarrow $ $\widehat{\mathcal{D}}_{\alpha }^{A}:=\ \widehat{%
\mathcal{D}}_{\alpha }-iq\mathbf{A}_{\alpha },$ where $i$ is the complex
unity.

The canonical equations of motion on $_{1}^{3}\mathcal{V}$ of the spinor fields $\psi $ and  $\overline{\psi }$ are postulated respectively, 
\begin{eqnarray}
\lbrack i\ \hbar \gamma ^{\alpha }\widehat{\mathcal{D}}_{\alpha }^{A}-\
m_{0}+\frac{3}{2}\hbar \ \ ^{3}\widehat{\mathbf{T}}^{\alpha }\gamma _{\alpha
}\gamma ^{5}]\ \psi &=&0,  \label{canddiracs} \\
i\hbar \ \widehat{\mathcal{D}}_{\alpha }^{A}\overline{\psi }\gamma ^{\alpha
}+\ m_{0}+\frac{3}{2}\hbar (\ ^{3}\widehat{\mathbf{T}}^{\alpha }\ \
^{\shortmid }\overline{\Psi }\gamma ^{5}\gamma _{\alpha }) &=&0.  \notag
\end{eqnarray}%
These generalized Dirac equations are constructed following the same principles as in GR but nonholonomically extended for canonical d-variables with nontrivial $\ ^{3}\widehat{\mathbf{T}}^{\alpha }$ (\ref{candtors}). For such canonical d-spinor configurations, we can define also a Dirac axial spin
vector h-current $\breve{s}^{\alpha }:=\frac{\hbar }{2}\ \overline{\psi }%
\gamma ^{\alpha }\gamma ^{5}\psi .$ The nonholonomic structure can be chosen
in certain forms when the canonical d-adapted Dirac equations (\ref{canddiracs}) generalize  respective Riemann-Cartan equations from \cite{cabral} when the torsion structure is a canonical
nonholonomic induced one $\widehat{\mathcal{T}}$. In above formulas, we can
consider arbitrary d-torsions, $\widehat{\mathcal{T}}\rightarrow \mathcal{T}%
=\{\mathbf{T}_{\beta \gamma }^{\alpha }\},$ but for the so called $f(Q)$
theories we can consider only the canonical nonholonomic one.

At the next step, we postulate geometrically (as in \cite{misner} but
considering canonical N-adapted generalizations) the Maxwell equations, 
\begin{equation}
\widehat{\mathbf{D}}_{\alpha }\widehat{\mathbf{F}}^{\alpha \beta }=q\widehat{%
\mathbf{j}}^{\beta }.  \label{candmeq}
\end{equation}%
In these formulas, the anti-symmetric $\widehat{\mathbf{F}}_{\alpha \beta
}:=[\widehat{\mathbf{D}}_{\alpha }-iq\mathbf{A}_{\alpha },\widehat{\mathbf{D}%
}_{\beta }-iq\mathbf{A}_{\beta }]$ is defined as the strength d-tensor of
the Abelian gauge field $\mathbf{A}_{\beta _{s}}$ and the current is
computed as a d-vector $\widehat{\mathbf{j}}^{\beta }:=\overline{\psi }\ 
\mathbf{\gamma }^{\beta }\psi .$ The modified electromagnetic equations (\ref{candmeq}) can be also derived by a generalized N-adapted variational calculus in "hat" variables (we omit such a tedious proof in this work).

Using the canonical Ricci s-tensor as in (\ref{cfeq4a}) and the
corresponding scalar curvature $\widehat{\mathcal{R}}s:=\mathbf{g}^{\alpha
\beta }\ \widehat{\mathbf{R}}_{\alpha \beta },$ we postulate such a system
of N-adapted gravitational equations on $_{1}^{3}\mathcal{V}:$%
\begin{equation}
\ \widehat{\mathbf{R}}_{\alpha \beta }-\frac{1}{2}\mathbf{g}_{\alpha \beta
}\ \widehat{\mathcal{R}}s=\ ^{DA}\widehat{\mathbf{T}}_{\alpha \beta }.
\label{candeinst}
\end{equation}%
In (\ref{candeinst}),  the energy (stress) - momentum tensor for Dirac-Maxwell matter is
chosen as in GR but in nonholonomic canonical variables:
\begin{eqnarray}
\ ^{\ ^{DA}}\widehat{\mathbf{T}}_{\alpha \beta } &=&\widehat{\mathbf{T}}%
_{\alpha \beta }^{[A]}+\widehat{\mathbf{T}}_{\alpha \beta }^{[D]},\mbox{ for}
\label{semt} \\
\widehat{\mathbf{T}}_{\alpha \beta }^{[A]} &=&2\widehat{\mathbf{F}}_{\alpha
\tau }\widehat{\mathbf{F}}_{\ \ \beta }^{\tau }-\frac{1}{2}\mathbf{g}%
_{\alpha \beta }\widehat{F}^{2},\mbox{ for }\widehat{F}^{2}=\widehat{\mathbf{%
F}}^{\alpha \beta }\widehat{\mathbf{F}}_{\alpha \beta };\mbox{ and }  \notag
\\
\widehat{\mathbf{T}}_{\alpha \beta }^{[D]} &=&-\frac{i}{2}[\overline{\psi }\ 
\mathbf{\gamma }_{\alpha }\ \widehat{\mathcal{D}}_{\beta }^{A}\psi +%
\overline{\psi }\ \mathbf{\gamma }_{\beta }\ \widehat{\mathcal{D}}_{\alpha
}^{A}\psi -\widehat{\mathcal{D}}_{\alpha }^{A}(\overline{\psi })\mathbf{%
\gamma }_{\beta }\psi -\ \widehat{\mathcal{D}}_{\beta }^{A}(\overline{\psi })%
\mathbf{\gamma }_{\alpha }\psi ].  \notag
\end{eqnarray}

The priority to work with nonholonomic canonical d-operators $\widehat{%
\mathcal{D}}$ and $\widehat{\mathbf{D}}$ is that they allow to decouple and
solve in certain general forms the canonical Einstein-Dirac (ED) equations
and to consider generalizations of such systems of nonlinear PDEs to
nonholonomic (metric or nonmetric) EDM equations and their solutions.

\subsection{Nonmetric EDM equations in canonical dyadic variables}

Let us formulate a nonmetric modification of the nonholonomic Einstein-Dirac-Maxwell (EDM) theory. For simplicity, we can consider only $f(Q)$ models written in canonical nonholonomic variables as in \cite{lb24epjc2}. In MAG, we can use a general affine d-connection 
$\mathbf{D}=\{\mathbf{\Gamma }_{\ \alpha \nu }^{\beta }\}$ and cotorsion $K_{\beta \alpha \nu }$
instead of the canonical ones in (\ref{candirac}) and define a general Dirac d-operator $\mathcal{D}=\{\mathcal{D}_{\alpha }=\mathbf{e}_{\alpha }-\mathbf{\Gamma }_{\alpha }\}$. Such a construction is not compatible under actions of covariant derivatives on gamma matrices quadratic formulas (\ref{gammamatr}) and transports along curves of the Dirac equation (\ref{diracgr}). In that nonlinear equation, a constant mass $m_{0}$ is considered. It can be zero or induced effectively by certain nonlinear interactions in MGTs. In this work, an important idea is that we can use the disformation relations (%
\ref{disf}) and express $\mathcal{D}=\ \widehat{\mathcal{D}} + \mathcal{Z}$ and $m_{0}\rightarrow m_{0}+\hat{M}(u)$. For Dirac spinors, the nonmetric modifications are encoded in $\hat{M}(u)$ which has to be computed for a respective class of exact/ parametric solutions of the nonmetric EDM equations which will be formulated below. So, the nonmetricity fields polarize, or induce, masses of fermions in certain locally anisotropic forms. Such physical effects can be checked in experimental form or using observational data.

The effective electromagnetic source $\ ^{e}\mathbf{j}^{\beta }$, for 
$\mathbf{j}^{\beta } \to \widehat{\mathbf{j}}^{\beta } + \ ^{e}\mathbf{j}^{\beta } $, see formulas (\ref{candmeq}); and the gravitational source $\ ^{e}\widehat{\mathbf{Y}}_{\alpha \beta }$, see (\ref{ceemt}),  encode nonmetric deformations of the respective canonical ones from (\ref{candeinst}) and (\ref{semt}), can be computed following such formulas:
\begin{enumerate}
\item The disformations (\ref{disf}) allow us to express the nonmetric
canonical deformations 
\begin{equation}
\ \mathbf{D}_{\beta }=\widehat{\mathbf{D}}_{\beta }+\widehat{\mathbf{Z}}%
_{\beta },\mbox{  for }\ \mathbf{\Gamma }_{\ \alpha \nu }^{\beta }=\widehat{%
\mathbf{\Gamma }}_{\ \alpha \nu }^{\beta }+\widehat{\mathbf{Z}}_{\ \alpha
\nu }^{\beta }.  \label{scanconp}
\end{equation}%
Using (\ref{scanconp}), we compute respectively the nonmetric deformations of the torsion and cotrorsion d-tensor 
\begin{equation}
\mathbf{T}_{\ \alpha \nu }^{\beta }=\widehat{\mathbf{T}}_{\ \alpha \nu
}^{\beta }+\ ^{Q}\widehat{\mathbf{T}}_{\ \alpha \nu }^{\beta }\mbox{ and }\ 
\mathbf{K}_{\ \alpha \nu }^{\beta }=\widehat{\mathbf{K}}_{\ \alpha \nu
}^{\beta }+\ ^{Q}\widehat{\mathbf{K}}_{\ \alpha \nu }^{\beta }.
\label{storscomp}
\end{equation}

\item For the nonmetric Dirac d-operator elongated by a $U(1)$ nonassociative gauge field, 
\begin{eqnarray}
\mathcal{D}_{\alpha }^{A}&:= &\ \mathcal{D}_{\alpha }-iq\mathbf{A}_{\alpha
}=\ \mathbf{e}_{\alpha }-\mathbf{\Gamma }_{\alpha }-iq\mathbf{A}_{\alpha }=%
\mathbf{e}_{\alpha }-\widehat{\mathbf{\Gamma }}_{\alpha }-\ ^{Q}\widehat{%
\mathbf{\Gamma }}_{\alpha }-iq\mathbf{A}_{\alpha }  \notag \\
&=&\ \widehat{\mathcal{D}}_{\alpha }^{A}-\ ^{Q}\widehat{\mathbf{\Gamma }}%
_{\alpha }=\ \widehat{\mathcal{D}}_{\alpha }-iq\mathbf{A}_{\alpha }-\ ^{Q}%
\widehat{\mathbf{\Gamma }}_{\alpha },  \label{qcandir}
\end{eqnarray}%
where $\widehat{\mathcal{D}}$ is the canonical Dirac d-operator (\ref{candirac}) on $_{1}^{3}\mathcal{V}.$

\item Nonmetric deformations result in locally anisotropic polarizations of the fermionic (electronic) masses, $m_{0}\rightarrow M(u)=m_{0}+\ ^{Q}M(u),$ for $\Psi =B(u)\ \psi $ computed as a $4\times 4$ matrix for any fixed point values, when 
\begin{equation}
\ ^{Q}M=-i\ \hbar \ \gamma ^{\alpha }\ ^{Q}\widehat{\mathbf{\Gamma }}%
_{\alpha }+m_{0}B-\frac{3}{2}\hbar \ \ ^{3}\widehat{\mathbf{T}}^{\alpha
}\gamma _{\alpha }\gamma ^{5}.  \label{anisotrm}
\end{equation}

\item For a respective electromagnetic strength field and a source on $_{1}^{3}\mathcal{V}$, we have 
\begin{eqnarray}
\mathbf{F}_{\alpha \beta }&:= &[\mathbf{D}_{\alpha }-iq\mathbf{A}_{\alpha },%
\mathbf{D}_{\beta }-iq\mathbf{A}_{\beta }]=[\widehat{\mathbf{D}}_{\alpha }+%
\widehat{\mathbf{Z}}_{\alpha }-iq\mathbf{A}_{\alpha },\widehat{\mathbf{D}}%
_{\beta }+\widehat{\mathbf{Z}}_{\beta }-iq\mathbf{A}_{\beta }]
\label{starsmaxstr} \\
&=&\ \widehat{\mathbf{F}}_{\alpha \beta }+\ ^{Q}\widehat{\mathbf{F}}_{\alpha
\beta },\ ^{Q}\widehat{\mathbf{F}}_{\alpha \beta }=\widehat{\mathbf{D}}%
_{\alpha }\widehat{\mathbf{Z}}_{\beta }-\widehat{\mathbf{D}}_{\beta }%
\widehat{\mathbf{Z}}_{\alpha }+\widehat{\mathbf{Z}}_{\alpha }\widehat{%
\mathbf{Z}}_{\beta }-\widehat{\mathbf{Z}}_{\beta }\widehat{\mathbf{Z}}%
_{\alpha },\mbox{ and }  \notag \\
\mathbf{j}^{\beta }&:=&\overline{\Psi }\ \mathbf{\gamma }^{\beta }\ \Psi =%
\widehat{\mathbf{j}}^{\beta }+\ ^{e}\mathbf{j}^{\beta },  \notag
\end{eqnarray}%
for a respective parametrization of $B(u)$ which allow us to represent $%
\widehat{\mathbf{j}}^{\beta }:=\overline{\psi }\ \mathbf{\gamma }^{\beta
}\psi $ and additional $Q$-deformations resulting in $\ ^{e}\mathbf{j}%
^{\beta }$.

\item Let us explain how to define the nonmetric deformations of the
energy-momentum sources consisting from the electromagnetic and Dirac fields 
\begin{eqnarray}
\ \ ^{DA}\widehat{\mathbf{Y}}_{\alpha \beta } &\rightarrow &\ \ ^{DA}%
\widehat{\yen }_{\alpha \beta }=\ \ \ ^{DA}\widehat{\mathbf{Y}}_{\alpha
\beta }+\ _{Q}^{DA}\widehat{\mathbf{Y}}_{\alpha \beta },\mbox{ where }
\label{qdefsourcm} \\
\ \ ^{DA}\widehat{\mathbf{Y}}_{\alpha \beta } &=&\widehat{\mathbf{Y}}%
_{\alpha \beta }^{[A]}+\widehat{\mathbf{Y}}_{\alpha \beta }^{[D]}=\ \ ^{DA}%
\widehat{\mathbf{T}}_{\alpha \beta }-\frac{1}{2}\mathbf{g}_{\alpha \beta }\
^{DA}\widehat{T}s,  \notag \\
&& \mbox{ with } \ ^{DA}\widehat{T}s :=\ \mathbf{g}^{\alpha \beta }\ ^{DA}%
\widehat{\mathbf{T}}_{\alpha \beta },\mbox{ for }\ ^{DA}\widehat{\mathbf{T}}%
_{\alpha \beta }\ (\ref{semt});  \notag \\
\ _{Q}^{DA}\widehat{\mathbf{Y}}_{\alpha \beta } &=&\ _{Q}\widehat{\mathbf{Y}}%
_{\alpha \beta }^{[A]}+\ _{Q}\widehat{\mathbf{Y}}_{\alpha \beta }^{[D]}=\ \
_{Q}^{DA}\widehat{\mathbf{T}}_{\alpha \beta }-\frac{1}{2}\mathbf{g}_{\alpha
\beta }\ \ _{Q}^{DA}\widehat{T}s,\mbox{
with }\ \ \ _{Q}^{DA}\widehat{T}s:=\ \mathbf{g}^{\alpha \beta }\ _{Q}^{DA}%
\widehat{\mathbf{T}}_{\alpha \beta }.  \notag
\end{eqnarray}%
In these formulas, $\ \ _{Q}^{DA}\widehat{\mathbf{T}}_{\alpha \beta }$ and 
$\ _{Q}^{DA}\widehat{\mathbf{Y}}_{\alpha \beta }$ are computed respectively
by introducing distortions $\mathbf{D}_{\beta }=\widehat{\mathbf{D}}_{\beta
}+\widehat{\mathbf{Z}}_{\beta }$ (\ref{scanconp}) and $\ \mathcal{D}_{\alpha
}^{A}=\ \widehat{\mathcal{D}}_{\alpha }^{A}-\ ^{Q}\widehat{\mathbf{\Gamma }}%
_{\alpha }$ (\ref{qcandir}). We also consider locally anisotropic mass $Q$%
-deformations $\ m_{0}\rightarrow M(u)=m_{0}+\ ^{Q}M(u)$ for $\ \Psi =B(u)\
\psi $ as in (\ref{anisotrm}) in formulas (\ref{starsmaxstr}) used for (\ref%
{semt}). So, $\ _{Q}^{DA}\widehat{\mathbf{Y}}_{\alpha \beta }$ includes the
terms with Dirac and Maxwell fields multiplied to some coefficients
containing $Q$-multiples. Here we note that we can compute distortions of
conventional matter sources using distortions of the formulas (\ref{gfeq2a}%
)-(\ref{emtlc}) with the LC-connection $\nabla $ used in modern MAG and $%
f(Q) $ cosmology cosmological works \cite%
{harko21,iosifidis22,khyllep23,ghil23,koussour23,jhao22,de22,lheis23}. To
study nonmetric deformations of EDM systems we have to modify the formula (%
\ref{emtlc}) for $\ ^{m}\mathcal{L}$ not depending in explicit form on $%
\Gamma _{\ \alpha \beta }^{\gamma }$ in a form involving electromagnetic and
spinor fields as in (\ref{qdefsourcm}). In such an approach, we work with
canonical nonholonomic variables which are important for finding
off-diagonal solutions for nonmetric and metric-compatible nonlinear systems
of PDEs. Nevertheless, the multi-connection character of MAG theories allows
us to postulate certain sources of electromagnetic and Dirac fields not
stating possible distortions in explicit form but using certain general $%
\mathbf{D}_{\beta }=\{\mathbf{\Gamma }_{\ \alpha \nu }^{\beta }\}$ and $%
\mathcal{D}_{\alpha }^{A},$ when 
\begin{eqnarray*}
\ \ ^{DA}\mathbf{T}_{\alpha \beta } &=&\mathbf{T}_{\alpha \beta }^{[A]}+%
\mathbf{T}_{\alpha \beta }^{[D]},\mbox{ where }\  \\
\mathbf{T}_{\alpha \beta }^{[A]} &=&2\mathbf{F}_{\alpha \tau }\mathbf{F}_{\
\ \beta }^{\tau }-\frac{1}{2}\mathbf{g}_{\alpha \beta }(F)^{2},\mbox{ for
}F^{2}=\mathbf{F}^{\alpha \beta }\mathbf{F}_{\alpha \beta }; \\
\ \widehat{\mathbf{T}}_{\alpha \beta }^{[D]} &=&-\frac{i}{2}[\overline{\Psi }%
^{\star }\ \mathbf{\gamma }_{\alpha }\mathcal{D}_{\beta }^{A}\ \Psi +%
\overline{\Psi }\ \mathbf{\gamma }_{\beta }\mathcal{D}_{\alpha }^{A}\
^{\shortmid }\Psi -\mathcal{D}_{\alpha }^{A}(\overline{\Psi })\mathbf{\gamma 
}_{\beta }\Psi -\mathcal{D}_{\beta }^{A}(\overline{\Psi })\mathbf{\gamma }%
_{\alpha }\Psi ].
\end{eqnarray*}%
Such geometric objects are not preserved under transports along curves on $%
_{1}^{3}\mathcal{V}$ in some forms compatible with the gamma matrix
splitting (\ref{gammamatr}). We need more assumptions to include such
objects in a system of gravitational and matter field equation on a
metric-affine manifold or to extract $f(Q)$ or GR configurations. It is not
clear how to prove general decoupling and integration properties of such
systems of nonlinear PDEs. So, we shall prefer in this work to use
nonholonomic canonical variables and matter field sources of type (\ref%
{qdefsourcm}).

\item Using canonical distortions of the Ricci d-tensor $\mathbf{R}ic=%
\widehat{\mathbf{R}}ic+\widehat{\mathbf{Z}}ic$ (\ref{driccidist}) for
nonmetric $f(Q)$ gravity (\ref{cfeq3a}), we can consider $\ _{Q}^{e}\widehat{%
\mathbf{Y}}_{\beta \gamma }\simeq -\widehat{\mathbf{Z}}ic_{_{\beta \gamma }}$
as an additional effective source to $\widehat{\yen }_{\alpha \beta }$ (\ref%
{ceemt}) of the modified Einstein equations (\ref{cfeq4a}). \ We formulate a
nonmetric EDM source by adding $\ ^{DA}\widehat{\mathbf{Y}}_{\alpha \beta
}+\ _{Q}^{DA}\widehat{\mathbf{Y}}_{\alpha \beta }$ (\ref{qdefsourcm}) and
considering a total source 
\begin{eqnarray}
\ \ _{Q}\widehat{\mathbf{J}}_{\alpha \beta } &=&\widehat{\mathbf{Y}}_{\alpha
\beta }^{[A]}+\widehat{\mathbf{Y}}_{\alpha \beta }^{[D]}+\ _{Q}\widehat{%
\mathbf{Y}}_{\alpha \beta }^{[A]}+\ _{Q}\widehat{\mathbf{Y}}_{\alpha \beta
}^{[D]}+\ _{Q}^{e}\widehat{\mathbf{Y}}_{\beta \gamma }  \label{totsnmedm} \\
&\rightarrow &\widehat{\mathbf{J}}_{\alpha \beta }=\widehat{\mathbf{Y}}%
_{\alpha \beta }^{[A]}+\widehat{\mathbf{Y}}_{\alpha \beta }^{[D]},%
\mbox{ for
}Q\rightarrow 0.  \notag
\end{eqnarray}
\end{enumerate}

For above assumptions 1-6, the nonmetric $f(Q)$-deformations of the
canonical EDM equations (\ref{candeinst}) for (\ref{semt}) can be written in
the form:%
\begin{eqnarray}
\lbrack i\ \hbar \gamma ^{\alpha }(\ \widehat{\mathcal{D}}_{\alpha }-iq%
\mathbf{A}_{\alpha }-\ ^{Q}\widehat{\mathbf{\Gamma }}_{\alpha }\ )-m_{0}\ -\
^{Q}M(u)]\ \Psi &=&0,  \label{cnmdir} \\
\ \widehat{\mathbf{D}}_{\alpha }\widehat{\mathbf{F}}^{\alpha \beta }+%
\widehat{\mathbf{D}}_{\alpha }\ ^{Q}\widehat{\mathbf{F}}^{\alpha \beta }
&=&q(\mathbf{j}^{\beta }+\ ^{e}\mathbf{j}^{\beta })  \label{cnamax} \\
\ \widehat{\mathbf{R}}_{\alpha \beta } &=&\ _{Q}\widehat{\mathbf{J}}_{\alpha
\beta }.  \label{cnmeinst}
\end{eqnarray}%
The structure of such systems of nonlinear PDEs is similar for other types
of nonmetric theories if we work in canonical dyadic variables for effective
models on nonholonomic Lorentz manifolds, or with generalizations on (co)
tangent Lorentz bundles. For those classes of theories, the
geometric/physical d-objects labeled with left "Q or e" in (\ref{cnmdir})-(%
\ref{cnmeinst}) are defined and computed using different principles than
those for $f(Q)$-deformations.

Finally, we note that nonmetric distortions with LC-configurations $\ \nabla 
$ can be extracted from (\ref{cnmdir}), (\ref{cnamax}), and (\ref{cnmeinst})
by imposing some zero torsion conditions, $\widehat{\mathbf{T}}_{\ \alpha
\beta }^{\gamma }=0,$ as we explain in details in \cite{lb24epjc2}. If we
impose such LC-conditions from the very beginning, we are not able to apply
the AFCDM and decouple and integrate in explicit form respective systems of
nonlinear PDEs. The priority to use "hat" variables for geometric
constructions is that they allow to decouple and integrate such nonmetric
EDM and other type physical important equations in general off-diagonal
form and respective nonmetric distortions.

\section{Nonmetric geometric flow evolution of nonholonomic EDM systems}

\label{sec3}Models of metric noncompatible geometric flows and nonmetric MGTs were studied in section 2 of \cite{lb24epjc1} and section 3 in \cite{lb24epjc2} considering $\tau $-running effective and matter field Lagrange densities not involving (nonmetric) spinor variables and Dirac operators. In such theories, $\tau $ is treated as a temperature-like geometric evolution parameter \cite{perelman1}). In this section, we generalize those results modelling nonmetric geometric flow of EDM structures when nonmetric $f(Q)$-distortions of canonical nonholonomic gravitational equations (\ref{candeinst}) with Dirac and Maxwell field source (\ref{semt}) are generated for a fixed $\tau =\tau _{0}$ as certain nonmetric Ricci solitons.

We emphasize that in our works \cite{svnonh08,sv12,vacaru20} on nonholonomic metric and nonmetric geometric flows and applications to MGTs we do not attempt to formulate and prove certain generalizations of the famous Poincar\'{e}--Thurston conjecture \cite{hamilton82,perelman1}. Rigorous mathematical rigorous proofs of the results for Ricci flows of Riemannian metrics consists of some hundred of pages as in monographs \cite{monogrrf1,monogrrf2,monogrrf3}. The topological and geometric analysis methods and main results depend on the type of affine connections we use in a model of nonmetric MGT. In general, we can construct an infinite number of nonmetric flow theories if we drop the condition of the metric compatibility. In such cases, we can speculate on nonmetric physical viable modifications if we consider, for instance, $f(Q)$-distortions of the geometric objects (primary data) on a Lorentz manifold and considering respective (target) F- and W-functionals (to be defined bellow). Using the nonmetric distorted W-functional as a "minus" entropy we can derive respective thermodynamic variables. 

In abstract geometric or N-adapted variational form, we can also derive (in equivalent form, from F- and W-functionals) respective nonmetric geometric flow equations. In \cite{lb24epjc1,lb24epjc2}, we proved that applying the AFCDM we can construct various classes of $\tau $-families of off-diagonal quasi-stationary and cosmological solutions in MGTs modelled as nonmetric Ricci solitons. We can decide if certain classes of such solutions describe physical viable models by analyzing corresponding physical properties and
evaluating corresponding nonmetric G. Perelman thermodynamic variables.

\subsection{Nonmetric geometric flows of EDM systems}

In this subsection, we elaborate on nonmetric geometric flow models of EDM systems formulated in canonical dyadic variables. We consider $\tau $-families of nonmetric geometric data 
\begin{equation}
(_{1}^{3}\mathcal{V}\mathbf{,N}(\tau ),\mathbf{g}(\tau ),\mathbf{D}(\tau
)=\nabla (\tau )+\mathbf{L}(\tau )=\widehat{\mathbf{D}}(\tau )+\widehat{%
\mathbf{L}}(\tau ),\ ^{g}\widehat{\mathcal{L}}(\tau )+\ ^{nDA}\widehat{%
\mathcal{L}}(\tau )),  \label{canonicalmafdata}
\end{equation}%
which are parameterized by a real parameter $\tau ,0\leq \tau \leq \tau _{1}, $ and when corresponding left labels are used ($g$ for gravitational fields and $nDA$ for nonmetric Dirac-Maxwell fields). The Lagrange densities for nonmetric gravitational field and matter fields, $\ ^{g}\widehat{\mathcal{L}}(\tau )$ and 
$\ ^{nDA}\widehat{\mathcal{L}}(\tau )$ in (\ref{canonicalmafdata}) are determined by certain additional parameterizations of
(\ref{actnmc}), 
\begin{equation}
\ ^{nDA}\widehat{\mathcal{L}}(\tau )=\ ^{A}\widehat{\mathcal{L}}(\tau )+ 
\ ^{D}\widehat{\mathcal{L}}(\tau ) + \ _{Q}^{A}\widehat{\mathcal{L}}(\tau )+
\ _{Q}^{D}\widehat{\mathcal{L}}(\tau ) + \ _{Q}^{e}\widehat{\mathcal{L}}(\tau ).
\label{mlagd}
\end{equation}%
This results in similar sums (with left labels $A,D,...,$ for corresponding distortions and nonholonomic distributions) of effective sources $\ \widehat{\mathbf{J}}_{\alpha \beta }$ (\ref{totsnmedm})  for nonmetric Einstein and Dirac fields. For geometric flow models, the geometric and physical objects for such theories depend, in general, on all spacetime coordinates. To simplify our notations, we shall write, for instance, $\mathbf{g}(\tau )$ instead of $\mathbf{g}(\tau ,u^{\beta }), \mathbf{D}(\tau )$ instead of 
$\mathbf{D}(\tau ,u^{\beta })$ if such abstract and parametric notations will not result in ambiguities. We suppose that  we can choose any form of effective matter with nonmetric Dirac spinor and electromagnetic fields (\ref{mlagd}). For any fixed value $\tau =\tau _{0}$,  a nonmetric geometric flow model defines a nonmetric EDM theory with effective field equations (\ref{cnmdir}), (\ref{cnamax}), and (\ref{cnmeinst}).

\subsubsection{$f(Q)$-deformed F- and W-functionals}

A fundamental step in elaborating various types of geometric flow theories and applications in modern physics consists in the definition and generalization of the so-called $\mathcal{F}$- and $\mathcal{W}$-functionals \cite{perelman1,svnonh08,sv12} from which the geometric flow equations can be proved in variational form. Such systems of nonlinear evolution systems of PDEs are called  as R. Hamilton or Hamilton-Friedan equations \cite{hamilton82,vacaru20}, see \cite{lb24epjc1,lb24epjc2,lb24grg} for recent nonmetric MGTs and noassociative EDMe developments. In a general form of metric-affine distortions, it is not clear how mathematically can be formulated and proved relativistic variants of such conjectures and generalizations. The constructions depend on the type of nonmetric models or respective nonholonomic structures for supersymmetric, nonassociative, noncommutative, Finsler like geometric and physical models. Nevertheless, generalizations of  $\mathcal{F}$- and $\mathcal{W}$-functionals allow to elaborate on geometric
thermodynamic models (see next subsection) which can be associated with respective classes of solutions of physically important systems on nonlinear PDEs using the AFCDM.

For nonmetric geometric flows of EDM systems (\ref{mlagd}) we postulate the modified G. Perelman's functionals: 
\begin{eqnarray}
\ _{Q}\widehat{\mathcal{F}}(\tau ) &=&\int_{t_{1}}^{t_{2}}\int_{\Xi _{t}}e^{-%
\widehat{\zeta }(\tau )}\sqrt{|\mathbf{g}(\tau )|}\delta ^{4}u[\widehat{%
\mathbf{R}}sc(\tau )+\ ^{nDA}\widehat{\mathcal{L}}(\tau )+|\widehat{\mathbf{D%
}}(\tau )\widehat{\zeta }(\tau )|^{2}],  \label{fperelmDA} \\
\ _{Q}\widehat{\mathcal{W}}(\tau ) &=&\int_{t_{1}}^{t_{2}}\int_{\Xi
_{t}}\left( 4\pi \tau \right) ^{-2}e^{-\widehat{\zeta }(\tau )}\sqrt{|%
\mathbf{g}(\tau )|}\delta ^{4}u[\tau (\widehat{\mathbf{R}}sc(\tau )+\ ^{nDA}%
\widehat{\mathcal{L}}(\tau )+|\widehat{\mathbf{D}}(\tau )\widehat{\zeta }%
(\tau )|^{2})+\widehat{\zeta }(\tau )-4],  \label{wfperelmDA}
\end{eqnarray}%
In the formulas, we can impose the normalization conditions%
\begin{equation}
\int_{t_{1}}^{t_{2}}\int_{\Xi _{t}}\left( 4\pi \tau \right) ^{-2}e^{-%
\widehat{\zeta }(\tau )}\sqrt{|\mathbf{g}|}d^{4}u=1,  \label{normcond}
\end{equation}%
where $\widehat{\zeta }(\tau )=\widehat{\zeta }(\tau ,u)$ is the normalizing function. In our approach, the F- and W- functionals are postulated in hat variables to derive further geometric flow evolution
equations which can be decoupled and integrated in certain general forms. The left abstract labels $Q$ and $nDA$ state that the nonmetricity fields are included in such forms that for LC-configurations with a fixed $\tau _{0} $ we obtain equations of type (\ref{gfeq2a}) - (\ref{emtlc}).

We can redefine the normalizing function, $\widehat{\zeta }(\tau )\rightarrow $ $\zeta (\tau ),$ from (\ref{fperelmDA}) and (\ref{wfperelmDA}), in  different forms which allow to generate, or absorb, respective distortions of affine connections and work with respective functionals 
$\ _{Q}\mathcal{F}(\tau ,\zeta ,\mathbf{g},\mathbf{D},\mathbf{R}sc,...)$ and 
$\ _{Q}\mathcal{W}(\tau ,\zeta ,\mathbf{g},\mathbf{D},\mathbf{R}sc,...).$ In such "non-hat" cases it will be not possible to generate nonmetric geometric and physical important systems of nonlinear PDEs which are integrable in certain general forms. The difference from the original F- and W-functionals \cite{perelman1} introduced for 3-d Riemannian $\tau $-flows $(g(\tau),\nabla (\tau ))$ is that in our works we study geometric flows of canonical geometric data $(\mathbf{g}(\tau ), \mathbf{N}(\tau ),\widehat{\mathbf{D}}(\tau )).$ For $Q$-deformations of EDM systems on nonholonomic Lorentz manifolds, this results in physically important systems of nonlinear PDEs which can be decoupled and integrated as in \cite{lb24epjc1,lb24epjc2}.

\subsubsection{Geometric flow equations and nonholonomic Ricci solitons of nonmetric EDM systems}

There are two possibilities to derive geometric flow equations from functionals 
$\ _{Q}\widehat{\mathcal{F}}(\tau )$ (\ref{fperelmDA}) and 
$\ _{Q}\widehat{\mathcal{W}}(\tau )$ (\ref{wfperelmDA}). In the first case \cite{svnonh08,sv12,vacaru20}, we can use $\widehat{\mathbf{D}}(\tau )$ instead of $\nabla (\tau )$ and reproduce in N-adapted distorted form all covariant differential and integral calculus from \cite{perelman1,monogrrf1,monogrrf2,monogrrf3}. Such proofs are on hundred of pages as in and depends on the type of nonmetric MGT.

In the second case, we use an abstract geometric formalism as in \cite{misner,lb24epjc1,lb24epjc2}. For such an approach, the geometric and physical objects and important physical equations can be derived by
corresponding distortions of geometric objects in (pseudo) Riemannian geometry to corresponding nonholonomic metric-affine ones. We can use nonholonomic canonical variables with $\nabla (\tau )\rightarrow \mathbf{D}(\tau )=\widehat{\mathbf{D}}(\tau )+\widehat{\mathbf{K}}(\tau )+\ ^{q}%
\widehat{\mathbf{Z}}(\tau ),$ see (\ref{cotors}) and (\ref{disf}). Respective generalizations of Ricci, torsion, and energy-momentum d-tensors (in our case, for nonmetric EDM systems) have to be defined. Considering some effective data $(\mathbf{g}=\{\mathbf{g}_{\mu \nu }=[g_{ij},g_{ab}]\},\mathbf{N}%
=\{N_{i}^{a}\},\widehat{\mathbf{D}},\ ^{nDA}\widehat{\mathcal{L}}(\tau )),$ see the parametrization for $\ ^{nDA}\widehat{\mathcal{L}}(\tau )$ (\ref{mlagd}), the nonmetric geometric flow evolution equations are postulated: 
\begin{eqnarray}
\partial _{\tau }g_{ij}(\tau ) &=&-2[\widehat{\mathbf{R}}_{ij}(\tau )-\ _{Q}%
\widehat{\mathbf{J}}_{ij}(\tau )];\ \partial _{\tau }g_{ab}(\tau )=-2[%
\widehat{\mathbf{R}}_{ab}(\tau )-\ _{Q}\widehat{\mathbf{J}}_{ab}(\tau )];
\label{ricciflowr2} \\
\widehat{\mathbf{R}}_{ia}(\tau ) &=&\widehat{\mathbf{R}}_{ai}(\tau )=0;%
\widehat{\mathbf{R}}_{ij}(\tau )=\widehat{\mathbf{R}}_{ji}(\tau );\widehat{%
\mathbf{R}}_{ab}(\tau )=\widehat{\mathbf{R}}_{ba}(\tau );\   \notag \\
\partial _{\tau }\widehat{\zeta }(\tau ) &=&-\widehat{\square }(\tau )[%
\widehat{\zeta }(\tau )]+\left\vert \widehat{\mathbf{D}}(\tau )[\widehat{%
\zeta }(\tau )]\right\vert ^{2}-\ \widehat{R}sc(\tau )+\ \ _{Q}\widehat{%
\mathbf{J}}_{\ \alpha }^{\alpha }(\tau ).  \notag
\end{eqnarray}%
Such systems of nonlinear PDEs consist of nonholonomic generalizations of the Hamilton-Friedan equations \cite{hamilton82,friedan2,friedan3}. In our case, they encode nonmetricity and describe metric-affine geometric flows of EDM systems. In formulas (\ref{ricciflowr2}), we use a generalized Laplace operator $\widehat{\square }(\tau )=\widehat{\mathbf{D}}^{\alpha }(\tau )%
\widehat{\mathbf{D}}_{\alpha }(\tau ),$ when the conditions $\widehat{\mathbf{R}}_{ia}=\widehat{\mathbf{R}}_{ai}=0$ for the Ricci tensor $\widehat{%
R}ic[\widehat{\mathbf{D}}]=\{\widehat{\mathbf{R}}_{\alpha \beta }=[\widehat{R%
}_{ij},\widehat{R}_{ia},\widehat{R}_{ai},\widehat{R}_{ab,}]\}$ have to be imposed if we want to keep the metric $\mathbf{g}(\tau )$ to be symmetric under respective nonholonomic and nonmetric Ricci flow evolution scenarios.

In the next sections, we shall construct in explicit form and study the physical implications of quasi-stationary solutions of nonmetric geometric flow equations (\ref{ricciflowr2}) for a $\tau $-families of d-metrics $\mathbf{g}(\tau )=[g_{i}(\tau ),g_{a}(\tau ),N_{i}^{a}(\tau )]$ (\ref{dm}), when the N-adapted coefficients $g_{\alpha }(\tau )$ do not depend on the time-like variable $y^{4}=t$ and can be parameterized in the form 
\begin{equation}
g_{i}(\tau )=e^{\psi {(\tau ,x^{j})}},g_{a}(\tau )=h_{a}(\tau
,x^{k},y^{3}),\ N_{i}^{3}=w_{i}(\tau
,x^{k},y^{3}),\,\,\,\,N_{i}^{4}=n_{i}(\tau ,x^{k},y^{3}).
\label{stationarydm}
\end{equation}%
Here we note that in \cite{lb24epjc2} we consider locally anisotropic
cosmological solutions which are can be generated as certain dual
configurations with $y^{3}\rightarrow y^{4}=t$ and changing of indices $%
3\rightarrow 4,4\rightarrow 3,$ when $g_{a}(\tau ,x^{k},t)$ and $%
N_{i}^{a}(\tau ,x^{k},t)$ depend explitly on the time-like variables but not
on $y^{3}.$

For generating quasi-stationary $\tau $-solutions in explicit form, we can
introduce effective sources $\ _{Q}\widehat{\mathbf{J}}_{\alpha }(\tau
)=diag[\ _{Q}^{h}\widehat{\mathbf{J}}(\tau ),\ _{Q}^{v}\widehat{\mathbf{J}}%
(\tau )]$ related to the source $\ _{Q}\widehat{\mathbf{J}}_{ab}(\tau )$ (%
\ref{ricciflowr2}) via N--adapted frame transforms, 
\begin{equation}
\ _{Q}\widehat{\mathbf{J}}_{\ \nu }^{\mu }(\tau )=diag[\ _{Q}\widehat{%
\mathbf{J}}_{\alpha }(\tau )]=\mathbf{e}_{\ \mu }^{\mu ^{\prime }}(\tau )%
\mathbf{e}_{\nu }^{\ \nu ^{\prime }}(\tau )[~\ _{Q}\widehat{\mathbf{J}}_{\mu
^{\prime }\nu ^{\prime }}(\tau )-\frac{1}{2}~\partial _{\tau }\mathbf{g}%
_{\mu ^{\prime }\nu ^{\prime }}(\tau )]=[~\ _{Q}^{h}\widehat{\mathbf{J}}%
(\tau ,{x}^{k})\delta _{j}^{i},\ _{Q}^{v}\widehat{\mathbf{J}}(\tau
,x^{k},y^{3})\delta _{b}^{a}].  \label{dsourcparam}
\end{equation}%
In (\ref{dsourcparam}), we consider $\tau $-families of vierbein transforms $%
\mathbf{e}_{\ \mu ^{\prime }}^{\mu }(\tau )=\mathbf{e}_{\ \mu ^{\prime
}}^{\mu }(\tau ,u^{\gamma })$ and their dual $\mathbf{e}_{\nu }^{\ \nu
^{\prime }}(\tau ,u^{\gamma })$, where $\mathbf{e}_{\ }^{\mu }=\mathbf{e}_{\
\mu ^{\prime }}^{\mu }du^{\mu ^{\prime }},$ and $[\ _{Q}^{h}\widehat{\mathbf{%
J}}(\tau ,{x}^{k}),\ _{Q}^{v}\widehat{\mathbf{J}}(\tau ,x^{k},y^{3})] $ can
be fixed as generating matter sources corresponding to $\ ^{nDA}\widehat{%
\mathcal{L}}(\tau )$ (\ref{mlagd}) and imposing nonholonomic frame
constraints on quasi-stationary distributions of (effective) matter fields.

Nonmetric Ricci solitons were defined and studied in \cite{lb24epjc1,lb24epjc2} as self-similar configurations for the corresponding nonmetric geometric flow equations. Fixing $\tau =\tau _{0}$ in (\ref{ricciflowr2}), we obtain the equations for the modified Einstein equations with Dirac-Maxwell sources (\ref{cnmeinst}), which can be written for quasi-stationary sources (\ref{dsourcparam}) in the form: 
\begin{align}
\widehat{\mathbf{R}}_{ij}& =\ _{Q}^{h}\widehat{\mathbf{J}}(\tau _{0},{x}%
^{k}),\ \widehat{\mathbf{R}}_{ab}=\ _{Q}^{v}\widehat{\mathbf{J}}(\tau
_{0},x^{k},y^{3}),  \label{canriccisolda} \\
\widehat{\mathbf{R}}_{ia}& =\widehat{\mathbf{R}}_{ai}=0;\widehat{\mathbf{R}}%
_{ij}=\widehat{\mathbf{R}}_{ji};\widehat{\mathbf{R}}_{ab}=\widehat{\mathbf{R}%
}_{ba}.  \notag
\end{align}%
For additional nonholonomic constraints, such equations define quasi-stationary solutions of the nonmetric Einstein equations (\ref{gfeq2b}) for $\nabla (\tau _{0})$ but involving contributions from $\ ^{nDA}%
\widehat{\mathcal{L}}(\tau _{0})$ (\ref{mlagd}).

\subsubsection{Generalizing G. Perelman thermodynamics for nonmetric EDM systems}

We can consider different types of normalization functions for functionals (\ref{fperelmDA}) and (\ref{wfperelmDA}), $\zeta (\tau )\rightarrow \widehat{\zeta }(\tau ),$ when 
\begin{equation}
\partial _{\tau }\zeta (\tau )+\widehat{\square }(\tau )[\zeta (\tau
)]-\left\vert \widehat{\mathbf{D}}\zeta (\tau )\right\vert ^{2}-\ ^{nDA}%
\widehat{\mathcal{L}}(\tau )=\partial _{\tau }\ \widehat{\zeta }(\tau )+%
\widehat{\square }(\tau )[\widehat{\zeta }(\tau )]-\left\vert \widehat{%
\mathbf{D}}\widehat{\zeta }(\tau )\right\vert ^{2}.  \label{normcondc1}
\end{equation}%
This is possible if we use for (\ref{normcond}) a $\widehat{\zeta }(\tau )$ satisfying the conditions 
\begin{equation}
\partial _{\tau }\widehat{\zeta }(\tau )=-\widehat{\square }(\tau )[\widehat{%
\zeta }(\tau )]+\left\vert \widehat{\mathbf{D}}\widehat{\zeta }(\tau
)\right\vert ^{2}-\widehat{\mathbf{R}}sc(\tau ).  \label{normcondc}
\end{equation}%
Such normalization functions, $\zeta (\tau )$ or $\widehat{\zeta }(\tau ),$ define different integration measures in topological type theories. In the geometric evolution or gravitational and matter field equations described by corresponding systems of nonlinear PDEs, we use different types of transforms 
$\zeta (\tau )\rightarrow \widehat{\zeta }(\tau )$ which allow us to absorb, or (inversely) distinguish different types of (effective) $\tau $-running Lagrange densities.

In this work, we model geometric flow evolution scenarios of nonmetric EDM systems. In terms of a corresponding integration measure, the W-functional (\ref{wfperelmDA}) can be written in the form 
\begin{equation}
\ _{Q}\widehat{\mathcal{W}}(\tau )=\int_{t_{1}}^{t_{2}}\int_{\Xi _{t}}\left(
4\pi \tau \right) ^{-2}e^{-\widehat{\zeta }(\tau )}\sqrt{|\ _{Q}\mathbf{g}%
(\tau )|}\delta ^{4}u[\tau (\ _{Q}\widehat{\mathbf{R}}sc(\tau )+|\ _{Q}%
\widehat{\mathbf{D}}(\tau )\widehat{\zeta }(\tau )|^{2}+\widehat{\zeta }%
(\tau )-4].  \label{wf1}
\end{equation}%
In this formula,  the effective nonmetric and matter sources $\ ^{nDA}\widehat{\mathcal{L}}(\tau )$ are encoded into geometric data if we consider an explicit class of solutions of (\ref{ricciflowr2}) related for $\tau =\tau _{0}$ to solutions of the nonmetric EDM equations (\ref{cnmdir}) - (\ref{cnmeinst}).
We use left $Q$-labels in (\ref{wf1}) (for instance, writing $\ _{Q}\mathbf{g}(\tau ))$ to emphasize that the nonmetric geometric objects are defined by certain classes of solutions of (\ref{ricciflowr2}), or (\ref{canriccisolda}).

On a metric-affine space $_{1}^{3}\mathcal{V}$ a canonical nonholonomic 2+2 decomposition is important for generating off-diagonal solutions but for elaborating thermodynamical models we have to consider an additional nonholonomic (3+1) splitting. Such double nonmetric spacetime fibrations allow us to introduce the statistical partition function 
\begin{equation}
\ _{Q}\widehat{Z}(\tau )=\exp [{\int_{\widehat{\Xi }}[-\widehat{\zeta }+2]\ \left( 4\pi \tau \right) ^{-2}e^{-\widehat{\zeta }}\ \delta \ _{Q}\widehat{\mathcal{V}}(\tau )}],  \label{spf}
\end{equation}%
where the volume element is defined and computed as 
\begin{equation}
\delta \ _{Q}\widehat{V}(\tau )=\sqrt{|\ _{Q}\mathbf{g}(\tau )|}\
dx^{1}dx^{2}\delta y^{3}\delta y^{4}\ .  \label{volume}
\end{equation}

We can modify the variational procedure provided in section 5 of \cite{perelman1}) for geometric flows of Riemannian metrics in a form to include $Q$-deformations and using $\ _{Q}\widehat{Z}(\tau )$ (\ref{spf}) and $\ _{Q}\widehat{\mathcal{W}}(\tau )$ (\ref{wf1}) on a closed region in $_{1}^{3}\mathcal{V}$. Such tedious calculations allow us to define and compute nonmetric distortions of G. Perelman thermodynamic variables. In an equivalent form, we can apply the abstract geometric formalism \cite{misner,lb24epjc1,lb24epjc2} for canonical nonholonomic variables and define nonmetric analogues of respective energy, entropy and quadratic fluctuations functionals: 
\begin{align}
\ \ _{Q}\widehat{\mathcal{E}}\ (\tau )& =-\tau ^{2}\int_{\widehat{\Xi }}\
\left( 4\pi \tau \right) ^{-2}\left( \widehat{\mathbf{R}}sc+|\ \widehat{%
\mathbf{D}}\ \widehat{\zeta }|^{2}-\frac{2}{\tau }\right) e^{-\widehat{\zeta 
}}\ \delta \ _{Q}\widehat{V}(\tau ),  \label{qthermvar} \\
\ \ \ _{Q}\widehat{S}(\tau )& =-\ _{Q}\widehat{\mathcal{W}}(\tau )=-\int_{%
\widehat{\Xi }}\left( 4\pi \tau \right) ^{-2}\left( \tau (\widehat{\mathbf{R}%
}sc+|\widehat{\mathbf{D}}\widehat{\zeta }|^{2})+\widehat{\zeta }-4\right)
e^{-\widehat{\zeta }}\ \delta \ _{Q}\widehat{V}(\tau ),  \notag \\
\ \ \ _{Q}\widehat{\sigma }(\tau )& =2\ \tau ^{4}\int_{\widehat{\Xi }}\left(
4\pi \tau \right) ^{-2}|\ \widehat{\mathbf{R}}_{\alpha \beta }+\widehat{%
\mathbf{D}}_{\alpha }\ \widehat{\mathbf{D}}_{\beta }\widehat{\zeta }_{[1]}-%
\frac{1}{2\tau }\ _{Q}\mathbf{g}_{\alpha \beta }|^{2}e^{-\widehat{\zeta }}\
\delta \ _{Q}\widehat{V}(\tau ).  \notag
\end{align}%
The nonmetric geometric thermodynamic variables (\ref{qthermvar}) can be considered for characterizing different classes of metric-affine theories. For instance, including Lagrange densities $\widehat{f}(\widehat{\mathbf{R}}sc)$ for nonmetric MGTs, or to consider nonmetric deformations of the Dirac and Maxwell fields. The quadratic fluctuation thermodynamic variable $\ _{Q}\widehat{\sigma }(\tau )$ can be written as a functional of $\widehat{\mathbf{R}}_{\alpha \beta }$ even 
$\ _{Q}\widehat{\mathcal{E}}\ (\tau )$ and $\ _{Q}\widehat{S}(\tau )$ are functionals of $\widehat{%
\mathbf{R}}sc$ if we correspondingly re-define the normalizing functions 
$\widehat{\zeta }\rightarrow \widehat{\zeta }_{[1]}$. We omit such details in this work because we shall not compute $\ _{Q}\widehat{\sigma }(\tau )$ for certain classes of solutions.

\subsubsection{Alternative models of nonmetric geometric flow thermodynamics}

A corresponding re-definition of the normalizing function $\zeta (\tau )\rightarrow 
\widehat{\zeta }(\tau )$ (for instance, in the form (\ref{normcondc})) allows us to introduce in the basic formulas for nonmetric geometric flows certain explicit dependencies on $Q$-deformed terms and
(effective) $\tau $-running sources, $\ ^{nDA}\widehat{\mathcal{L}}(\tau )$ (\ref{mlagd}) and 
$\ _{Q}\widehat{\mathbf{J}}_{\alpha \beta }(\tau ),$ see (\ref{totsnmedm}). In a more general form, we can work directly with an arbitrary $\tau $-family of d-connections $\mathbf{D}(\tau )$ and define,
for instance, "alternative" thermodynamic variables: 
\begin{align}
\ \ \mathcal{E}\ (\tau )& =-\tau ^{2}\int_{\Xi }\ \left( 4\pi \tau \right)
^{-2}\left( \mathbf{R}sc+|\ \mathbf{D}\ \zeta |^{2}-\frac{2}{\tau }\right)
e^{-\zeta }\ \delta V(\tau ),  \label{qthermvarma} \\
\ \ \ S(\tau )& =-\ \mathcal{W}(\tau )=-\int_{\Xi }\left( 4\pi \tau \right)
^{-2}\left( \tau (\mathbf{R}sc+|\mathbf{D}\zeta |^{2})+\zeta -4\right)
e^{-\zeta }\ \delta V(\tau ),  \notag \\
\ \ \ \sigma (\tau )& =2\ \tau ^{4}\int_{\Xi }\left( 4\pi \tau \right)
^{-2}|\ \mathbf{R}_{\alpha \beta }+\mathbf{D}_{\alpha }\mathbf{D}_{\beta
}\zeta _{\lbrack 1]}-\frac{1}{2\tau }\mathbf{g}_{\alpha \beta
}|^{2}e^{-\zeta }\ \delta V(\tau ),  \notag
\end{align}%
where $\mathbf{R}sc(\tau ),\mathbf{R}_{\alpha \beta }(\tau ),\delta V(\tau )$ etc. are some defined by corresponding families of metric-affine geometric data $(\mathbf{g}_{\alpha \beta }(\tau ),\mathbf{D}(\tau )).$ Such metric-affine geometric flow and thermodynamic theories can be also formulated. Computing and analyzing the corresponding thermodynamic data $\left[ \ _{Q}\widehat{\mathcal{E}}(\tau ),\ _{Q}\widehat{\mathcal{S}}(\tau),\ _{Q}\widehat{\sigma }(\tau )\right] $ (\ref{qthermvar}) and $\left[ \ 
\mathcal{E}(\tau ),\mathcal{S}(\tau ),\sigma (\tau )\right] $ (\ref{qthermvarma}), we can speculate which model is more realistic (for some minimal entropies, minimal or maximal effective energy, less quadratic fluctuations etc.). This is possible if we introduce canonical nonholonomic variables and corresponding distortions $\mathbf{D}(\tau )=\widehat{\mathbf{D}}(\tau )+\widehat{\mathbf{K}}(\tau )+\ ^{q}\widehat{\mathbf{Z}}(\tau )$ as for (\ref{cotors}) and (\ref{disf}). Contrary, we are not able to decouple and integrate in certain general forms the metric-affine geometric flow equations corresponding to (\ref{qthermvarma}) and even to compute the corresponding variables. So, to work with canonical hat variables is both a conceptual issue and of elaborating a geometric technique with computational priorities.

Another important priority of the canonical thermodynamic geometric flow variables (\ref{qthermvar})  is that they can be computed in explicit form using respective classes of nonlinear symmetries for quasi-stationary
solutions (see next subsection), when certain effective $\tau $-running cosmological constants can be introduced. Such results were obtained and studied for nonmetric geometric flow and MGTs in our partner works \cite{lb24epjc1,lb24epjc2} (for off-diagonal quasi-stationary or locally anisotropic cosmological configurations), when we used canonical nonholonomic variables, but not for nonmetric EDM systems. In this work, the nonmetric geometric flow thermodynamics is formulated and studied for new classes of quasi-stationary solutions encoding nonmetric EDM data.

Finally, we note that fixing the temperature $\tau _{0}$ in (\ref{qthermvar}), we can compute the thermodynamic variables $\left[ \ \ _{Q}\widehat{\mathcal{E}}(\tau _{0}),\ 
\ _{Q}\widehat{\mathcal{S}}(\tau _{0}),\ \ _{Q}\widehat{\sigma }(\tau _{0})\right] $ for respective nonmetric Ricci-Dirac-Maxwell solitons (\ref{canriccisolda}). Certain classes of general off-diagonal solutions can be not well-defined in the general form if, for instance, $\ _{Q}\widehat{\mathcal{S}}(\tau _{0})<0.$ Such solutions can be physically viable if there are some more special sub-classes of
nonholonomic distributions/distortions which allow to formulate and compute physically viable configurations  (with relativistic causality, well thermodynamic behaviour and effective energetic regimes). As we explained in  \cite{lb24epjc1,lb24epjc2} for respective examples, in some spacetime regions, the nonmetric $Q$-deformations may result in un-physical models, but be well-defined for other nonholonomic conditions and interactions with matter fields (in our case, nonmetric Dirac-Maxwell ones) because of different sign contributions. Such issues have to be investigated in explicit form for any class of exact/parametric solutions of nonmetric EDM systems with stated boundary conditions for quasi-stationary solutions; or certain prescribed initial conditions for locally anisotropic cosmological configurations.

\subsection{Quasi-stationary solutions for nonmetric EDM systems}

In \cite{lb24grg} we constructed quasi-stationary solutions and analyzed the physical properties of  BH and  WH solutions in nonassociative R-flux modified EDM theory. The goal of this subsection is to show how applying the AFCDM we can construct similar solutions encoding nonmetric deformations of EDM quasi-stationary configurations in GR. The new classes of solutions will be generated for the system of nonlinear PDEs (\ref{ricciflowr2}), or (\ref{cnmdir}) - (\ref{cnmeinst}), in a form to include nonmetric Dirac and Maxwell configurations from \cite{lb24epjc1,lb24epjc2}. Here we note that such classes of nonmetric EDM solutions are generic off-diagonal and may depend on all spacetime coordinates and on a geometric flow evolution parameter $\tau $. Technically, to generate solutions with (at least) one spacetime Killing symmetry it is more simple. In this work, we consider quasi-stationary solutions with Killing symmetry on $\mathbf{e}_{4}=\partial _{4}$ when the geometric d-objects do not depend on the time-like coordinate $u^{4}=y^{4}=t.$ We explained in detail in \cite{lb24epjc2,lb24grg} how using abstract geometric principles, the quasi-stationary solutions can be transformed into locally anisotropic cosmological ones, with Killing symmetry on $\mathbf{e}_{3}=\partial _{3}$ using duality properties, various nonlinear symmetries and transforms of formulas for ($\mathbf{e}_{4}\leftrightarrow \mathbf{e}_{3} $ etc.).

The details and proofs on general decoupling properties and constructing quasi-stationary solutions of nonmetric metric-affine geometric flow equations and modified Einstein equations can be found in Appendix A of \cite{lb24epjc1}. Explicit examples of off-diagonal solutions for nonmetric solitonic hierarchies and nonmetric WH solitonic deformations are provided in sections 3.2 and 3.3 of that work. In this subsection, we show how those ansatz and nonlinear symmetries can be extended to generate quasi-stationary nonmetric EDM configurations. Necessary results of subsection 3.3 of \cite{lb24grg} are in such way modified that allow  computing in general form solutions for $f(Q)$-deformations of the Dirac-Maxwell fields and respective effective masses and currents using formulas (\ref{cnmdir}) and (\ref{cnamax}).

\subsubsection{The off-diagonal ansatz and nonmetric EDM generating functions and sources}

The ansatz for generating quasi-stationary solutions of nonassociative geometric flow equations (\ref{ricciflowr2}) can be chosen in the form: 
\begin{eqnarray}
d\ ^{q}s^{2}(\tau ) &=&\ ^{q}g_{i}(\tau ,x^{k})(dx^{i})^{2}+\ ^{q}g_{a}(\tau
,x^{i},y^{3})(\ ^{q}\mathbf{e}^{a}(\tau ))^{2},\mbox{where }  \notag \\
\ ^{q}\mathbf{e}^{a}(\tau ) &=&dy^{a}+\ ^{q}N_{k}^{a}(\tau
,x^{i},y^{3})dx^{k},\   \label{ans1rf}
\end{eqnarray}%
where the label "q" states that corresponding coefficients are considered for a quasi-stationary configuration. By tedious straightforward computations, we can verify that parameterizations of $\tau $-families of
d-metrics $\ ^{q}\mathbf{g}_{\alpha }(\tau )=(\ ^{q}g_{i}(\tau ), \ ^{q}g_{a}(\tau ))$ of type (\ref{dm}) allow us to generate quasi-stationary solutions for effective d-source parameterizations 
$[\ _{Q}^{h}\widehat{\mathbf{J}}(\tau ,{x}^{k}),\ _{Q}^{v}\widehat{\mathbf{J}}(\tau,x^{k},y^{3})]$ (\ref{dsourcparam}). For such configurations, the coefficients of the respective families of d-metrics and N-connection coefficients in (\ref{ans1rf}) are computed following such formulas: 
\begin{eqnarray}
\ ^{q}g_{1}(\tau ) &=&\ ^{q}g_{2}(\tau )=e^{\psi (\hbar ,\kappa ;\tau
,x^{k})};  \label{qstcoefh} \\
\ ^{q}g_{3}(\tau ) &=&\frac{[\partial _{3}(\ _{J}\Phi (\tau ))]^{2}}{4(\
_{Q}^{v}\widehat{\mathbf{J}}(\tau ))^{2}\{g_{4}^{[0]}(\tau )-\int dy^{3}%
\frac{\partial _{3}[(\ _{J}\Phi (\tau ))^{2}]}{4(\ \ _{Q}^{v}\widehat{%
\mathbf{J}}(\tau ))}\}};\ ^{q}g_{4}(\tau )=g_{4}^{[0]}(\tau )-\int dy^{3}%
\frac{\partial _{3}[(\ _{J}\Phi (\tau ))^{2}]}{4(\ _{Q}^{v}\widehat{\mathbf{J%
}}(\tau ))};  \label{qcoefv}
\end{eqnarray}%
\begin{equation}
\ ^{q}N_{k}^{3}(\tau )=\frac{\partial _{k}(\ _{J}\Phi )}{\partial _{3}(\
_{J}\Phi )};\ \ ^{q}N_{k}^{4}(\tau )=\ _{1}n_{k}(\tau )+\ _{2}n_{k}(\tau
)\int dy^{3}\frac{\partial _{3}[(\ _{J}\Phi )^{2}]}{4(\ _{Q}^{v}\widehat{%
\mathbf{J}})^{2}|g_{4}^{[0]}(\tau )-\int dy^{3}\frac{\partial _{3}[(\
_{J}\Phi )^{2}]}{4(\ _{Q}^{v}\widehat{\mathbf{J}})}|^{5/2}}.  \label{qncoef}
\end{equation}%
These coefficients, can be chosen of necessary smooth class or to define
singular configurations. In (\ref{qstcoefh}), $\psi (\tau )$ are solutions
of a respective $\tau $-family of 2-d Poisson equations, 
\begin{equation}
\partial _{11}^{2}\psi (x^{k})+\partial _{22}^{2}\psi (\tau ,x^{k})=2\ \
_{Q}^{h}\widehat{\mathbf{J}}(\tau ,{x}^{k}).  \label{poisson}
\end{equation}%
The coefficients (\ref{qcoefv}) and (\ref{qncoef}) are determined in general
form by respective generating and integration functions with $\tau $%
-parametric dependence: 
\begin{eqnarray}
\mbox{generating functions: \quad } &&\ _{J}\Phi (\tau )\simeq \ _{J}\Phi
(\tau ,x^{k},y^{3});  \label{integrfunctrf} \\
\mbox{generating sources: \quad} &&\ \ _{Q}^{h}\widehat{\mathbf{J}}\mathcal{%
(\tau )}\simeq \ _{Q}^{h}\widehat{\mathbf{J}}(\tau ,x^{k});\ \ _{Q}^{v}%
\widehat{\mathbf{J}}\mathcal{(\tau )}\simeq \ _{Q}^{v}\widehat{\mathbf{J}}%
(\tau ,x^{k},y^{3});  \notag \\
\mbox{ integration functions: \quad} &&\ \ g_{4}^{[0]}(\tau )\simeq
g_{4}^{[0]}(x^{k}),\ _{1}n_{k}\mathcal{(\tau )}\simeq \ _{1}n_{k}(\tau
,x^{j}),\ _{2}n_{k}\mathcal{(\tau )}\simeq \ _{2}n_{k}(\tau ,x^{j_{1}}), 
\notag
\end{eqnarray}%
where $\ _{Q}^{h}\widehat{\mathbf{J}}$ and $\ \ _{Q}^{v}\widehat{\mathbf{J}}$
encode nonmetric distortions of the EMD systems.

We shall write in brief $\ ^{q}\mathbf{g}_{\alpha }(\tau )=(\ ^{q}g_{i}(\tau),\ ^{q}g_{a}(\tau ))$ for a quasi-stationary d-metric (\ref{ans1rf}) with coefficients generated by formulas (\ref{qstcoefh})--(\ref{integrfunctrf}). If necessary, for instance, $\ ^{q}\mathbf{g}_{\alpha }[\ _{J}\Phi (\tau )]$ will be used to emphasize that the generating functions are defined by effective sources (\ref{dsourcparam}).

\subsubsection{Nonlinear symmetries and $\protect\tau $-running effective cosmological constants}

The quasi-stationary d-metrics (\ref{ans1rf}) and N-connections  (\ref{qncoef}) possess certain very important nonlinear symmetries which allow us to introduce effective $\tau $-running cosmological constants $\ _{Q}\Lambda (\tau )=[\ _{Q}^{h}\Lambda (\tau ),\ _{Q}^{v}\Lambda (\tau )]$. For nonmetric configurations, such details are provided in \cite{lb24epjc1,lb24epjc2} (for nonmetric matter fluid sources). To define such symmetries we consider nonholonomic and nonmetric geometric flow deformations of families of prime d-metrics $\mathbf{\mathring{g}}(\tau )$ into a corresponding family of target d-metrics $\mathbf{g}(\tau )$, 
\begin{equation}
\mathbf{\mathring{g}}(\tau )\rightarrow \mathbf{g}(\tau )=[g_{\alpha }(\tau
)=\eta _{\alpha }(\tau )\mathring{g}_{\alpha }(\tau ),N_{i}^{a}(\tau )=\eta
_{i}^{a}(\tau )\mathring{N}_{i}^{a}(\tau )].  \label{offdiagdefr}
\end{equation}%
In these formulas, the gravitational $\eta $-polarization functions $\eta _{\alpha }(\tau )$ and 
$\eta _{i}^{a}(\tau )$ are be constrained by the condition that $\mathbf{g}(\tau )$ defines certain 
$\tau $-families of exact/parametric solutions. We note that the coefficients of $\mathbf{\mathring{g}}(\tau )$ can be arbitrary ones, or chosen as certain physically important solutions of some (modified) Einstein equations in an MGT or GR theory.

We parameterize the nonlinear transforms (\ref{offdiagdefr}) in the form: 
\begin{eqnarray}
&&(\ _{J}\Phi (\tau ),\ _{Q}^{v}\widehat{\mathbf{J}}(\tau ))\leftrightarrow (%
\mathbf{g}(\tau ),\ _{Q}^{v}\widehat{\mathbf{J}}(\tau ))\leftrightarrow
(\eta (\tau )\mathring{g}_{\alpha }(\tau )\sim (\zeta _{\alpha }(\tau
)(1+\varepsilon \chi _{\alpha }(\tau ))\mathring{g}_{\alpha }(\tau ),\
_{Q}^{v}\widehat{\mathbf{J}}(\tau ))\leftrightarrow  \notag \\
&&(\ _{\Lambda }\Phi (\tau ),\ _{Q}^{v}\Lambda (\tau ))\leftrightarrow (%
\mathbf{g}(\tau ),\ _{Q}^{v}\Lambda (\tau ))\leftrightarrow (\eta (\tau )%
\mathring{g}_{\alpha }(\tau )\sim (\zeta _{\alpha }(\tau )(1+\varepsilon
\chi _{\alpha }(\tau ))\mathring{g}_{\alpha _{s}}(\tau ),\ _{Q}^{v}\Lambda
(\tau )),  \label{nonlinsymr}
\end{eqnarray}%
where $\ _{Q}\Lambda (\tau _{0})=[\ _{Q}^{h}\Lambda (\tau _{0}),\
_{Q}^{v}\Lambda (\tau _{0})]$ can be fixed for nonmetric Ricci
soliton symmetries (\ref{canriccisolda}) and $\varepsilon $ is a small
parameter, $0\leq \varepsilon <1$. An ansatz (\ref{ans1rf}) with coefficients
(\ref{qstcoefh})-(\ref{qncoef}) is generated by nonlinear transforms (\ref%
{nonlinsymr}) if 
\begin{eqnarray}
\partial _{3}[(\ _{J}\Phi )^{2}] &=&-\int dy^{3}~\ _{Q}^{v}\widehat{\mathbf{J%
}}\ \partial _{3}g_{4}\simeq -\int dy^{3}~\ _{Q}^{v}\widehat{\mathbf{J}}\
\partial _{3}(\eta _{4}\mathring{g}_{4})\simeq -\int dy^{3}\ _{Q}^{v}%
\widehat{\mathbf{J}}\ \partial _{3}[\zeta _{4}(\tau )(1+\varepsilon \chi
_{4}(\tau ))\ \mathring{g}_{4}],  \notag \\
(\Phi )^{2} &=&-4\ _{Q}^{v}\Lambda (\tau )g_{4}\simeq -4\ _{Q}^{v}\Lambda
(\tau )\eta _{4}\ \mathring{g}_{4}\simeq -4\ \ _{Q}^{v}\Lambda (\tau )\
\zeta _{4}(\tau )(1+\varepsilon \chi _{4}(\tau ))\ \mathring{g}_{4}.
\label{nonlinsymrf}
\end{eqnarray}%
The nonlinear symmetries (\ref{nonlinsymr}) with $\ [\ _{J}\Phi (\tau ),\
_{Q}^{v}\widehat{\mathbf{J}}(\tau )]\rightarrow \lbrack \Phi (\tau ),\ \
_{Q}^{v}\Lambda (\tau )]$ (\ref{nonlinsymrf}) \ allow us to redefine the
nonmetric generating data and construct different types of off-diagonal
solutions. Such solutions can encode effective and matter field sources in
explicit form, or to re-distribute nonmetric EDM modifications into
off-diagonal terms with effective $\tau $-running cosmological constants.

To distinguish different types of modifications, we can consider conventional sums of effective cosmological constants corresponding to effective nonmetric Lagrange densities (\ref{mlagd}), 
\begin{equation}
\ \ _{Q}\Lambda (\tau )=\ ^{A}\Lambda (\tau )+\ ^{D}\Lambda (\tau )+\
_{Q}^{A}\Lambda (\tau )+\ \ _{Q}^{D}\Lambda (\tau )+\ _{Q}^{e}\Lambda (\tau
).  \label{effectcosmnedm}
\end{equation}%
For instance, an effective $\tau $-running cosmological constant 
$\ _{Q}^{v}\Lambda (\tau )= \ ^{D}\Lambda (\tau )+ \ _{Q}^{D}\Lambda (\tau)+\ _{Q}^{e}\Lambda (\tau )$ describes the geometric flow evolution of nonmetric Einstein-Dirac systems with conventional zero Maxwell fields.

Using the conventions (\ref{nonlinsymrf}) and (\ref{effectcosmnedm}), the
nonmetric geometric flow equations (\ref{ricciflowr2}) can be written in an
equivalent functional form as 
\begin{equation}
\ \widehat{\mathbf{R}}_{\ \ \gamma }^{\beta }(\tau ,\Phi (\tau ),\ _{Q}%
\widehat{\mathbf{J}}(\tau ))={\delta }_{\ \ \gamma }^{\beta }\ \ _{Q}\Lambda
(\tau ),  \label{ricciflowr2cosmc}
\end{equation}%
where ${\delta }_{\ \ \gamma }^{\beta }\ _{Q}\Lambda (\tau )=[{\delta }_{\ \
j}^{i}\ \ _{Q}^{h}\Lambda (\tau ),{\delta }_{\ \ b}^{a}\ _{Q}^{v}\Lambda
(\tau )].$ Such a system of nonlinear PDEs with effective cosmological
constants allows to application a more "simple" techniques to generate solutions and
to compute respective thermodynamic variables. Here we note that nonlinear
transforms of type $(\eta _{\alpha }(\tau )\mathring{g}_{\alpha }(\tau ))$
allow to describe non-perturbative deformations of some prime metrics when $%
\mathring{g}_{\alpha }(\tau )$ which are additionally defined by some
physically important constants (for instance, the BH mass and charge). For
small parametric deformations with $(\zeta _{\alpha }(\tau )(1+\varepsilon
\chi _{\alpha }(\tau ))\mathring{g}_{\alpha }(\tau ),\ _{Q}\Lambda (\tau )),$
such solutions can be self-consistently embedded into a nonlinear nonmetric
spacetime vacuum. We shall consider different types of generating functions 
\begin{equation}
\ ^{q}\mathbf{g}_{\alpha }[\ _{J}\Phi (\tau )]=\ ^{q}\mathbf{g}_{\alpha }[\Phi (\tau )]\simeq \ ^{q}\mathbf{g}_{\alpha }[\eta _{4}(\tau )]\simeq \
^{q}\mathbf{g}_{\alpha }[\chi _{4}(\tau )]  \label{paramdsol}
\end{equation}%
for parameterizing quasi-stationary solutions $\ ^{q}\mathbf{g}_{\alpha}(\tau )$ (\ref{ans1rf}) with coefficients (\ref{qstcoefh})-(\ref{qncoef}) of the nonmetric geometric flow equations (\ref{ricciflowr2}) or (\ref{ricciflowr2cosmc}).

\subsubsection{Nonmetric induced mass anisotropies and effective electromagnetic sources}

Let us explain how we can compute respective $f(Q)$ distortions of the mass of Dirac fields and modifications of the spinor wave functions and of the Maxwell electromagnetic fields and sources:

We introduce $\tau $-families of d-metric and N-connection coefficients (\ref{qstcoefh}), (\ref{qcoefv}) and (\ref{qncoef}) of a quasi-stationary ansatz (\ref{ans1rf}) in respective formulas for $\widehat{\mathbf{D}}=\mathbf{D}-\widehat{\mathbf{L}}$ (\ref{disf}) and $\widehat{\mathbf{D}}$  
(\ref{coefcandcon}) Then, the results are used for computing the distortion of the canonical Dirac d-operator $\widehat{\mathcal{D}}_{\alpha}^{A}\rightarrow \mathcal{D}_{\alpha }^{A}=\mathbf{e}_{\alpha }-\widehat{\mathbf{\Gamma }}_{\alpha }-\ ^{Q}\widehat{\mathbf{\Gamma }}_{\alpha }-iq%
\mathbf{A}_{\alpha },$ see (\ref{candirac}) and (\ref{qcandir}). We compute also the respective nonmetric distortions of the fermionic mass, $m_{0}\rightarrow M(\tau )=m_{0}+\ ^{Q}M(\tau ),$ see formulas (\ref{anisotrm}), when 
\begin{equation}
\ ^{Q}M(\tau ,x^{k},y^{3})=-i\ \hbar \ \gamma ^{\alpha }(\tau ,x^{k},y^{3})\
^{Q}\widehat{\mathbf{\Gamma }}_{\alpha }(\tau ,x^{k},y^{3})+m_{0}B(\tau
,x^{k},y^{3})-\frac{3}{2}\hbar \ \ ^{3}\widehat{\mathbf{T}}^{\alpha }(\tau
,x^{k},y^{3})\gamma _{\alpha }(\tau ,x^{k},y^{3})\gamma ^{5}.
\label{anisotropqs}
\end{equation}%
Such nonmetric quasi-stationary polarizations of $m_{0}$ can be observed experimentally and computed explicitly if the generating functions/sources and integrating functions (\ref{integrfunctrf}) are prescribed 
explicitly for a realistic experiment. We can consider also $f(Q)$-effects inducing a fermionic mass if $m_{0}=0$. Here note that to observe directly the corresponding nonmetric modified $\tau $-families of wave functions $\Psi (\tau ,x^{k},y^{3})=B(\tau ,x^{k},y^{3})\ \psi (x^{k},y^{3})$ is not possible. But we can compute in certain parametric forms the corresponding families $4\times 4$ matrices $B(\tau ,x^{k},y^{3})$ if, for instance, is defined $\psi (x^{k},y^{3})$ by a Dirac solution in GR and when $\Psi (\tau,x^{k},y^{3})$  is constrained to define a $\tau $-family of solutions of (\ref{cnmdir}). So, we conclude that nonmetric $Q$-deformations result in quasi-stationary variations of the masses of Dirac fermions, $\ ^{Q}M(\tau
,x^{k},y^{3})$ (\ref{anisotropqs}) because of nontrivial terms $\ ^{Q}\widehat{\mathbf{\Gamma }}_{\alpha }(\tau ,x^{k},y^{3})$.\footnote{This is enough for this paper (when we do not study general
classes of solutions for nonmetric Dirac equations, which will be considered in our future works).}

We emphasize that physical effects with nontrivial $\ _{\nabla }^{Q}M(\tau )$ are possible also for LC-configurations with $\ ^{3}\widehat{\mathbf{T}}^{\alpha }(\tau ,x^{k},y^{3})=0$ and $\widehat{\mathbf{Z}}=0,$ see (\ref{cdist}). The explicit forms of the solutions of nonmetric Dirac equations
under geometric flow evolution on a quasi-stationary $\ _{1}^{3}\mathcal{V}$ depend on the data (\ref{integrfunctrf}) and prescribed effective cosmological $\tau $-running cosmological constants for nonlinear symmetries (\ref{nonlinsymrf}) and (\ref{effectcosmnedm}).

In a similar form, introducing the d-metric and N-connection coefficients (\ref{qstcoefh}), (\ref{qcoefv}) and (\ref{qncoef}), for a quasi-stationary ansatz (\ref{ans1rf}), in respective formulas for $\widehat{\mathbf{D}}=\mathbf{D}-\widehat{\mathbf{L}}$ (\ref{disf}) and $\widehat{\mathbf{D}}$  (\ref{coefcandcon}), we can compute $\mathbf{F}_{\alpha \beta}=\ \widehat{\mathbf{F}}_{\alpha \beta }+\ ^{Q}\widehat{\mathbf{F}}_{\alpha \beta }$ (\ref{starsmaxstr}). In such cases, 
$\ ^{Q}\widehat{\mathbf{F}}_{\alpha \beta }(\tau ,x^{k},y^{3})$ define a $\tau $-family of solutions
of nonmetric distorted Maxwell equations (\ref{cnamax}). Such nonmetric electromagnetic effects can be observed on a quasi-stationary $\ _{1}^{3}\mathcal{V}$. We also compute $\tau $-parametric nonmetric distortions $\ ^{e}\mathbf{j}^{\beta }(\tau ,x^{k},y^{3})$ of the fermionic sources, when 
$\mathbf{j}^{\beta }(x^{k},y^{3}):=\overline{\psi }\ \mathbf{\gamma }^{\beta }\psi $ can be chosen for a EDM solution in GR but $\overline{\Psi }\  \mathbf{\gamma }^{\beta }\ \Psi =\mathbf{j}^{\beta }+\ ^{e}\mathbf{j}^{\beta }$ is defined both for $Q$-deformed (\ref{cnmdir}) and (\ref{cnamax}). The
explicit form of such quasi-stationary solutions of the Dirac-Maxwell, or only Maxwell, fields under geometric evolution depend also on the data (\ref{integrfunctrf}) and prescribed effective cosmological $\tau $-running
cosmological constants for nonlinear symmetries (\ref{nonlinsymrf}) and (\ref{effectcosmnedm}). It is not the purpose of this work to provide tedious technical computations of some explicit 
$\ ^{Q}\widehat{\mathbf{F}}_{\alpha \beta }(\tau )$ and $\ ^{e}\mathbf{j}^{\beta }(\tau )$ because they can be encoded in general form into normalization functions as in (\ref{normcondc1}) and "dispersed" in off-diagonal terms for corresponding solutions of (\ref{ricciflowr2cosmc}). We also omit to provide some examples of nonmetric deformations of the Dirac-Maxwell equations which will be studied our future works. 

Nonmetric effects with induced effective electromagnetic sources can be modelled  also using LC configurations if distorting  $Q$-terms are not trivial. We do not need such values for computing nonmetric thermodynamic variables (\ref{qthermvar}), which for quasi-stationary solutions are determined by effective $\tau $-running cosmological constants as we shall prove in the next subsection. In a general context, we can consider that nonmetric $Q$-deformations of a quasi-stationary EDM system induce respective effective sources $\ ^{e}\mathbf{j}^{\beta }(\tau ,x^{k},y^{3})$ (\ref{starsmaxstr}) like in the classical electrodynamics for locally anisotropic media. In this work, we consider  metric-affine spacetimes  endowed with nonmetric Dirac fermion fields.

\subsubsection{Computing thermodynamic variables for nonmetric quasi-stationary EDM solutions}

Taking a general quasi-stationary solution $\ ^{q}\mathbf{g}_{\alpha }[\Phi]
\simeq \ ^{q}\mathbf{g}_{\alpha }[\eta _{4}]$ (\ref{paramdsol}) of (\ref{ricciflowr2cosmc}) (determined by coefficients (\ref{qstcoefh})--(\ref{qncoef})), we can compute the canonical thermodynamic variables $\ _{Q}
\widehat{\mathcal{E}}(\tau )$ and $\ _{Q}\widehat{\mathcal{S}}(\tau )$ (\ref{qthermvar}) using dependencies on $\tau $-families canonical Ricci scalars $\widehat{\mathbf{R}}sc(\tau )$ expressed in terms of $\tau $-running effective cosmological constants $\ _{Q}\Lambda (\tau )$. For simplicity, we do not present in this work more cumbersome technical results for $\ _{Q} \widehat{\sigma }(\tau )$ with formulas involving the canonical Ricci d-tensor.  

We begin with the volume element $\delta \ _{Q}^{q}\widehat{V}(\tau )=\sqrt{| \ ^{q}\mathbf{g}(\tau )|}\ dx^{1}dx^{2}\delta y^{3}\delta y^{4}$ (\ref%
{volume}) and%
\begin{equation*}
\widehat{\mathbf{R}}sc(\tau )=2\ \ _{Q}\Lambda (\tau )=2[\ \ _{Q}^{h}\Lambda
(\tau )+\ \ _{Q}^{v}\Lambda (\tau )],
\end{equation*}%
see formulas (\ref{effectcosmnedm}) and (\ref{ricciflowr2cosmc}). To simplify the formulas and computations, we chose certain frame/coordinate systems when the normalizing functions have the properties when $\widehat{\mathbf{D}}_{\alpha }\ {\widehat{\zeta }}=0$ and ${\widehat{\zeta }}\approx 0.$ We obtain respectively for the statistical partition function and thermodynamic variables 
\begin{eqnarray}
\ _{Q}^{q}\widehat{Z}(\tau ) &=&\exp [\int_{\widehat{\Xi }}\frac{1}{8\left(
\pi \tau \right) ^{2}}\ \delta  \ _{Q}^{q}\mathcal{V}(\tau )], \ _{Q}^{q}%
\widehat{\mathcal{E}}\ (\tau )=-\tau ^{2}\int_{\widehat{\Xi }}\ \frac{1}{%
8\left( \pi \tau \right) ^{2}}[  \ _{Q}^{h}\Lambda (\tau )+ 
\ _{Q}^{v}\Lambda (\tau ) -\frac{1}{\tau }]\ \delta \ ^{q}\mathcal{V}(\tau ),  \notag \\
 \ _{Q}^{q}\widehat{S}(\tau ) &=&-\ \ _{Q}^{q}\widehat{W}(\tau )=-\int_{%
\widehat{\Xi }}\frac{1}{8\left( \pi \tau \right) ^{2}}[\tau ( \ _{Q}^{h}\Lambda (\tau )+ \ _{Q}^{v}\Lambda (\tau ))\ -2]\delta \ _{Q}^{q}\mathcal{V}(\tau ). \label{thermvar1}
\end{eqnarray}%
To simplify computations and formulas in (\ref{thermvar1}) we consider
trivial integration functions $\ _{1}n_{k}=0$ and $\ _{2}n_{k}=0$ (such
conditions change for arbitrary frame/coordinate transforms).

Using the formulas (\ref{nonlinsymr}) and (\ref{nonlinsymrf}) for nonlinear
symmetries, we express 
\begin{equation}
\ \Phi (\tau )=2\sqrt{|[\ _{Q}^{v}\Lambda (\tau )]\ g_{4}(\tau )|}=\ 2\sqrt{%
|\ \ _{Q}^{v}\Lambda (\tau )\  \eta _{4}(\tau )\ \mathring{g}_{4}(\tau )|}%
\simeq 2\sqrt{|\ \ _{Q}^{v}\Lambda (\tau )\ \zeta _{4}(\tau )\ \mathring{g}%
_{4}|}[1-\frac{\varepsilon }{2}\chi _{4}(\tau )],  \notag
\end{equation}%
for $\ [\ _{J}\Phi (\tau ),\ _{Q}^{v}\widehat{\mathbf{J}}(\tau)]\rightarrow \lbrack \Phi (\tau ),
 \ _{Q}^{v}\Lambda (\tau )].$  This allows us to compute 
\begin{eqnarray*}
\ \delta \ \ _{Q}^{q}\mathcal{V} &=&\delta \mathcal{V}[\tau ,\ _{Q}\widehat{%
\mathbf{J}}(\tau ),\ _{Q}\Lambda (\tau );\psi (\tau ),\ g_{4}(\tau )]=\delta 
\mathcal{V}(\ _{Q}\widehat{\mathbf{J}}(\tau ),\ _{Q}\Lambda (\tau ),\eta
_{4}(\tau )\ \mathring{g}_{4}) \\
&=&\frac{1}{\sqrt{|\ _{Q}^{h}\Lambda (\tau )\times \ \ _{Q}^{v}\Lambda (\tau
)|}}\ \delta \ _{\eta }\mathcal{V},\mbox{ where }\ \delta \ _{\eta }\mathcal{%
V}=\ \delta \ _{\eta }^{h}\mathcal{V}\times \delta \ _{\eta }^{v}\mathcal{V}.
\end{eqnarray*}%
In these formulas, we use the functionals: 
\begin{eqnarray}
\delta \ _{\eta }^{h}\mathcal{V} &=&\delta \ _{\eta }^{h}\mathcal{V}[\
_{Q}^{h}\widehat{\mathbf{J}}(\tau ),\eta _{1}(\tau )\ \mathring{g}_{1}]
\label{volumfuncts} \\
&=&e^{\widetilde{\psi }(\tau )}dx^{1}dx^{2}=\sqrt{|\ _{Q}^{h}\widehat{%
\mathbf{J}}(\tau )|}e^{\psi (\tau )}dx^{1}dx^{2},\mbox{ for }\psi (\tau )%
\mbox{ being a
solution of  }(\ref{poisson}),  \notag \\
\delta \ _{\eta }^{v}\mathcal{V} &=&\delta \ _{\eta }^{v}\mathcal{V}[\
_{Q}^{v}\widehat{\mathbf{J}}(\tau ),\eta _{4}(\tau ),\ \mathring{g}_{4}] 
\notag \\
&=&\frac{\partial _{3}|\ \eta _{4}(\tau )\ \mathring{g}_{4}|^{3/2}}{\ \sqrt{%
|\int dy^{3}\ \ _{Q}^{v}\widehat{\mathbf{J}}(\tau )\{\partial _{3}|\ \eta
_{4}(\tau )\ \mathring{g}_{4}|\}^{2}|}}[dy^{3}+\frac{\partial _{i}\left(
\int dy^{3}\ \ _{Q}^{v}\widehat{\mathbf{J}}(\tau )\partial _{3}|\ \eta
_{4}(\tau )\ \mathring{g}_{4}|\right) dx^{i}}{\ \ \ _{Q}^{v}\widehat{\mathbf{%
J}}(\tau )\partial _{3}|\ \eta _{4}(\tau )\mathring{g}_{4}|}]dt.  \notag
\end{eqnarray}%
Integrating such products of $h$- and $v$-forms on a closed hypersurface $%
\widehat{\Xi },$ we obtain a running nonmetric spacetime volume functional 
\begin{equation}
\ _{\eta }^{\shortmid }\mathcal{\mathring{V}[}\ ^{q}\mathbf{g}(\tau
)]=\int_{\ \widehat{\Xi }}\delta \ _{\eta }\mathcal{V}(\ _{Q}^{v}\widehat{%
\mathbf{J}}(\tau ),\ \eta _{\alpha }(\tau ),\mathring{g}_{\alpha }).
\label{volumf1}
\end{equation}%
This functional depends on hypersurface data, effective sources for nonmetric EDM systems, primary d-metrics and generating functions $\eta _{4}(\tau )$ and effective cosmological constants $\ _{Q}\Lambda (\tau ).$

Using the volume functional (\ref{volumf1}), the formulas for nonmetric thermodynamic variables (\ref{thermvar1}) can be written: 
\begin{eqnarray}
\ _{Q}^{q}\widehat{Z}(\tau ) &=&\exp [\frac{1}{8\pi ^{2}\tau ^{2}}\ _{\eta
}^{\shortmid }\mathcal{\mathring{V}}[\ ^{q}\mathbf{g}(\tau )],\ _{Q}^{q}%
\widehat{\mathcal{E}}\ (\tau )=\ \frac{1-\tau \ (\ \ _{Q}^{h}\Lambda (\tau
)+\ \ _{Q}^{v}\Lambda (\tau ))}{8\pi ^{2}\tau }\ \ _{\eta }^{\shortmid }%
\mathcal{\mathring{V}[}\ ^{q}\mathbf{g}(\tau )],  \label{thermvar2} \\
\ \ \ \ _{Q}^{q}\widehat{S}(\tau ) &=&-\ \ _{Q}^{q}\widehat{W}(\tau )=\frac{%
2-\tau (\ \ _{Q}^{h}\Lambda (\tau )+\ \ _{Q}^{v}\Lambda (\tau ))}{8\pi
^{2}\tau ^{2}}\ _{\eta }^{\shortmid }\mathcal{\mathring{V}[}\ ^{q}\mathbf{g}%
(\tau )].  \notag
\end{eqnarray}%
In these formulas, the explicit form of $\ _{\eta }^{\shortmid }\mathcal{\mathring{V}} 
[\ ^{q}\mathbf{g}(\tau )]$ depends on the type of generating and integrating data (\ref{integrfunctrf}) defining a respective class of quasi-stationary solutions of (\ref{ricciflowr2}) or (\ref{ricciflowr2cosmc}). It can be computed if we prescribe such data for a fixed system of reference/ coordinates. The multiples in (\ref{thermvar2}) depending on effective $\tau $-running cosmological constant $\ _{Q}\Lambda (\tau )$
allow us to determine  certain important thermodynamic properties of 
$ \ _{Q}^{q}\widehat{\mathcal{E}}\ (\tau )$ and $\ _{Q}^{q}\widehat{S}(\tau )$ even 
$\ _{\eta }^{\shortmid }\mathcal{\mathring{V}[}\ ^{q}\mathbf{g}(\tau )]$ is considered as a general quasi-stationary functional. For instance, we can compute $\ _{Q}^{q}\widehat{S}(\tau )$ when $\ _{Q}^{e}\Lambda (\tau )=0$ and $\ _{Q}^{e}\Lambda (\tau )\neq 0$ in $\ _{Q}^{v}\Lambda (\tau )$ 
(\ref{effectcosmnedm}) and decide which quasi-stationary configurations (nonmetric or metric ones) possess more small entropic values. Other examples can be considered for $\ _{Q}^{q}\widehat{\mathcal{E}}\ (\tau )$ computed with $\ _{Q}^{D}\Lambda (\tau )=0$ and $\ _{Q}^{D}\Lambda (\tau )\neq 0$ in 
$\ _{Q}^{v}\Lambda (\tau )$. Comparing such dependencies, we can decide if certain nonmetric deformations of the Dirac fields are more/less convenient energetically under nonmetric geometric flows. In a similar form, we can compute the thermodynamic variables (\ref{thermvar2}) for any variant 
$\ ^{A}\Lambda (\tau )=0,$ or $\neq 0;\ ^{D}\Lambda (\tau )=0,$ or$\neq 0;\ $ and $_{Q}^{A}\Lambda (\tau )=0,$ or $\neq 0,$ and decide which quasi-stationary configurations under respective metric or nonmetric
geometric flows and respective Dirac-Maxwell interactions are more, or less thermodynamically convenient.

 Finally, we emphasize that for $\tau =\tau _{0},$ the formulas (\ref{thermvar2}) can be used for defining statistical/ thermodynamic characteristics of nonmetric Ricci soliton quasi-stationary configurations,
i.e. off-diagonal quasi-stationary solutions of nonmetric EDM equations (\ref{canriccisolda}).

\section{Physically important solutions for nonmetric geometric EDM flows}

\label{sec4}

In this section, we construct and analyze three examples of generic off-diagonal solutions for $\tau $-running nonmetric quasi-stationary EDM systems. They describe nonmetric evolution and nonmetric distortions of BH, WH and   BT configurations.

\subsection{Nonmetric deformations of Kerr de Sitter solutions to spheroidal EDM configurations}

Nonholonomic off-diagonal deformations of the Kerr and Schwarzschild - (anti) de Sitter (KdS) and other types of BH metrics were studied in a series of works 
\cite{sv00a,vp,vs01a,vs01b,vmon05,v10,bubuianu17,vacaru18,bubuianu19,partner03,partner04}. Those papers were on developments and applications of the AFCDM in various supersymmetric, string, noncommutative and nonassociative, or Finsler like modifications of GR and geometric information flow theories. For spherical rotating configurations of KdS in GR, the corresponding solutions can be described by certain families of rotating diagonal metrics involving, or not, warping effects of curvature, see details in \cite{ovalle21}. The goal of this subsection is to analyse how rotating BHs can be nonmetrically
EDM deformed under geometric evolution to parametric quasi-stationary d-metrics of type (\ref{ans1rf}). The conditions for  rotoid deformations are formulated in explicit form.

\subsubsection{Prime off-diagonal metric for KdS solutions encoding Dirac-Maxwell fields in GR}

Let us consider a prime d-metric $\mathbf{\mathring{g}}(\tau _{0})=[%
\mathring{g}(\tau _{0})=\breve{g}_{\alpha }(r,\varphi ,\theta ),\mathring{N}%
_{i}^{a}(\tau _{0})=\breve{N}_{i}^{a}(r,\varphi ,\theta )]$ for spherical
coordinates parameterized in the form $x^{1}=r,x^{2}=\varphi ,y^{3}=\theta
,y^{4}=t,$ when the quadratic element (\ref{ans1rf}) is written: 
\begin{equation}
d\breve{s}^{2}=\breve{g}_{\alpha }(r,\varphi ,\theta )(\mathbf{\breve{e}}%
^{\alpha })^{2}.  \label{offdiagpm1}
\end{equation}%
This d-metric will be subjected to nonlinear transforms to a target d-metric $\ ^{q}\mathbf{g}(\tau )$ (\ref{offdiagdefr}) for generating nonmetric quasi-stationary EDM configurations under geometric flow evolution. In (\ref{offdiagpm1}), the nontrivial coefficients of the d-metric and N-connection  are 
\begin{equation*}
\breve{g}_{1}=\frac{\breve{\rho}^{2}}{\triangle _{\Lambda }},\breve{g}_{2}=%
\frac{\sin ^{2}\theta }{\breve{\rho}^{2}}[\Sigma _{\Lambda }-\frac{%
(r^{2}+a^{2}-\triangle _{\Lambda })^{2}}{a^{2}\sin ^{2}\theta -\triangle
_{\Lambda }}],\breve{g}_{3}=\breve{\rho}^{2},\breve{g}_{4}=\frac{a^{2}\sin
^{2}\theta -\triangle _{\Lambda }}{\breve{\rho}^{2}},\breve{N}_{2}^{4}=%
\breve{n}_{2}=-a\sin \theta \frac{r^{2}+a^{2}-\triangle _{\Lambda }}{%
a^{2}\sin ^{2}\theta -\triangle _{\Lambda }}.
\end{equation*}%
For $\Sigma _{\Lambda }=(r^{2}+a^{2})^{2}-\triangle _{\Lambda }a^{2}\sin ^{2}\theta ,
\triangle _{\Lambda }=r^{2}-2Mr+a^{2}-\frac{\Lambda _{0}}{3}r^{4},\breve{\rho}^{2}=r^{2}+a^{2}\cos ^{2}\theta ,$ and $a=J/M=const,$ where $J$ is the angular momentum and $M$ is the total mass of the system, and the cosmological constant $\Lambda _{0}>0,$ the d-metric (\ref{offdiagpm1}) defines a new type of KdS solution \cite{ovalle21}. In this paper, we use a different system of notations. Such solutions are different from the standard KdS metrics, see details in \cite{misner,hawking73,wald82,kramer03}, called also $\Lambda $-vacuum solutions. This is because the corresponding scalar curvature $R(r,\theta )=4\widetilde{\Lambda }(r,\theta )=4\Lambda _{0}\frac{r^{2}}{\breve{\rho}^{2}}\neq 4\Lambda _{0}$ is certain way polarized on coordinates $(r,\theta ).$ A warped effect when the curvature is warped every where except the equatorial plane is also present,and the effective polarization of the cosmological constant also shows a rotational effect on the vacuum energy. Such effects disappear for $r\gg a.$

We note that we can consider similarly  any primary d-metric of type  $\mathbf{\mathring{g}}(\tau,r,\varphi ,\theta ))$ which can be generalized to quasi-stationary solutions of the  EDM equations using the AFCDM. In \cite{lb24epjc1,lb24epjc2}, the geometric method was developed for constructing physically important solutions in metric-affine and (in particular) in $f(Q)$ gravity. In the first paper, we studied  WHs and solitonic hierarchies as DE and DM configurations in $F(R,T,Q,T_m)$ gravity. The second mentioned partner work was devoted to generic off-diagonal solutions for nonmetric geometric flows describing quasicrystalline topological phases for DE and DM in $f(Q)$ cosmology. In this article, we generate  a simple class of solutions describing nonmetric deformations of KdS solutions (\ref{offdiagpm1}).  Even in such cases, the nonmetric flow deformations or nonmetric Dirac spinor - Maxwell contributions modify substantially the prime metrics with spherical symmetry. In the first simplest case of rotoid configurations with gravitational polarizations (see formulas (\ref{rotoid}) below) they also encode nonmetric EDM data. We can't perform further simplifications of solutions if we want to compute some nontrivial effects of nonmetricity and Dirac spinor contributions. Such nonmetric geometric flows or quasi-stationary configurations are generic off-diagonal and can be treated as certain possible DE (for polarization of effective cosmological constants). Using certain configurations with (nonmetric) gravitational polarization of BH physical constants, or small ellipsoid deformation of horizons, we can model certain additional DM interactions. Only the AFCDM allows us to construct in explicit analytic form such generic off-diagonal exact and parametric solutions. Other method described in \cite{kramer03} with further developments in MGTs do not allow to decouple and solve in certain general off-diagonal forms nonmetric geometric flow or nonmetric modified Einstein equations.  We motivated in details the priority of the AFCDM, providing rigorous proofs of results and reviewed various applications in modern accelerating cosmology and astrophysics  \cite{vmon05,v13,v14,lb24epjc1,lb24epjc2,lb24grg,bubuianu17,vacaru18,bubuianu19,partner03,partner04}.

An off-diagonal prime d-metric $\mathbf{\mathring{g}}$ (\ref{offdiagpm1}) can be considered as a rotating version of the Schwarzschild de Sitter metric representing a new solution describing the exterior of a BH with
cosmological constant. To define a BH solution, the prime d-metric involves certain bonds for $M(a,\Lambda _{0}).$ The corresponding upper, $M_{\max}:=M_{+}$ and lower, $M_{\min }:=M,$ bounds are computed as $18\Lambda _{0}M_{\pm }^{2}=1+12\Lambda _{0}a^{2}\pm (1-4\Lambda _{0}a^{2})^{3/2}.$
Such conditions define a LC configuration for the standard Einstein equations (\ref{gfeq2a}) with zero torsion and zero nonmetricity and for a fluid type energy momentum tensor (\ref{emtlc}), when 
\begin{equation}
\breve{T}_{\alpha \beta }(r,\theta )=diag[p_{r},p_{\varphi }=p_{\theta
},p_{\theta }=\rho -2\Lambda _{0}r^{2}/\mathring{\rho}^{2},\rho =-p_{r}=%
\widetilde{\Lambda }^{2}/\Lambda _{0}].  \label{efmt1}
\end{equation}%
Such configurations consist of an example of LC Ricci soliton configuration defined as metric compatible example of nonholonomic Ricci flow equations (\ref{ricciflowr2cosmc}) determined by a  subclass of nonlinear symmetries (\ref{nonlinsymr}). We can consider certain effective cosmological constants 
$\ _{Q}\Lambda (\tau _{0})=[\ _{Q}^{h}\Lambda (\tau _{0}),\ _{Q}^{v}\Lambda (\tau _{0})]$ related to (\ref{efmt1}). This class of rotating diagonal quasi-stationary configurations defines metric EDM configurations if $\ _{Q}\Lambda (\tau _{0})$ is extended to encode effective Dirac and Maxwell
sources in the form (\ref{effectcosmnedm}), but with vanishing both torsion and nonmetricity. In such cases, we have certain nontrivial values of anisotropic polarized mass $^{Q}\breve{M}(\tau,x^{k},y^{3})=m_{0}B(\tau,x^{k},y^{3}),$ see formula (\ref{anisotropqs}) even if we prescribe, for simplicity, a zero of the electromagnetic field. We shall explain in the next subsections how such quasi-stationary Kerr-de Sitter - DE can be computed for nonmetric geometric flow generalizations.

\subsubsection{Quasi-stationary nonmetric EDM polarizations of KdS configurations}

In this subsection, we construct quasi-stationary off-diagonal deformations of the prime Kerr solutions (\ref{offdiagpm1}) generating the solutions for nonmetric geometric flow systems (\ref{ricciflowr2}). Such generic off-diagonal solutions can involve, in general, $\tau $-families of vacuum polarizations both of effective cosmological constants and d-metric coefficients depending on all space coordinates $(r,\varphi
,\theta ).$ The dependencies of target quasi-stationary d-metrics on space coordinates are more general than those for prime d-metrics (\ref{efmt1}) with coefficients as functions on $(r,\theta ).$ This class of target
solutions will be of type (\ref{ans1rf}), when the coefficients (\ref{qstcoefh})-(\ref{qncoef}) encode EDM configurations and subjected to nonlinear symmetries (\ref{nonlinsymrf}). The generating $\tau $-sources are
taken in the form%
\begin{equation*}
_{Q}\mathbf{\breve{J}}(\tau )=\ \ ^{m}\mathbf{\breve{Y}}_{\alpha \beta
}(\tau ,r,\varphi )+\ _{Q}\widehat{\mathbf{J}}(\tau ,r,\varphi ,\theta ),
\end{equation*}%
where $\ _{Q}\widehat{\mathbf{J}}=[\ _{Q}^{h}\widehat{\mathbf{J}}(\tau,r,\varphi ),
\ _{Q}^{v}\widehat{\mathbf{J}}(\tau ,r,\varphi ,\theta )]$ encodes, for respective coordinates, nonmetric deformations of the EDM fields as in (\ref{dsourcparam}). The total generating source $\ _{Q}\mathbf{%
\breve{J}}(\tau )=[\ _{Q}^{h}\mathbf{\breve{J}}(\tau ),\ _{Q}^{v}\mathbf{\breve{J}}(\tau )]$ includes terms (we can consider a formal $\tau $-running or a fixed $\tau _{0}$) with 
$\ ^{m}\widehat{\mathbf{Y}}_{\alpha \beta }=\ ^{m}\mathbf{\breve{Y}}_{\alpha \beta }$ \ as in (\ref{ceemt}) defined by $\breve{T}_{\alpha \beta }(r,\theta )$ (\ref{efmt1}) considered instead of 
$\ ^{m}T_{\alpha \beta }$ (\ref{emtlc}). This means that for constructing such systems of nonlinear PDEs we use an effective Lagrange density $\ ^{nDA}\mathcal{\breve{L}}(\tau )=\mathcal{\breve{L}}(\tau )+
\ ^{nDA}\widehat{\mathcal{L}}(\tau ),$ where $\ ^{m}\mathcal{L}(\tau )=\mathcal{\breve{L}}(\tau )$ corresponds to $\breve{T}_{\alpha \beta }$ and the term for nonmetric Dirac-Maxwell effective Lagrange density is $\ ^{nDA}\widehat{\mathcal{L}}(\tau )$ (\ref{mlagd}). The respective generating sources are
parameterized in the form 
\begin{equation}
\ _{Q}\mathbf{\breve{J}}(\tau )=\ _{Q}\mathbf{\breve{J}}_{\ \ \beta
}^{\alpha }(\tau ,r,\varphi ,\theta )=[\ ^{h}\breve{J}(\tau ,r,\varphi
)\delta _{\ \ j}^{i},\ ^{v}\breve{J}(\tau ,r,\varphi ,\theta )\delta _{\ \
b}^{a}].  \label{dsourcparamkerr}
\end{equation}%
As result of nonlinear symmetries (\ref{nonlinsymrf}), the Kerr terms 
$\ ^{m}\mathbf{\breve{Y}}_{\alpha \beta }$ result in an additional contribution 
$\ ^{m}\breve{\Lambda}(\tau )$ to the effective cosmological $\tau $-running
cosmological constants (\ref{effectcosmnedm}), when 
\begin{eqnarray}
\ \ _{Q}\breve{\Lambda}(\tau ) &=&[\ \ _{Q}^{h}\breve{\Lambda}(\tau ),\ \
_{Q}^{v}\breve{\Lambda}(\tau )]=\ ^{m}\breve{\Lambda}(\tau )+\ \ _{Q}\Lambda
(\tau )  \label{effectcosmnedmker} \\
&=&\ ^{m}\breve{\Lambda}(\tau )+\ ^{A}\Lambda (\tau )+\ ^{D}\Lambda (\tau
)+\ _{Q}^{A}\Lambda (\tau )+\ \ _{Q}^{D}\Lambda (\tau )+\ _{Q}^{e}\Lambda
(\tau ).  \notag
\end{eqnarray}

The solutions for the target quasi-stationary d-metrics can be written in terms of $\eta $-polarization functions for d-metrics of the form (\ref{offdiagdefr}), when 
\begin{eqnarray}
d\widehat{s}^{2} &=&\widehat{g}_{\alpha \beta }(\tau ,r,\varphi ,\theta
)du^{\alpha }du^{\beta }=e^{\psi (\tau ,r,\varphi )}[(dx^{1}(r,\varphi
))^{2}+(dx^{2}(r,\varphi ))^{2}]  \label{nkernewnmrf} \\
&&-\frac{[\partial _{\theta }(\eta _{4}\ \breve{g}_{4})]^{2}}{|\int d\theta
\ \ ^{v}\breve{J}\partial _{\theta }(\eta _{4}\ \breve{g}_{4})|\ \eta _{4}%
\breve{g}_{4}}\{dy^{3}+\frac{\partial _{i}[\int d\theta \ \ ^{v}\breve{J}\
\partial _{3}(\eta _{4}\breve{g}_{4})]}{\ \ ^{v}\breve{J}\partial _{\theta
}(\eta _{4}\breve{g}_{4})}dx^{i}\}^{2}  \notag \\
&&+\eta _{4}\breve{g}_{4}\{dt+[\ _{1}n_{k}(\tau ,r,\varphi )+\
_{2}n_{k}(\tau ,r,\varphi )\int d\theta \frac{\lbrack \partial _{\theta
}(\eta _{4}\breve{g}_{4})]^{2}}{|\int d\theta \ \ ^{v}\breve{J}\partial
_{3}(\eta _{4}\breve{g}_{4})|\ (\eta _{4}\breve{g}_{4})^{5/2}}]dx^{k}\}^{2} 
\notag
\end{eqnarray}%
These off-diagonal solutions are determined by corresponding $\tau $%
-families of generating functions \newline $\eta _{4}(\tau )=\eta _{4}(\tau
,r,\varphi ,\theta )$ and respective integration functions like $\ _{1}n_{k}(\tau )=$ $\ _{1}n_{k}(\tau ,r,\varphi )$ and $\ _{2}n_{k}(\tau )=\
_{2}n_{k}(\tau ,r,\varphi ).$

Any quasi-stationary $\tau $-family of d-metrics (\ref{nkernewnmrf}) can be
characterized by nonlinear symmetries of type (\ref{nonlinsymrf}), 
\begin{eqnarray}
\partial _{\theta }[\ _{J}\Psi ^{2}(\tau )] &=&-\int d\theta \ \ ^{v}\breve{J%
}(\tau )\partial _{\theta }h_{4}\simeq -\int d\theta \ \ ^{v}\breve{J}(\tau
)\partial _{\theta }(\eta _{4}\ \breve{g}_{4})\simeq -\int d\theta \ \ ^{v}%
\breve{J}(\tau )\partial _{\theta }[\zeta _{4}(1+\kappa \ \chi _{4})\ \breve{%
g}_{4}],  \label{nlims2} \\
\ _{J}\Phi (\tau ) &=&|\ \ _{Q}^{v}\breve{\Lambda}(\tau )|^{-1/2}\sqrt{|\int
d\theta \ \ \ ^{v}\breve{J}(\tau )\ (\Phi ^{2})^{\ast }|},  \notag \\
\Phi ^{2}(\tau ) &=&-4\ \ _{Q}^{v}\breve{\Lambda}(\tau )h_{4}\simeq -4\ \
_{Q}^{v}\breve{\Lambda}(\tau )\eta _{4}\breve{g}_{4}\simeq -4\ \ _{Q}^{v}%
\breve{\Lambda}\ (\tau )\zeta _{4}(1+\kappa \chi _{4})\ \breve{g}_{4}. 
\notag
\end{eqnarray}%
These formulas are provided for the parametrization of generating sources and effective cosmological constants (\ref{dsourcparamkerr}) and (\ref{effectcosmnedmker}). Such symmetries depend on the type of nonmetric EDM configurations. Using the AFCDM, for instance, in \cite{bubuianu17,bubuianu19,partner03} we constructed and studied physical properties of d-metrics when the K(a)dS and other type BH solutions involve nonholonomic deformations on $y^{3}=\varphi .$ The effective sources $\ ^{v}\breve{J}(\tau )$ can be parameterized in respective forms related to certain extra dimension (super) string contributions. They may be related to nonassociative and/or noncommutative terms, certain types of generalized Finsler and modified dispersion deformations, or other type of nonmetric MGTs \cite{lb24epjc1,lb24epjc2}. We proved that such locally anisotropic configurations can be stable, or stabilized by imposing corresponding nonholonomic constraints. Similar methods and results allow to construct nonmetric quasi-stationary solutions with effective gravitational $\eta $-polarizations when $y^{3}=\theta .$

\subsubsection{Off-diagonal small parametric nonmetric (ellipsoidal) deformations of KdS d-metrics}

Considering gravitational polarization functions with small $\varepsilon $-deformations as in (\ref{nonlinsymr}), we can provide a more clear physical interpretation of $\tau $-families of quasi-stationary d-metrics (\ref{nkernewnmrf}) and respective nonlinear symmetries (\ref{nlims2}). We avoid off-diagonal frame/coordinate deformations with singularities if we consider a new system of coordinates resulting in nontrivial terms for $\breve{N}_{i}^{a}.$ Let us we denote $\breve{N}_{i}^{3}=\breve{w}_{i}(r,\varphi
,\theta )$ (such values can be zero in certain rotation frames) and $\breve{N}_{i}^{4}=\breve{n}_{i}(r,\varphi ,\theta ),$ with a nontrivial 
$\breve{n}_{2}=-a\sin \theta (r^{2}+a^{2}-\triangle _{\Lambda })/(a^{2}\sin ^{2}\theta
-\triangle _{\Lambda })$ as we considered above.

Applying the AFCDM with $\varepsilon $-deformations, we construct a $\tau $-family of quasi-stationary d-metrics with nonmetric $\varepsilon $-deformations (\ref{offdiagdefr}) determined by $\chi $-generating
functions: 
\begin{equation*}
d\ \widehat{s}^{2}(\tau )=\widehat{g}_{\alpha \beta }(\tau ,r,\varphi
,\theta ;\psi ,g_{4};\ ^{v}\breve{J})du^{\alpha }du^{\beta }=e^{\psi
_{0}}(1+\varepsilon \ ^{\psi }\chi (\tau ))[(dx^{1}(r,\varphi
))^{2}+(dx^{2}(r,\varphi ))^{2}]
\end{equation*}%
\begin{eqnarray*}
&&-\{\frac{4[\partial _{\theta }(|\zeta _{4}(\tau )\breve{g}_{4}|^{1/2})]^{2}%
}{\breve{g}_{3}|\int d\theta \lbrack \ ^{v}\breve{J}(\tau )\partial
_{3}(\zeta _{4}(\tau )\breve{g}_{4})]|}-\varepsilon \lbrack \frac{\partial
_{\theta }(\chi _{4}(\tau )|\zeta _{4}(\tau )\breve{g}_{4}|^{1/2})}{%
4\partial _{\theta }(|\zeta _{4}(\tau )\breve{g}_{4}|^{1/2})}-\frac{\int
d\theta \{\ ^{v}\breve{J}(\tau )\partial _{\theta }[(\zeta _{4}(\tau )\breve{%
g}_{4})\chi _{4}(\tau )]\}}{\int d\theta \lbrack \ ^{v}\breve{J}(\tau
)\partial _{\theta }(\zeta _{4}(\tau )\breve{g}_{4})]}]\}\breve{g}_{3} \\
&&\{d\theta +[\frac{\partial _{i}\ \int d\theta \ ^{v}\breve{J}(\tau )\
\partial _{\theta }\zeta _{4}(\tau )}{(\breve{N}_{i}^{3})\ ^{v}\breve{J}%
(\tau )\partial _{\theta }\zeta _{4}(\tau )}+\varepsilon (\frac{\partial
_{i}[\int d\theta \ \ ^{v}\breve{J}(\tau )\ \partial _{\theta }(\zeta
_{4}(\tau )\chi _{4}(\tau ))]}{\partial _{i}\ [\int d\theta \ \ ^{v}\breve{J}%
(\tau )\partial _{\theta }\zeta _{4}(\tau )]}-\frac{\partial _{\theta
}(\zeta _{4}(\tau )\chi _{4}(\tau ))}{\partial _{\theta }\zeta _{4}(\tau )})]%
\breve{N}_{i}^{3}dx^{i}\}^{2}
\end{eqnarray*}%
\begin{eqnarray}
&&+\zeta _{4}(\tau )(1+\varepsilon \ \chi _{4}(\tau ))\ \breve{g}_{4}\{dt+[(%
\breve{N}_{k}^{4})^{-1}[\ _{1}n_{k}(\tau )+16\ _{2}n_{k}(\tau )[\int d\theta 
\frac{\left( \partial _{\theta }[(\zeta _{4}(\tau )\breve{g}%
_{4})^{-1/4}]\right) ^{2}}{|\int d\theta \partial _{\theta }[\ ^{v}\breve{J}%
(\tau )(\zeta _{4}(\tau )\breve{g}_{4})]|}]  \label{offdnceleps1} \\
&&+\varepsilon \frac{16\ _{2}n_{k}(\tau )\int d\theta \frac{\left( \partial
_{\theta }[(\zeta _{4}(\tau )\breve{g}_{4})^{-1/4}]\right) ^{2}}{|\int
d\theta \partial _{\theta }[\ ^{v}\breve{J}(\tau )(\zeta _{4}(\tau )\breve{g}%
_{4})]|}(\frac{\partial _{\theta }[(\zeta _{4}(\tau )\breve{g}%
_{4})^{-1/4}\chi _{4}(\tau ))]}{2\partial _{\theta }[(\zeta _{4}(\tau )%
\breve{g}_{4})^{-1/4}]}+\frac{\int d\theta \partial _{\theta }[\ ^{v}\breve{J%
}(\tau )(\zeta _{4}(\tau )\chi _{4}(\tau )\breve{g}_{4})]}{\int d\theta
\partial _{\theta }[\ ^{v}\breve{J}(\tau )(\zeta _{4}(\tau )\breve{g}_{4})]})%
}{\ _{1}n_{k}(\tau )+16\ _{2}n_{k}(\tau )[\int d\theta \frac{\left( \partial
_{\theta }[(\zeta _{4}(\tau )\breve{g}_{4})^{-1/4}]\right) ^{2}}{|\int
d\theta \partial _{\theta }[\ ^{v}\breve{J}(\tau )(\zeta _{4}(\tau )\breve{g}%
_{4})]|}]}]\breve{N}_{k}^{4}dx^{k}\}^{2}.  \notag
\end{eqnarray}%
In above formulas, the polarization functions 
$\zeta _{4}(\tau ,r,\varphi,\theta )$ and $\chi _{4}(\tau ,r,\varphi ,\theta )$ can be prescribed to be
of a necessary smooth class when $\chi _{4}(\tau )$ are considered as families of generating functions. The d-metrics (\ref{offdnceleps1}) describe small $\varepsilon $-parametric deformations of the of KdS
d-metrics (\ref{offdiagpm1})  when the coefficients get additional anisotropy on $\varphi $-coordinate being determined by effective nonmetric EDM sources.

A more special class of off-diagonal solutions of type (\ref{offdnceleps1}) can be generated as ellipsoidal deformations on $\theta $ if we chose 
\begin{equation}
\chi _{4}(\tau ,r,\varphi ,\theta )=\underline{\chi }(\tau ,r,\varphi )\sin
(\omega _{0}\theta +\theta _{0}),  \label{rotoid}
\end{equation}%
where $\underline{\chi }(\tau ,r,\varphi )$ are smooth functions and $\omega _{0}$ and $\theta _{0}$ are some constants. For such $\tau $-families of generating polarization functions and $\zeta _{4}(\tau ,r,\varphi ,\theta )\neq 0,$ we can approximate 
\begin{equation*}
(1+\varepsilon \ \chi _{4})\ \breve{g}_{4}\simeq a^{2}\sin ^{2}\theta
-\triangle _{\Lambda }+\varepsilon \ \chi _{4}=0.
\end{equation*}%
Considering small values $a$ and $\frac{\Lambda _{0}}{3},$ and fixing $\tau =\tau _{0},$ we obtain $r=2M/(1+\varepsilon \ \chi _{4}).$ This parametric equation define a rotoid configuration with the eccentricity parameter $\varepsilon $ and generating function $\chi _{4}(\tau _{0},r,\varphi ,\theta
)$ as in (\ref{rotoid}). 

In general, we can consider polarization functions $\chi _{4}(\tau,r,\varphi ,\theta )$ when KdS BH are embedded into nontrivial nonholonomic quasi-stationary backgrounds under nonmetric geometric flow evolution. The nonholonomic conditions can be imposed in such a way that the BH configurations
are preserved as conventional h- and v-distributions encoding nonmetric EDM interactions. For small ellipsoidal deformations of type (\ref{rotoid}), we model black ellipsoid, BE, objects. They can be prescribed to be stable \cite{vmon05} (see also references therein) by choosing corresponding classes of nonholonomic constraints. Imposing respective classes of generating and integration functions subjected to zero torsion conditions, $\widehat{\mathbf{T}}_{\ \alpha \beta }^{\gamma }=0,$ we extract LC configurations. In
such cases, the scalar curvatures are of type $R(\tau ,r,\varphi ,\theta)\simeq \Lambda (\tau ,r,\varphi ,\theta )$ as can be described by nonlinear symmetries (\ref{nlims2}). The phenomenon of warped curvature described in  \cite{ovalle21} can be preserved for some subclasses of nonmetric and nonholonomic deformations. The nonmetric gravitational vacuum became more complex with an effective matter tensor 
$\ _{Q}\mathbf{\breve{J}}(\tau )=\ ^{m}\mathbf{\breve{Y}}_{\alpha \beta }(\tau ,r,\varphi )+\ _{Q}%
\widehat{\mathbf{J}}(\tau ,r,\varphi ,\theta )$ parameterized in a form (\ref{dsourcparamkerr}).

\subsubsection{Computing thermodynamic variables for nonmetric EDM deformations of KdS solutions}

\label{sscompthermvar}If we impose conditions of type (\ref{rotoid}), we can generate classes of off-diagonal solutions with an ellipsoidal type horizon. In such cases, we can use the  concept of Bekenstein-Hawking entropy \cite{bek2,haw2} and define a corresponding thermodynamic model. Such constructions were performed in certain nonassociative forms on 8-d phase space \cite{partner03,partner04}. If we follow the nonmetric nonholonomic approach from this work, we can take the commutative coefficients from
nononassociative thermodynamic variables, change corresponding parameterizations of physical constants and consider effective nonmetric EDM sources. We do not reproduce those formulas for 4-d nonholonomic distortions.

General classes of $\tau $-running off-diagonal solutions, for instance, of type (\ref{nkernewnmrf}) or (\ref{offdnceleps1}) do not possess closed horizons and do not involve any duality/ holographic properties. So, the Bekenstein-Hawking paradigm is not applicable to general nonmetric quasi-stationary and locally anisotropic solutions. To characterize the physical properties of generic solutions encoding nonmetric EDM data we change the thermodynamic paradigm using the concept of G. Perelman entropy \cite{perelman1}. For nonmetric geometric flow and MGTs, such models of statistical thermodynamics were elaborated in \cite{lb24epjc1,lb24epjc2} as a generalization of nonholonomic geometric thermodynamic models from 
\cite{sv12,svnonh08,vacaru20,partner04}.

The formulas for nonmetric thermodynamic variables (\ref{thermvar2}) can be computed for KdS-EDM solutions (\ref{offdnceleps1}): 
\begin{eqnarray*}
\ _{Q}^{q}\widehat{Z}(\tau ) &=&\exp [\frac{1}{8\pi ^{2}\tau ^{2}}\ _{\eta
}^{\shortmid }\mathcal{\mathring{V}}[ \widehat{g}_{\alpha \beta }(\tau)], \ _{Q}^{q}\widehat{\mathcal{E}}\ (\tau )=\ \frac{1-\tau \ (\ _{Q}^{h}%
\breve{\Lambda}(\tau )+\ \ _{Q}^{v}\breve{\Lambda}(\tau ))}{8\pi ^{2}\tau }\
\ _{\eta }^{\shortmid }\mathcal{\mathring{V}}[ \widehat{g}_{\alpha \beta
}(\tau )], \\
\ \ \ \ _{Q}^{q}\widehat{S}(\tau ) &=&-\ \ _{Q}^{q}\widehat{W}(\tau )=\frac{%
2-\tau (\ \ _{Q}^{h}\breve{\Lambda}(\tau )+\ \ _{Q}^{v}\breve{\Lambda}(\tau
))}{8\pi ^{2}\tau ^{2}}\ _{\eta }^{\shortmid }\mathcal{\mathring{V}[}\ 
\widehat{g}_{\alpha \beta }(\tau )].
\end{eqnarray*}%
For such variables, we consider generating sources $\ _{Q}\mathbf{\breve{J}}(\tau )$ (\ref{dsourcparamkerr}) and $\ _{Q}\breve{\Lambda}(\tau )$ and effective cosmological constants (\ref{effectcosmnedmker}). The explicit form of the volume functional (\ref{volumf1}) for $^{q}\mathbf{g}(\tau )=\{\ \widehat{g}_{\alpha \beta }(\tau )\},\ _{\eta }^{\shortmid }\mathcal{%
\mathring{V}[}\ ^{q}\mathbf{g}(\tau )],$ depends on the type of generating and integrating data (\ref{integrfunctrf}) for $^{q}\mathbf{g}(\tau )=\{\ \widehat{g}_{\alpha \beta }(\tau )\}$ defining a respective class of solutions (\ref{nkernewnmrf}) or (\ref{offdnceleps1}).

It is important to compare our solutions of nonmetric KdS configurations with the solutions for nonmetric gravity theories from \cite{dambrosio21}. The equations (4.6) and (4.7) of that work also contain nonmetric generalized Einstein equations with effective cosmological constants and energy-momentum tensors, but not EDM generalizations. Those authors constructed and analyzed a series of beyond-GR solutions encoding nonmetricity, for instance, how to generate Schwarzschild - de Sitter - Nordstr\"{o}m metrics; certain examples of off-diagonal perturbative corrections to the Schwarzschild metrics; approximate and exact vacuum solutions etc. If we chose any variant of the metrics from \cite{dambrosio21} (with corresponding nonlinear coordinate transforms to a variant which allows us to apply the AFCDM)  as a prime d-metric instead of $\breve{g}_{\alpha }(r,\varphi ,\theta )$  (\ref{offdiagpm1}), we can perform similar computations as in above subsections 4.1.1 - 4.1.4 but for not KdS configurations. This way, we can compute nonmetric EDM deformations of any solutions from that work and even compute the corresponding nonmetric geometric flow thermodynamic variables.  We do not reproduce such technical results in this work but note that nonmetric geometric flow and/or EDM corrections result in generic off-diagonal configurations and nontrivial cosmological constants. In principle, we can consider some small parametric nonmetric nonholonomic deformations of the Schwarzschild metric as certain very special  configurations which, in many cases, are not stable. This is not a crucial physical problem for MGTs when typically the BH uniqueness theorem and topological criteria can't be proven in rigorous forms as in GR. Unstable solutions also play an important role in nonlinear physics, for instance, describing evolution processes, phase space transitions, anisotropic kinetic processes etc. The AFCDM allows us to prove that nonmetric geometric flow and/or EDM deformations of any type of solutions (KdS or another type ones form \cite{dambrosio21}) result in generic off-diagonal configurations characterized by new types of nonlinear symmetries  as we explained in section 3.2.2. So, we deviate substantially from the Schwarzschild solution going beyond GR with nonmetric EDM deformations even in the vacuum case. This results in a new nonmetric geometric flow, or nonmetric Ricci soliton,  thermodynamic paradigm as we prove in section 3.1.3.   

\subsection{Nonmetric locally anisotropic EDM wormholes}

Nonholonomic off-diagonal deformations of WH solutions to locally anisotropic configurations were studied in \cite{v13,v14}. In this subsection, we apply similar methods for constructing new classes of $\tau $%
-families of quasi-stationary nonmetric EDM solutions derived for primary WH metrics in GR.

Let us consider a Morris-Thorne WH solution \cite{morris88} defined by a quadratic line element 
\begin{equation*}
d\mathring{s}^{2}=(1-\frac{b(r)}{r})^{-1}dr^{2}+r^{2}d\theta ^{2}+r^{2}\sin
^{2}\theta d\varphi ^{2}-e^{2\Phi (r)}dt^{2}.
\end{equation*}%
In this formula, we use spherical polar coordinates $u^{\alpha }=(r,\theta ,\varphi ,t)$ where $e^{2\Phi (r)}$ is a red-shift function and $b(r)$ is the shape function. In a similar form, the usual Ellis-Bronnikov (EB) WH
configuration is defined for $\Phi (r)=0$ and $b(r)=\ _{0}b^{2}/r.$ Such values characterize a zero tidal WH  with $\ _{0}b$ being the throat radius. Reviews and new results can be found in \cite{kar94,roy20,souza22}.

A generalized EB metric is characterized additionally by even integers $2k$ (with $k=1,2,...$) when $r(l)=(l^{2k}+\ _{0}b^{2k})^{1/2k}$ is a proper radial distance coordinate (tortoise) and the cylindrical angular coordinate $\phi \in \lbrack 0,2\pi )$ is called parallel. In such coordinates, $-\infty <l<\infty $ which is different from the cylindrical radial coordinate $\rho , $ when $0\leq \rho <\infty .$ This allows us to define a prime metric 
\begin{eqnarray}
d\mathring{s}^{2} &=&dl^{2}+r^{2}(l)d\theta ^{2}+r^{2}(l)\sin ^{2}\theta
d\varphi ^{2}-dt^{2},\mbox{ for }  \label{pmwhd} \\
dl^{2} &=&(1-\frac{b(r)}{r})^{-1}dr^{2}\mbox{ and }b(r)=r-r^{3(1-k)}(r^{2k}-%
\ _{0}b^{2k})^{(2-1/k))}.  \notag
\end{eqnarray}%
To apply the AFCDM  we consider a new system curved of coordinates with trivial N-connection coefficients $\check{N}_{i}^{a}=$ $\check{N}_{i}^{a}(u^{\alpha }(l,\theta ,\varphi ,t))$ and $\check{g}_{\beta
}(u^{j}(l,\theta ,\varphi ),u^{3}(l,\theta ,\varphi )).$ Such a system has to be chosen in any form to avoid off-diagonal deformations with coordinate and frame coefficient singularities. For instance, we consider new
coordinates $u^{1}=x^{1}=l,u^{2}=\theta ,$ and $u^{3}=y^{3}=\varphi +\
^{3}B(l,\theta ),u^{4}=y^{4}=t+\ ^{4}B(l,\theta ),$ resulting in local dual vertical frames 
\begin{eqnarray*}
\mathbf{\check{e}}^{3} &=&d\varphi =du^{3}+\check{N}_{i}^{3}(l,\theta
)dx^{i}=du^{3}+\check{N}_{1}^{3}(l,\theta )dl+\check{N}_{2}^{3}(l,\theta
)d\theta , \\
\mathbf{\check{e}}^{4} &=&dt=du^{4}+\check{N}_{i}^{4}(l,\theta
)dx^{i}=du^{4}+\check{N}_{1}^{4}(l,\theta )dl+\check{N}_{2}^{4}(l,\theta
)d\theta ,
\end{eqnarray*}%
for $\mathring{N}_{i}^{3}=-\partial \ ^{3}B/\partial x^{i}$ and $\mathring{N}%
_{i}^{4}=-\partial \ ^{4}B/\partial x^{i}.$ As a result, we can express the
WH solution (\ref{pmwhd}) as a prime d-metric, 
\begin{equation}
d\mathring{s}^{2}=\check{g}_{\alpha }(l,\theta ,\varphi )[\mathbf{\check{e}}%
^{\alpha }(l,\theta ,\varphi )]^{2},  \label{pmwh}
\end{equation}%
where $\check{g}_{1}=1,\check{g}_{2}=r^{2}(l),\check{g}_{3}=r^{2}(l)\sin
^{2}\theta $ and $\check{g}_{4}=-1.$

Nonmetric off-diagonal deformations of (\ref{pmwh}) to target d-metrics of type  (\ref{ans1rf}) can be generated by any nontrivial (effective) source $\ _{Q}\widehat{\mathbf{J}}=[\ _{Q}^{h}\widehat{\mathbf{J}}(\tau ,l,\theta ),\ _{Q}^{v}\widehat{\mathbf{J}}(\tau ,l,\theta ,\varphi )].$ Such effective sources  $\ ^{m}\widehat{\mathbf{Y}}_{\alpha \beta }=\ ^{m}\mathbf{%
\breve{Y}}_{\alpha \beta }$ \ (\ref{ceemt}) are defined by $\breve{T}%
_{\alpha \beta }(l,\theta )=\Lambda $ in (\ref{efmt1}) and can be used
for deforming WH solutions using coordinates $(l,\theta ,\varphi )$), 
instead of $\ ^{m}T_{\alpha \beta }$ (\ref{emtlc}.  So, for constructing
and finding quasi-stationary WH nonmetric deformations as solutions of the
systems of nonlinear PDEs the nonmetric geometric flow equations (\ref%
{ricciflowr2}) or (\ref{ricciflowr2cosmc}), we can use any  effective
Lagrange density $\ ^{nDA}\mathcal{\breve{L}}(\tau ,l,\theta ,\varphi )=%
\mathcal{\breve{L}}(\tau ,l,\theta ,\varphi )+\ ^{nDA}\widehat{\mathcal{L}}%
(\tau ,l,\theta ,\varphi ).$ We note that $\ ^{m}\mathcal{L}(\tau )=\mathcal{%
\breve{L}}(\tau ,l,\theta ,\varphi )$ corresponds to $\breve{T}_{\alpha
\beta }=\Lambda $ and the term for nonmetric Dirac-Maxwell effective
Lagrange density is $\ ^{nDA}\widehat{\mathcal{L}}(\tau ,l,\theta ,\varphi )$
(\ref{mlagd}). 

For above parameterizations of the EMD sources, off-diagonal quasi-stationary deformations of WH  (\ref{pmwh}) are generated by introducing nontrivial sources $\ _{Q}^{h}\widehat{\mathbf{J}}(\tau
,l,\theta )$ and $\ \ _{Q}^{v}\widehat{\mathbf{J}}(\tau ,l,\theta ,\varphi )=\ \ _{Q}^{wh}J(\tau ).$ Such effective nonmetric sources are  related to nonlinear symmetries of type (\ref{nonlinsymrf}) to a $\tau $-running effective cosmological constant $\Lambda (\tau ),$ when $\Lambda (\tau _{0})=
$ $\Lambda .$  In explicit form, such symmetries are defined by formulas: 
\begin{eqnarray}
\partial _{\varphi }[\ _{J}\Phi ^{2}(\tau )] &=&-\int d\varphi \ \
_{Q}^{wh}J(\tau )\partial _{\varphi }h_{4}\simeq -\int d\varphi \ \
_{Q}^{wh}J(\tau )\partial _{\varphi }(\eta _{4}\ \breve{g}_{4})\simeq -\int
d\theta \ \varphi \ \ _{Q}^{wh}J(\tau )\partial _{\varphi }[\zeta
_{4}(1+\kappa \ \chi _{4})\ \breve{g}_{4}],  \notag \\
\ _{J}\Phi (\tau ) &=&|\ \Lambda (\tau )|^{-1/2}\sqrt{|\int d\varphi \ \
_{Q}^{wh}J(\tau )\ (\Phi ^{2})^{\ast }|},  \label{nonlinsymwh} \\
\Phi ^{2}(\tau ) &=&-4\ \Lambda (\tau )h_{4}\simeq -4\ \Lambda (\tau )\eta
_{4}\breve{g}_{4}\simeq -4\ \Lambda (\tau )\zeta _{4}(1+\kappa \chi _{4})\ 
\breve{g}_{4}.  \notag
\end{eqnarray}%
The generating functions for such nonlinear symmetries are parameterized as%
\begin{equation*}
\ _{J}\Phi (\tau )=\ _{J}\Phi (\tau ,l,\theta ,\varphi ),\ \Phi (\tau )=\
\Phi (\tau ,l,\theta ,\varphi ),\eta _{4}(\tau )=\ \eta _{4}(\tau ,l,\theta
,\varphi ),\zeta _{4}(\tau )=\ \zeta _{4}(\tau ,l,\theta ,\varphi ),\chi
_{4}(\tau )=\ \chi _{4}(\tau ,l,\theta ,\varphi ).
\end{equation*}

Using $\eta $-polarization functions, we derived such $\tau $-families of target quasi-stationary metrics 
\begin{eqnarray}
d\widehat{s}^{2} &=&\ _{Q}^{wh}g_{\alpha \beta }(l,\theta ,\varphi ;\psi
,\eta _{4};\Lambda ,\ _{Q}^{wh}J,\ \check{g}_{\alpha })du^{\alpha }du^{\beta
}  \notag \\
&=&e^{\psi (\tau ,l,\theta )}[(dx^{1}(\tau ,l,\theta ))^{2}+(dx^{2}(\tau
,l,\theta ))^{2}]  \label{whpolf} \\
&&-\frac{[\partial _{\varphi }(\eta _{4}\ \check{g}_{4})]^{2}}{|\int
d\varphi \ \ \ _{Q}^{wh}J\partial _{\varphi }(\eta _{4}\ \breve{g}_{4})|\
\eta _{4}\ \check{g}_{4}}\{dy^{3}+\frac{\partial _{i}[\int d\varphi \ \ \
_{Q}^{wh}J\ \partial _{\varphi }(\eta _{4}\ \check{g}_{4})]}{\ \ \
_{Q}^{wh}J\partial _{\varphi }(\eta _{4}\ \check{g})}dx^{i}\}^{2}+\eta _{4}%
\breve{g}_{4}  \notag \\
&&\{dt+[\ _{1}n_{k}(\tau ,l,\theta )+\ _{2}n_{k}(\tau ,l,\theta )\int
d\varphi \frac{\lbrack \partial _{\varphi }(\eta _{4}\ \breve{g}_{4})]^{2}}{%
|\int d\varphi \ \ \ _{Q}^{wh}J\partial _{\varphi }(\eta _{4}\ \breve{g}%
_{4})|\ (\eta _{4}\ \breve{g}_{4})^{5/2}}]dx^{k}\}.  \notag
\end{eqnarray}%
This class of solutions are determined by respective generating function $%
\eta _{4}(\tau )=\eta _{4}(\tau ,l,\theta ,\varphi )$ and integration
functions $\ _{1}n_{k}(\tau ,l,\theta )$ and $\ _{2}n_{k}(\tau ,l,\theta ).$
The function $\psi (\tau )=\psi (\tau ,l,\theta )$ are solutions of a $\tau $%
-family of 2-d Poisson equations $\partial _{11}^{2}\psi (\tau )+\partial
_{22}^{2}\psi (\tau )=2\ \ _{Q}^{h}\widehat{\mathbf{J}}(\tau ,l,\theta ).$

We emphasize that, in general,  target d-metrics (\ref{whpolf}) do not
describe WH like locally anisotropic object for general classes of
generating and integrating data. Locally anisotropic WHs are generated  if
we consider  small parametric quasi-stationary deformations of prime metrics
of type (\ref{pmwh}). Such nonmetric EDM deformations are described in terms
of polarization functions $\ \chi _{4}(\tau ,l,\theta ,\varphi )$ when the
quadratic linear elements are computed 
\begin{equation*}
d\ \widehat{s}^{2}=\ _{Q}^{wh}g_{\alpha \beta }(l,\theta ,\varphi ;\psi
,\chi _{4};\Lambda ,\ _{Q}^{wh}J,\ \check{g}_{\alpha })du^{\alpha }du^{\beta
}=e^{\psi _{0}(l,\theta )}[1+\kappa \ ^{\psi (l,\theta )}\chi (l,\theta
)][(dx^{1}(l,\theta ))^{2}+(dx^{2}(l,\theta ))^{2}]
\end{equation*}%
\begin{eqnarray*}
&&-\{\frac{4[\partial _{\varphi }(|\zeta _{4}\ \breve{g}_{4}|^{1/2})]^{2}}{%
\breve{g}_{3}|\int d\varphi \{\ \ _{Q}^{wh}J\partial _{\varphi }(\zeta _{4}\ 
\breve{g}_{4})\}|}-\kappa \lbrack \frac{\partial _{\varphi }(\chi _{4}|\zeta
_{4}\breve{g}_{4}|^{1/2})}{4\partial _{\varphi }(|\zeta _{4}\ \breve{g}%
_{4}|^{1/2})}-\frac{\int d\varphi \{\ _{Q}^{wh}J\partial _{\varphi }[(\zeta
_{4}\ \breve{g}_{4})\chi _{4}]\}}{\int d\varphi \{\ _{Q}^{wh}J\partial
_{\varphi }(\zeta _{4}\ \breve{g}_{4})\}}]\}\ \breve{g}_{3} \\
&&\{d\varphi +[\frac{\partial _{i}\ \int d\varphi \ \ _{Q}^{wh}J\ \partial
_{\varphi }\zeta _{4}}{(\check{N}_{i}^{3})\ \ _{Q}^{wh}J\partial _{\varphi
}\zeta _{4}}+\kappa (\frac{\partial _{i}[\int d\varphi \ \ _{Q}^{wh}J\
\partial _{\varphi }(\zeta _{4}\chi _{4})]}{\partial _{i}\ [\int d\varphi \
\ \ _{Q}^{wh}J\partial _{\varphi }\zeta _{4}]}-\frac{\partial _{\varphi
}(\zeta _{4}\chi _{4})}{\partial _{\varphi }\zeta _{4}})]\check{N}%
_{i}^{3}dx^{i}\}^{2}
\end{eqnarray*}%
\begin{eqnarray}
&&+\zeta _{4}(1+\kappa \ \chi _{4})\ \breve{g}_{4}\{dt+[(\check{N}%
_{k}^{4})^{-1}[\ _{1}n_{k}+16\ _{2}n_{k}[\int d\varphi \frac{\left( \partial
_{\varphi }[(\zeta _{4}\ \breve{g}_{4})^{-1/4}]\right) ^{2}}{|\int d\varphi
\partial _{\varphi }[\ _{Q}^{wh}J(\zeta _{4}\ \breve{g}_{4})]|}]  \notag \\
&&+\kappa \frac{16\ _{2}n_{k}\int d\varphi \frac{\left( \partial _{\varphi
}[(\zeta _{4}\ \breve{g}_{4})^{-1/4}]\right) ^{2}}{|\int d\varphi \partial
_{\varphi }[\ _{Q}^{wh}J(\zeta _{4}\ \breve{g}_{4})]|}(\frac{\partial
_{\varphi }[(\zeta _{4}\ \breve{g}_{4})^{-1/4}\chi _{4})]}{2\partial
_{\varphi }[(\zeta _{4}\ ^{cy}g)^{-1/4}]}+\frac{\int d\varphi \partial
_{\varphi }[\ _{Q}^{wh}J(\zeta _{4}\chi _{4}\ \breve{g}_{4})]}{\int d\varphi
\partial _{\varphi }[\ _{Q}^{wh}J(\zeta _{4}\ \breve{g}_{4})]})}{\
_{1}n_{k}+16\ _{2}n_{k}[\int d\varphi \frac{\left( \partial _{\varphi
}[(\zeta _{4}\ \breve{g}_{4})^{-1/4}]\right) ^{2}}{|\int d\varphi \partial
_{\varphi }[\ _{Q}^{wh}J(\zeta _{4}\ \breve{g}_{4})]|}]}]\check{N}%
_{k}^{4}dx^{k}\}^{2}.  \label{whpolf1}
\end{eqnarray}%
In  formulas (\ref{whpolf})  and  (\ref{whpolf1}), $\tau $-parametric
dependencies are similar to those from (\ref{nonlinsymwh}). 

We can model elliptic WH deformations as a particular case of d-metrics (%
\ref{whpolf1}) for a fixed $\tau _{0}$ if we chose a generating function of
type%
\begin{equation*}
\chi _{4}(l,\theta ,\varphi )=\underline{\chi }(l,\theta )\sin (\omega
_{0}\varphi +\varphi _{0})
\end{equation*}%
defining ellipsoidal  $\varphi $-anisotropic deformations. Here we note that
a different family of solutions of type (\ref{whpolf}) and (\ref{whpolf1})
can be constructed if we change the order of angular coordinates in the
primary and target d-metrics, $\theta \leftrightarrow \varphi .$ We omit
details on such applications of the AFCDM.

The thermodynamic variables for nonmetric WHs and EDM deformations (\ref%
{thermvar2}) can be computed for WH solutions considering, for instance, for 
$\ _{Q}^{h}\breve{\Lambda}(\tau )=\ \ _{Q}^{v}\breve{\Lambda}(\tau )=\Lambda
(\tau ),$ when 
\begin{eqnarray*}
\ _{Q}^{q}\widehat{Z}(\tau ) &=&\exp [\frac{1}{8\pi ^{2}\tau ^{2}}\ _{\eta
}^{\shortmid }\mathcal{\mathring{V}}[ \ _{Q}^{wh}g_{\alpha \beta }(\tau)],
\ _{Q}^{q}\widehat{\mathcal{E}}\ (\tau )=\ \frac{1-2\tau \Lambda (\tau )%
}{8\pi ^{2}\tau }\ \ _{\eta }^{\shortmid }\mathcal{\mathring{V}[}\ \
_{Q}^{wh}g_{\alpha \beta }(\tau )], \\
\ \ \ \ _{Q}^{q}\widehat{S}(\tau ) &=&-\ \ _{Q}^{q}\widehat{W}(\tau )=\frac{%
1-\tau \Lambda (\tau )}{4\pi ^{2}\tau ^{2}}\ _{\eta }^{\shortmid }\mathcal{%
\mathring{V}[}\ _{Q}^{wh}g_{\alpha \beta }(\tau )].
\end{eqnarray*}
The explicit form of the volume functional (\ref{volumf1}) encoding nonmetric EDM deformations of WHs  depends on the type of prescribed generating and integrating data (\ref{integrfunctrf}) for 
$\ _{Q}^{wh}g_{\alpha \beta }(\tau )$ in (\ref{whpolf}) or (\ref{whpolf1}).  

\subsection{Nonmetric EDM toroid configurations and BT}

Nonholonomic off-diagonal deformations of toroidal BHs were studied in \cite{v01t,vs01b}. Those classes of solutions  provided generalizations and alternative solutions for BT and black ring generalizations in GR and MGTs \cite{lemos01,peca98,emparan02,emparan08,astorino17}. We cite 
\cite{astorino17} for a recent review of results by other authors using diagonal toroid ansatz.  In this subsection, we study an example when the AFCDM is applied for generating nonmetric EDM quasi-stationary off-diagonal solutions using prime BT metrics.  For simplicity, we  consider  small parametric deformations when the physical interpretation of new classes of solutions is very similar to the holonomic and diagonalizable metric ansatz.

As a prime metric, we consider a quadratic line element (see details in section 3.1 of \cite{astorino17}) 
\begin{eqnarray}
d\tilde{s}^{2} &=&f^{-1}(\tilde{r})d\tilde{r}^{2}+\tilde{r}^{2}(\tilde{k}%
_{1}^{2}dx^{2}+\tilde{k}_{2}^{2}dy^{2})-f(\tilde{r})d\tilde{t}^{2}
\label{prmtor1} \\
&=&\tilde{g}_{\alpha }(\tilde{x}^{1})(d\tilde{u}^{\alpha })^{2},\mbox{ for }%
f(\tilde{r})=-\epsilon ^{2}b^{2}-\tilde{\mu}/\tilde{r}-\Lambda \tilde{r}%
^{2}/3.  \notag
\end{eqnarray}%
The coordinates in this metric $\tilde{g}=\{\tilde{g}_{\alpha }\}$ are related via re-scaling parameter $\epsilon $ to toroidal "normalized" coordinates: $r$ is a radial coordinate, with $\theta =2\pi k_{1}x$ and $ 
\varphi =2\pi k_{2}y$ (when $x,y\in \lbrack 0,1]$). The re-scaling conditions are 
\begin{equation*}
k_{1}=\epsilon \tilde{k}_{1},k_{2}=\epsilon \tilde{k}_{2},\mu \rightarrow 
\frac{\mu }{(2\pi )^{3}}=\tilde{\mu}/\epsilon ^{3};r\rightarrow \frac{r}{%
2\pi }=\tilde{r}/\epsilon ,t\rightarrow 2\pi t=\epsilon \tilde{r}.
\end{equation*}%
In (\ref{prmtor1}), the parameter $b$ is a coupling constant for the energy-momentum tensor for the nonlinear SU(2) sigma model 
\begin{equation}
T_{\mu \nu }=\frac{b^{2}\epsilon ^{2}}{8\pi G\tilde{r}^{2}}[f(\tilde{r}%
)\delta _{\mu }^{4}\delta _{\nu }^{4}-f^{-1}(\tilde{r})\delta _{\mu
}^{1}\delta _{\nu }^{1}],  \label{emtsm}
\end{equation}%
where the integration constant  $\mu $ being can be fixed as a mass parameter. The value $\epsilon =0$ allows to recover in a formal way certain toroidal vacuum solutions studied  \cite{lemos01,peca98}. The toroidal configurations (\ref{prmtor1}) is for an exact static solution of the Einstein equations  for the LC connection and energy-momentum tensor (\ref{emtsm}). It defines  an AdS BH with a toroidal horizon in 4-d
Einstein gravity combined with a nonlinear $\sigma $-model.

To apply the AFCDM is convenient to consider frame transforms to an off-diagonal parametrization of (\ref{prmtor1}) to a form with nonzero N-connection coefficients $\tilde{N}_{i}^{a}= \tilde{N}_{i}^{a}(u^{\alpha }(\tilde{r},x,y,t))$ and $\tilde{g}_{\alpha \beta }(u^{j}(\tilde{r}%
,x,y),u^{3}(\tilde{r},x,y)).$ Such coefficients have a trivial N-connection curvature being  defined in any form which does not involve singular frame transforms and off-diagonal deformations. For instance, we introduce new coordinates $u^{1}=x^{1}=\tilde{r},u^{2}=x,$ and $u^{3}=y^{3}=y+\ ^{3}B(%
\tilde{r},x),u^{4}=y^{4}=t+\ ^{4}B(\tilde{r},x),$ when 
\begin{eqnarray*}
\mathbf{\tilde{e}}^{3} &=&dy=du^{3}+\tilde{N}_{i}^{3}(\tilde{r}%
,x)dx^{i}=du^{3}+\tilde{N}_{1}^{3}(\tilde{r},x)dr+\tilde{N}_{2}^{3}(\tilde{r}%
,x)dz, \\
\mathbf{\tilde{e}}^{4} &=&dt=du^{4}+\tilde{N}_{i}^{4}(\tilde{r}%
,x)dx^{i}=du^{4}+\tilde{N}_{1}^{4}(\tilde{r},x)dr+\tilde{N}_{2}^{4}(\tilde{r}%
,x)dz,
\end{eqnarray*}%
for $\tilde{N}_{i}^{3}=-\partial \ ^{3}B/\partial x^{i}$ and $\tilde{N}_{i}^{4}=-\partial \ ^{4}B/\partial x^{i}.$ Using such nonlinear coordinates, we transform the diagonal metric (\ref{prmtor1}) into a
toroidal d-metric 
\begin{equation}
d\tilde{s}^{2}=\tilde{g}_{\alpha }(\tilde{r},x,y)[\mathbf{\tilde{e}}^{\alpha
}(\tilde{r},x,y)]^{2},  \label{prmtor2}
\end{equation}%
where $\tilde{g}_{1}=f^{-1}(x^{1}),\tilde{g}_{2}=(x^{1})^{2}\tilde{k}%
_{1}^{2},\tilde{g}_{3}=(x^{2})^{2}\tilde{k}_{2}^{2}$ and $\tilde{g}%
_{4}=f(x^{1}).$

We generate nonmetric EDM locally anisotropic toroidal configurations if we construct small parametric quasi-stationary deformations of prime metrics of type (\ref{prmtor2}) defined by an effective source%
\begin{equation}
\ _{Q}^{tor}\mathbf{J}(\tau )=-\Lambda (\tau )\mathbf{g}+\ _{Q}\widehat{%
\mathbf{J}}(\tau )  \label{torsourc}
\end{equation}%
for $\ _{Q}\widehat{\mathbf{J}}=[\ _{Q}^{h}\widehat{\mathbf{J}}(\tau ,\tilde{%
r},x),\ _{Q}^{v}\widehat{\mathbf{J}}(\tau ,\tilde{r},x,y)]$ as in (\ref%
{dsourcparam}) but in toroid coordinates.  The $\tau $-running cosmological constant $\Lambda (\tau )$ in (\ref{torsourc})  can be nontrivial, zero, or absorbed by nonmetric EDM effective sources.   Nonlinear symmetries $[\ _{J}\Phi (\tau ),\ _{Q}^{v}\widehat{\mathbf{J}}(\tau )]\rightarrow \lbrack
\Phi (\tau ),\ \ _{Q}^{v}\Lambda (\tau )]$ (\ref{nonlinsymrf}), with 
$\ _{Q}\Lambda (\tau )=[\ _{Q}^{h}\Lambda (\tau ),\ _{Q}^{v}\Lambda (\tau )]$ (\ref{effectcosmnedm}), for $\tau $-families of quasi-stationary solutions of type (\ref{ans1rf}) are  described in explicit form by such formulas: 
\begin{eqnarray}
\partial _{y}[\ _{J}\Phi ^{2}(\tau ,\tilde{r},x,y)] &=&-\int dy[-\Lambda
(\tau )+\ _{Q}^{v}\widehat{\mathbf{J}}(\tau )]\partial _{y}g_{4}\simeq -\int
dy\ [-\Lambda (\tau )+\ _{Q}^{v}\widehat{\mathbf{J}}(\tau ,\tilde{r},x,y)]\
\partial _{y}[\eta _{4}(\tau ,\tilde{r},x,y)\ \tilde{g}_{4}(\tilde{r})]  \notag   \\
&\simeq &-\int dy\ [-\Lambda (\tau )+\ _{Q}^{v}\widehat{\mathbf{J}}(\tau ,%
\tilde{r},x,y)]\ \partial _{y}[\zeta _{4}(\tau ,\tilde{r},x,y)(1+\kappa \
\chi _{4}(\tau ,\tilde{r},x,y))\ \ \tilde{g}_{4}(\tilde{r})],  \label{nsymtor} \\
\ _{J}\Phi (\tau ,\tilde{r},x,y) &=&|\ \ \Lambda +\ _{Q}^{v}\Lambda (\tau
)|^{-1/2}\sqrt{|\int dy\ \ [-\Lambda (\tau )+\ _{Q}^{v}\widehat{\mathbf{J}}%
(\tau ,\tilde{r},x,y)]\ \ \partial _{y}(\Phi ^{2})|},  \notag \\
(\Phi (\tau ,\tilde{r},x,y))^{2} &=&-4\ \Lambda \tilde{g}_{4}(\tau ,\tilde{r}%
,x,y)\simeq -4\ (\ \Lambda +\ \ _{Q}^{v}\Lambda (\tau ))\eta _{4}(\tau ,%
\tilde{r},x,y)\ \ \tilde{g}_{4}(\tilde{r})  \notag \\
&\simeq &-4(\ \Lambda +\ _{Q}^{v}\Lambda (\tau ))\ \zeta _{4}(\tau ,\tilde{r}%
,x,y)(1+\kappa \chi _{4}(\tau ,\tilde{r},x,y))\ \tilde{g}_{4}(\tilde{r}). 
\notag
\end{eqnarray}

Using (\ref{nsymtor}), the small parametric nonmetric EDM deformations described in terms of 
$\chi $-polarization functions are defined by such  $\tau $-families of nonholonomic toroidal solutions: 
\begin{eqnarray*}
d\ \widehat{s}^{2}(\tau ) &=&\ ^{tor}g_{\alpha \beta }(\tilde{r},x,y;\psi
(\tau ),\eta _{4}(\tau );\ \Lambda (\tau )+\ _{Q}^{v}\Lambda (\tau ),\ \
_{Q}^{v}\widehat{\mathbf{J}}(\tau ),\ \tilde{g}_{\alpha })du^{\alpha
}du^{\beta } \\
&=&e^{\psi _{0}(\tau ,\tilde{r},x)}[1+\kappa \ ^{\psi (\tau ,\tilde{r}%
,x)}\chi (\tau ,\tilde{r},x)][(dx^{1}(\tau ,\tilde{r},x))^{2}+(dx^{2}(\tau ,%
\tilde{r},x))^{2}]-
\end{eqnarray*}%
\begin{eqnarray*}
&&\{\frac{4[\partial _{y}(|\zeta _{4}(\tau )\ \tilde{g}_{4}|^{1/2})]^{2}}{%
\tilde{g}_{3}|\int dy\{\ ^{tor}\Upsilon \partial _{y}(\zeta _{4}(\tau )\ 
\tilde{g}_{4})\}|}-\kappa \lbrack \frac{\partial _{y}(\chi _{4}(\tau )|\zeta
_{4}(\tau )\tilde{g}_{4}|^{1/2})}{4\partial _{y}(|\zeta _{4}(\tau )\tilde{g}%
_{4}|^{1/2})}
\\ &&
-\frac{\int dy\{[-\Lambda (\tau )+\ _{Q}^{v}\widehat{\mathbf{J}}(\tau ,%
\tilde{r},x,y)]\partial _{y}[(\zeta _{4}(\tau )\ \tilde{g}_{4})\chi
_{4}(\tau )]\}}{\int dy\{[-\Lambda (\tau )+\ _{Q}^{v}\widehat{\mathbf{J}}%
(\tau ,\tilde{r},x,y)]\partial _{y}(\zeta _{4}(\tau )\ \tilde{g}_{4})\}}]\}\ 
\tilde{g}_{3}
 %\\ &&
\{dy+[\frac{\partial _{i}\ \int dy\ [-\Lambda (\tau )+\ _{Q}^{v}\widehat{%
\mathbf{J}}(\tau ,\tilde{r},x,y)]\ \partial _{y}\zeta _{4}}{(\tilde{N}%
_{i}^{3})\ [-\Lambda (\tau )+\ _{Q}^{v}\widehat{\mathbf{J}}(\tau ,\tilde{r}%
,x,y)]\partial _{y}\zeta _{4}} \\
&&+\varepsilon (\frac{\partial _{i}[\int dy\ \ [-\Lambda (\tau )+\ _{Q}^{v}%
\widehat{\mathbf{J}}(\tau ,\tilde{r},x,y)]\ \partial _{y}(\zeta _{4}(\tau
)\chi _{4}(\tau ))]}{\partial _{i}\ [\int dy\ [-\Lambda (\tau )+\ _{Q}^{v}%
\widehat{\mathbf{J}}(\tau ,\tilde{r},x,y)]\partial _{y}\zeta _{4}(\tau )]}-%
\frac{\partial _{y}(\zeta _{4}(\tau )\chi _{4}(\tau ))}{\partial _{y}\zeta
_{4}(\tau )})]\tilde{N}_{i}^{3}dx^{i}\}^{2}+ \zeta _{4}(\tau )(1+\varepsilon \ \chi _{4}(\tau ))\ 
\end{eqnarray*}%
\begin{eqnarray}
&&\times \tilde{g}_{4}  \{dt+[(%
\tilde{N}_{k}^{4})^{-1}[\ _{1}n_{k}(\tau )+16\ _{2}n_{k}(\tau )[\int dy\frac{%
\left( \partial _{y}[(\zeta _{4}(\tau )\ \tilde{g}_{4})^{-1/4}]\right) ^{2}}{%
|\int dy\partial _{y}[[-\Lambda (\tau )+\ _{Q}^{v}\widehat{\mathbf{J}}(\tau ,%
\tilde{r},x,y)](\zeta _{4}(\tau )\ \tilde{g}_{4})]|}]+  \notag \\
&&\varepsilon \frac{16\ _{2}n_{k}(\tau )\int dy\frac{\left( \partial
_{y}[(\zeta _{4}(\tau )\ \tilde{g}_{4})^{-1/4}]\right) ^{2}}{|\int
dy\partial _{y}[[-\Lambda (\tau )+\ _{Q}^{v}\widehat{\mathbf{J}}(\tau ,%
\tilde{r},x,y)](\zeta _{4}(\tau )\ \tilde{g}_{4})]|}(\frac{\partial
_{y}[(\zeta _{4}(\tau )\ \tilde{g}_{4})^{-1/4}\chi _{4})]}{2\partial
_{y}[(\zeta _{4}(\tau )\ \tilde{g}_{4})^{-1/4}]}}{\ _{1}n_{k}(\tau )+16\
_{2}n_{k}(\tau )[\int dy\frac{\left( \partial _{y}[(\zeta _{4}(\tau )\ 
\tilde{g}_{4})^{-1/4}]\right) ^{2}}{|\int dy\partial _{y}[\ [-\Lambda (\tau
)+\ _{Q}^{v}\widehat{\mathbf{J}}(\tau ,\tilde{r},x,y)](\zeta _{4}\ \tilde{g}%
_{4})]|}]}]  \label{tortpol2} \\
&&\frac{+\frac{\int dy\partial _{y}[[-\Lambda (\tau )+\ _{Q}^{v}\widehat{%
\mathbf{J}}(\tau ,\tilde{r},x,y)](\zeta _{4}(\tau )\chi _{4}(\tau )\ \tilde{g%
}_{4})]}{\int dy\partial _{y}[[-\Lambda (\tau )+\ _{Q}^{v}\widehat{\mathbf{J}%
}(\tau ,\tilde{r},x,y)](\zeta _{4}(\tau )\ \tilde{g}_{4})]})}{\
_{1}n_{k}(\tau )+16\ _{2}n_{k}(\tau )[\int dy\frac{\left( \partial
_{y}[(\zeta _{4}(\tau )\ \tilde{g}_{4})^{-1/4}]\right) ^{2}}{|\int
dy\partial _{y}[\ [-\Lambda (\tau )+\ _{Q}^{v}\widehat{\mathbf{J}}(\tau ,%
\tilde{r},x,y)](\zeta _{4}\ \tilde{g}_{4})]|}]}]\tilde{N}_{k}^{4}dx^{k}%
\}^{2}.  \notag
\end{eqnarray}%
This  formula define certain sectional  elliptic deformations if we chose $\tau $-families of generating functions 
\begin{equation*}
\chi _{4}(\tau ,\tilde{r},x,y)=\underline{\chi }(\tau ,\tilde{r},x)\sin
(\omega _{0}y+y_{0}).
\end{equation*}%
These are toroid configurations with ellipsoidal deformations on $y$
coordinate. Above formulas and solutions can be re-defined for a different
family of quasi-stationary d-metrics with off-diagonal deformations on $x$%
-coordinate if we change the  coordinates $x\leftrightarrow y.$ 

The formulas for nonmetric EDM thermodynamic variables (\ref{thermvar2}) can
be computed for toroidal solutions (\ref{tortpol2}): 
\begin{eqnarray*}
\ _{Q}^{tor}\widehat{Z}(\tau ) &=&\exp [\frac{1}{8\pi ^{2}\tau ^{2}}\ _{\eta
}^{\shortmid }\mathcal{\mathring{V}[}\ ^{tor}g_{\alpha \beta }(\tau )],\
_{Q}^{tor}\widehat{\mathcal{E}}\ (\tau )=\ \frac{1-\tau \ (2\Lambda (\tau
)+\ _{Q}^{h}\breve{\Lambda}(\tau )+\ \ _{Q}^{v}\breve{\Lambda}(\tau ))}{8\pi
^{2}\tau }\ \ _{\eta }^{\shortmid }\mathcal{\mathring{V}[}\ \
^{tor}g_{\alpha \beta }(\tau )], \\
\ \ \ \ _{Q}^{tor}\widehat{S}(\tau ) &=&-\ \ _{Q}^{tor}\widehat{W}(\tau )=%
\frac{2-\tau (2\Lambda (\tau )+\ \ _{Q}^{h}\breve{\Lambda}(\tau )+\ \
_{Q}^{v}\breve{\Lambda}(\tau ))}{8\pi ^{2}\tau ^{2}}\ _{\eta }^{\shortmid }%
\mathcal{\mathring{V}[}\ \ ^{tor}g_{\alpha \beta }(\tau )].
\end{eqnarray*}%
The explicit form of the volume functional of type (\ref{volumf1}), $\
_{\eta }^{\shortmid }\mathcal{\mathring{V}[}\ \ ^{tor}g_{\alpha \beta }(\tau)],$ depends on the type of generating and integrating data (\ref{integrfunctrf}) written in toroidal variables.

Finally, we note that we can consider that various classes of deformations may transform a toroid prime d-metric into target "spaghetti" WHs, with additional black ellipsoids and other type quasi-stationary configurations. They may have  different sections, curved and waved hypersurfaces, possible interruptions, singularities etc. embedded into locally anisotropic and nonmetric gravitational vacuum media and EDM deformations. The geometry of such $\tau $-families of d-objects is deformed by respective prime metrics and prescribed generating functions and sources and integration functions and assumptions on nonlinear symmetries of off-diagonal gravitational and matter field interactions. The physical meaning of respective nonmetric EDM  models are also determined by corresponding types of geometric data and boundary/asymptotic conditions which are considered for a corresponding family of solutions.

\section{Conclusions and Discussion}

\label{sec5} 

Let us give an overview of the main results  and discuss their implications. This is the first paper in geometry and physics devoted to nonmetric generalizations of the theory of spinors and finding solutions of nonmetric Einstein-Dirac-Maxwell (EDM) equations.  It also presents a status report on recent developments and applications  of the anholonomic frame and connection deformation method (AFCDM) for constructing off-diagonal solutions in nonmetric modified gravity theories (MGTs) and nonmetric  geometric flow evolution models. 

\vskip4pt
We have shown that is possible to define a nonmetric theory of  spinors and EDM equations in  self-consisted forms using distortions of connections and canonical nonholonomic variables. Such a result {\bf (the first main one)} allows  the application of the  AFCDM for generating off-diagonal quasi-stationary solutions for physically important systems of  nonlinear PDEs.  It is important for solving the objective {\bf (the second main result)} on elaborating nonmetric  EDM theories  in certain forms which possess general decoupling  and integrability properties.

\vskip4pt
Typical off-diagonal solutions for nonmetric EDM systems do not possess any hyper-surface, holographic and duality properties which would allow thermodynamic interpretation in the frameworks of the Bekenstein--Hawking paradigm. This is an indication that to elaborate a nonmetric EDM generalization of the geometric flow theory {\bf (the third main result)} is of crucial importance because it involves as nonmetric Ricci solitons various quasi-stationary or locally anisotropic cosmological configurations. Moreover, such nonmetric Ricci flow models allow us to define generalized G. Perelman thermodynamic variables which can be computed in explicit form for any class of solutions in metric and nonmetric MGTs. This consists of the {\bf fourth main result} (for models of nonmetric geometric flow evolution of EDM systems) of our work. 

\vskip4pt
Based on examples of quasi-stationary solutions constructed in section \ref{sec4}, we show that our approach to nonmetric geometric flows of EDM systems results in important physical solutions {\bf (the fifth main result)}. Our AFCDM allows us to find nonmetric versions of BH, WH, and BT solutions encoding nontrivial EDM effects.  For this work, we also compute nonmetric locally anisotropic polarization of the  Dirac fermions and additional nonmetric induced currents in generalized Maxwell equations. Here, we emphasize that for such solutions, the nonmetric EDM contributions are encoded into off-diagonal components of metrics, nonlinear connections, generating functions and effective
sources of phase space matter, and effective currents of $U(1)$ fields. 

\vskip4pt
The main results support our {\bf main Hypothesis} (stated in Introduction) on formulating mathematical self-consistent and physically viable models of classical and quantum nonmetric modifications of GR using the formalism of distortion of connections and off-diagonal deformations of metric structures. Finally, we conclude that the results and methods of this work can be used for constructing quasi-stationary and cosmological solutions describing nonmetric Einstein-Yang-Mills-Higgs systems. They have perspectives in elaborating new classes of dark energy and dark matter models and nonmetric geometric and quantum information flow theories.  This is a task for our future partner works. 

\vskip5pt
 {\bf Acknowledgement:}\  The author thanks for the kind collaboration of his co-authors (L.   Bubuianu,  E. Nurlan,  J. O. Seti,   E.  V. Veliev and A. Zamysheva) of partner works \cite{lb24epjc1,lb24epjc2,lb24grg} on classical nonmetric geometric flows,  gravity and MGTs.   The research program on "nonassociative and nonmetric gravity"  was supported by  Prof. Douglas Alexander Singleton, as the  host of a Fulbright visit to the USA, and Prof. Dieter L\"{u}st, as the host of a scientist-at-risk program at CAS LMU Munich, Germany.  This work is performed in the framework of the author's volunteer research activity in Ukraine hosted by Prof. Valery I. Zhdanov from the Taras Shevchenko National University of Kyiv, Astronomical
Observatory.

\end{document}